\pdfoutput=1
 \documentclass[twocolumn,preprint2]{aastex63}
\usepackage[figuresright]{rotating}
\usepackage{tabularx}
\newcommand{\kms}{{\rm km\,s}^{-1}}

\newcommand{\HI}{H\,{\small{I}} }

\newcommand{\mjyb}{{\rm mJy\,beam}^{-1}}

\newcommand{\dotmin}{\rlap.{'}}
\newcommand{\dotdeg}{\rlap.{^\circ}}

\shorttitle{FASHI 21\,cm \HI absorption galaxies}
\shortauthors{Zhang et al.}

\begin{document}

\title{FASHI: An untargeted survey of the 21\,cm \HI absorption galaxies with FAST}

\correspondingauthor{Chuan-Peng Zhang}
\email{cpzhang@nao.cas.cn}

\author[0000-0002-4428-3183]{Chuan-Peng Zhang}
\affiliation{National Astronomical Observatories, Chinese Academy of Sciences, Beijing 100101, China}
\affiliation{Guizhou Radio Astronomical Observatory, Guizhou University, Guiyang 550000, China}

\author{Ming Zhu}
\author{Peng Jiang}
\affiliation{National Astronomical Observatories, Chinese Academy of Sciences, Beijing 100101, China}
\affiliation{Guizhou Radio Astronomical Observatory, Guizhou University, Guiyang 550000, China}

\author{Cheng Cheng}
\affiliation{Chinese Academy of Sciences South America Center for Astronomy, National Astronomical Observatories, CAS, Beijing 100101, China}

\author{Jin-Long Xu}
\author{Nai-Ping Yu}
\author{Xiao-Lan Liu}
\author{Bo Zhang}
\affiliation{National Astronomical Observatories, Chinese Academy of Sciences, Beijing 100101, China}
\affiliation{Guizhou Radio Astronomical Observatory, Guizhou University, Guiyang 550000, China}




\begin{abstract}
The \textbf{F}AST \textbf{A}ll \textbf{S}ky \textbf{H\,{\footnotesize{I}}} survey (FASHI) will cover the entire observable sky ($\sim$22000 square degrees) with the Five-hundred-meter Aperture Spherical radio Telescope (FAST). With the currently released data, we perform an untargeted survey of 21\,cm \HI absorption galaxies at redshift $z\lesssim0.09$ over an area of about 10000 square degrees. We have detected 51 \HI absorbers, including 21 previously known and 30 new ones. The probability of occurrence for the \HI absorbers in all \HI galaxies is 1/1078. The radio flux densities of the FASHI absorbers are mainly concentrated in the range of $S_{\rm 1.4GHz}=10\sim100$ mJy, but also as low as $2.6\pm0.4$\,mJy. We find that the host galaxies of the associated \HI absorbers have relatively high star formation rates, and there is a negative correlation between the H\,I column density and the stellar mass in the host galaxy. Consequently, FAST has significantly improved the capabilities and performance for \HI absorption observations and has provided a true untargeted survey of 21\,cm \HI absorption galaxies for such studies.
\end{abstract}

\keywords{\HI line emission (690), Extragalactic radio sources (508), Radio telescopes (1360), Redshift surveys (1378)}
\section{Introduction} \label{sec:intro}

Studies using the 21\,cm \HI emission line have played a key role in understanding the kinematics, structure, and evolution of galaxies in general \citep[e.g.,][]{Giovanelli2015,Haynes2018,Zhang2024fashi}. However, the 21\,cm \HI absorption line is a more powerful tracer than the \HI emission line on many scales, from the parsec scales near the central black hole to the tens of kpc structures tracing interactions and mergers of galaxies \citep{Morganti2018}. This is because the intensity of an \HI absorption line depends on the intensity of the background continuum source, independent of redshift. Furthermore, \HI absorption observations can be used to study the \HI gas near an active nucleus out to much higher redshifts than is possible using \HI emission, since the strong radio continuum emission is often associated with the central activity \citep[e.g.,][]{Gereb2014,Gereb2015,Darling2011,Curran2016,Curran2018,Aditya2018}. Low redshift \HI absorption line studies are also of great importance. We can study these intriguing cases of spatially resolved \HI absorption in low redshift \HI absorbers. The 21\,cm \HI absorption surveys can also potentially identify primordial atomic gas, infalling or outflowing gas in the vicinity of galaxies or spiral disks. Therefore, \HI absorption line observations are a good way to study the evolution of the \HI properties of galaxies and related questions.

\HI absorption observations have been made with all available radio telescopes, from single dish to interferometer \citep[e.g.,][]{Dickey1986,Gallimore1999,Mahony2013,Zwaan2015,Wu2015,Aditya2018,Su2022,Hu2023,Yu2023,Aditya2024}, but a number of limitations have affected \HI absorption observations in the past \citep{Morganti2018}. One is the limited velocity range that can be covered by an observation while maintaining good spectral resolution. Another major limitation has been the limited coverage and sensitivity at lower frequencies, making it difficult to study high redshift objects. The third is the current lack of an untargeted survey of a very large volume of the Universe. Targeted observations focus primarily on extragalactic radio sources with sufficiently high fluxes. Consequently, they can only yield continuum-flux biased \HI absorber samples, underrepresenting galaxies with weak radio emission, but the weak radio sources can also give rise to noticeable absorption features \citep{Sadler2007}. Untargeted surveys could largely remove the effects of many of the selection biases that plague most studies today and sometimes make it difficult to draw firm conclusions. \citet{Darling2011} conducted an untargeted pilot survey for 21\,cm \HI absorption lines in a 517 deg$^2$ section of the ALFALFA survey at $z<0.058$, but found no new absorption line systems. \citet{Hu2023} presented an untargeted search for the extragalactic 21\,cm \HI absorption lines with the Commensal Radio Astronomy FasT Survey (CRAFTS), but only two new and three known \HI absorbers were detected. The First Large Absorption Survey in \HI \citep[FLASH;][]{Allison2022} is a wide-field survey for \HI absorption which is mainly covering the southern sky in the redshift range $0.42 < z < 1.00$ \citep[e.g.,][]{Su2022,Aditya2024,Yoon2024}. The MeerKAT Absorption Line Survey (MALS) was designed to perform an unbiased census of the \HI and OH absorption lines at $0<z<2$ \citep{Gupta2016}. Therefore, further high-sensitivity, broadband, large-scale, and unbiased \HI absorption surveys are important and needed to overcome the above difficulty.

 \begin{figure*}[htp]
 \centering
 \includegraphics[width=0.99\textwidth, angle=0]{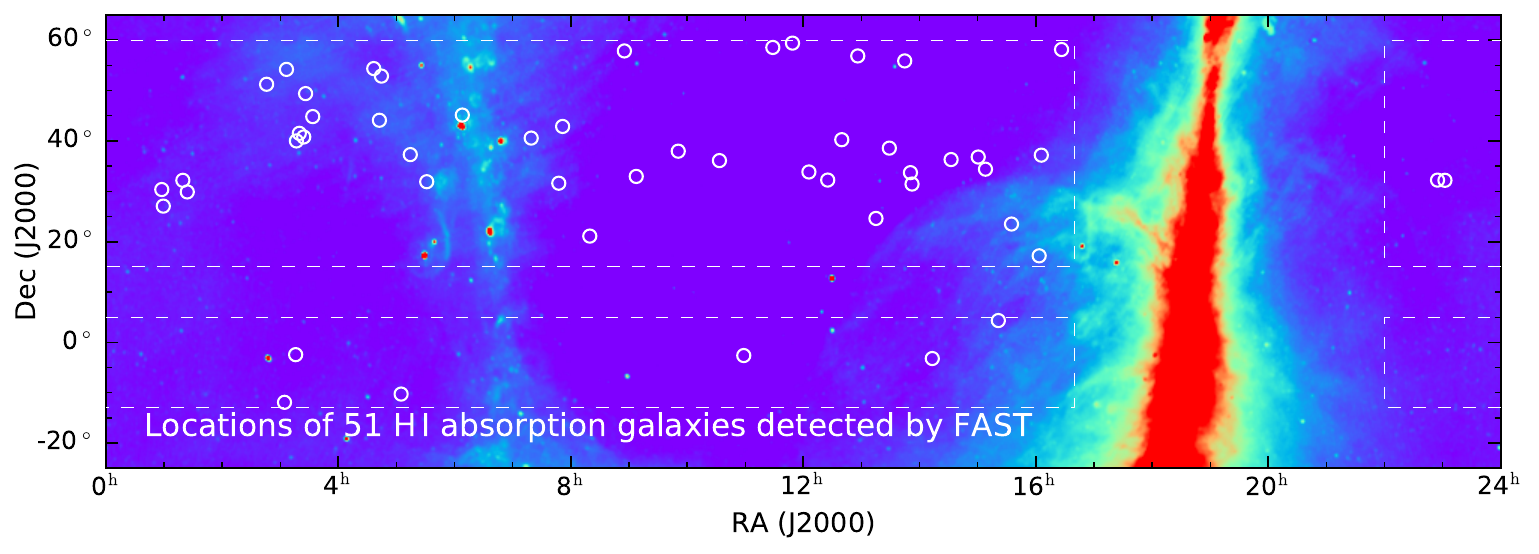}
 \caption{Locations for 51 H\,{\scriptsize{I}} absorption galaxies detected by FASHI over an area of about 10000 square degrees. The circles indicate their positions in the equatorial coordinate system. The background image is the all-sky 21-cm radio continuum of Stockert 25\,m \citep{Reich1982} and Villa Elisa 30\,m \citep{Reich1986,Reich2001}. The white dashed rectangles indicate roughly the area currently covered by FAST. }
 \label{Fig:location}
 \end{figure*}

The Five-hundred-meter Aperture Spherical radio Telescope (FAST) All Sky \HI survey (FASHI) aims to cover the entire sky visible to FAST, between declinations of $-14^\circ$ and $+66^\circ$, within a frequency range of $1.0-1.5$\,GHz \citep{Nan2011,Jiang2019,Jiang2020}. The survey currently has a typical map rms of $\sim$0.76\,$\mjyb$ with a velocity resolution of 6.4\,$\kms$ for \HI line \citep{Zhang2024fashi}, while the Arecibo Legacy Fast ALFA Survey (ALFALFA) has a lower detection sensitivity of $\sim$1.86\,$\mjyb$ after smoothing to a velocity resolution of 10\,$\kms$ \citep{Haynes2018}. This suggests that the FASHI could detect more \HI absorbers than the ALFALFA. Furthermore, the Ultra-Wide Bandwidth (UWB) receiver on FAST can simultaneously cover a frequency range of 500-3300\,MHz, making FAST a powerful instrument for observing the high redshift \HI absorbers in extragalactic objects \citep{Zhang2023uwb}. Therefore, FAST with 19-beam and UWB receivers could be used to carry out such surveys mentioned above.

In this paper we mainly report an untargeted survey of 21\,cm \HI absorption galaxies with FAST over an area of about 10000 square degrees (see Figure\,\ref{Fig:location}). We discovered 51 \HI absorbers, including 21 previously known and 30 new ones at redshift $z\lesssim0.09$\footnote{The H\,I absorbers with $z\gtrsim0.09$ will be released soon in the next work.}. Currently, there are about 200 known \HI absorbers in the literature \citep[e.g.,][]{Morganti2018,Zhangbo2021,Su2022,Hu2023,Yu2023,Aditya2024}. FAST has significantly improved the capabilities and performance for \HI absorption observations compared to previous observations. Section\,\ref{sec:data_reduc} shows the FAST observations, data reduction, and source identification. Section\,\ref{sec:result} presents the characterization of the detected \HI absorbers and lists their parameters. Section\,\ref{sec:discu} discusses the properties of the detected \HI absorbers and the host galaxies. Section\,\ref{sec:summary} is a summary and talks about future work. Throughout this paper, we assume a $\Lambda$CDM cosmology with $H_{0}$ = 75\,$\kms$\,Mpc$^{-1}$, $\Omega_{\rm M} = 0.3$, and $\Omega_{\rm \Lambda} = 0.7$.

\section{Observations and Data Processing}
\label{sec:data_reduc}

\subsection{Observations}

The 21\,cm \HI absorption spectral data come from the FASHI project \citep{Zhang2024fashi}. The FAST 19-beam receiver was employed to efficiently cover the FAST sky. The aperture of FAST is 500\,m, while the effective aperture is approximately 300\,m, resulting in a beam size of $\sim$2.9$'$ at 1.4\,GHz. The full band of 500\,MHz in the spectral line backend has a frequency coverage of 1000 to 1500\,MHz with a total of 64k channels. The corresponding frequency resolution is 7.63\,kHz or 1.6\,$\kms$ at 1.4\,GHz for the raw data. The pointing error is less than or equal to $15''$. To calibrate intensity, a high-noise signal with an amplitude of $\sim$11\,K was injected into the entire FASHI observation for a period of 32 or 64 seconds. The factor of degrees per flux unit (DPFU) is relevant for 19 beams, with a value of $\rm DPFU=13\sim17$\,K\,Jy$^{-1}$ at a frequency of 1400\,MHz, as detailed in \citet{Nan2011,Jiang2019,Jiang2020}. The FASHI project is predominantly performed using the drift scan mode. Throughout the survey, the telescope's azimuth arm was positioned on the meridian at a pre-assigned J2000 declination, with a spacing of $21\dotmin65$ between primary drift centers. The feed array was rotated by $23\dotdeg4$ to enable super-Nyquist sampling of Earth-rotation drift-scan tracks of individual beams in the declination of J2000 coordinates, with sampling rates smaller than half the FWHM.

\subsection{Data reduction}

Currently, we only use the datasets in the frequency range of above 1305.5\,MHz (or $z\lesssim0.09$) due to serious RFI and relatively low sensitivity at low frequency ranges \citep[see][]{Zhang2022rfi}. The FASHI data are reduced using the FAST spectral data reduction pipeline, \texttt{HiFAST} \citep{Jing2024}. The \texttt{HiFAST} pipeline combines data reduction packages, including antenna temperature correction, baseline correction, RFI mitigation, standing wave correction, gridding, flux correction, and fits cube generation. In baseline correction, we utilized the Asymmetrically Reweighted Penalized Least Squares algorithm \citep[\texttt{arPLS};][]{Baek2015} to adjust the spectral baseline. Several frequency periods ($\sim$1\,MHz, $\sim$2\,MHz, and $\sim$0.04\,MHz) of standing waves are inhabited in the FASHI spectral data. The spectral quality can be significantly improved if the standing wave is completely corrected. Standing waves are predominantly fitted and subtracted using the \texttt{SW} package  in the \texttt{HiFAST} pipeline \citep{Jing2024}. The data have been smoothed to a spectral resolution of $\sim$6.4\,$\kms$ per channel and the fits cubes have been gridded to a pixel scale of 1\,arcmin. A heliocentric velocity correction has been applied to the data due to the Doppler effect. A more detailed observation setup and data reduction procedure is presented in the FASHI paper by \citet{Zhang2024fashi}.

\begin{figure*}[htp]
 \centering
 \includegraphics[height=0.24\textwidth, angle=0]{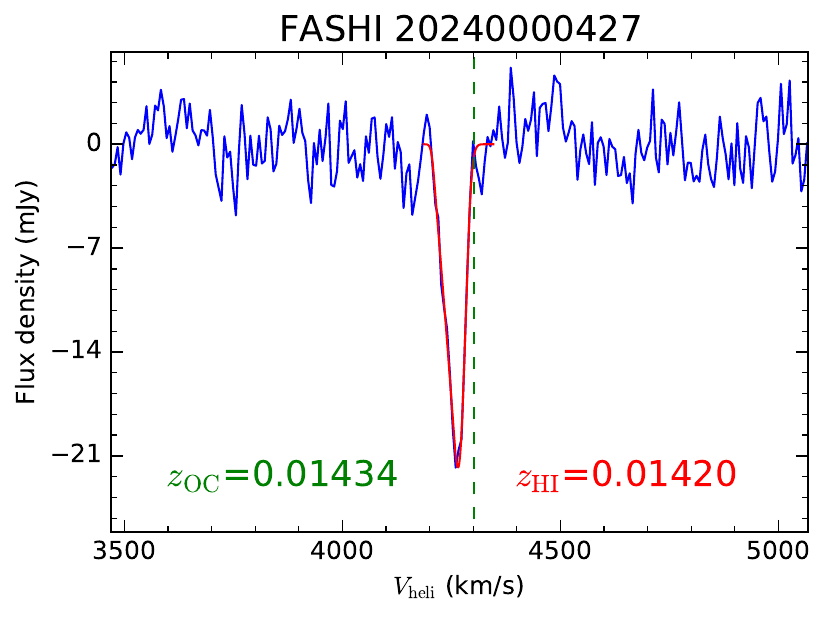}
 \includegraphics[height=0.27\textwidth, angle=0]{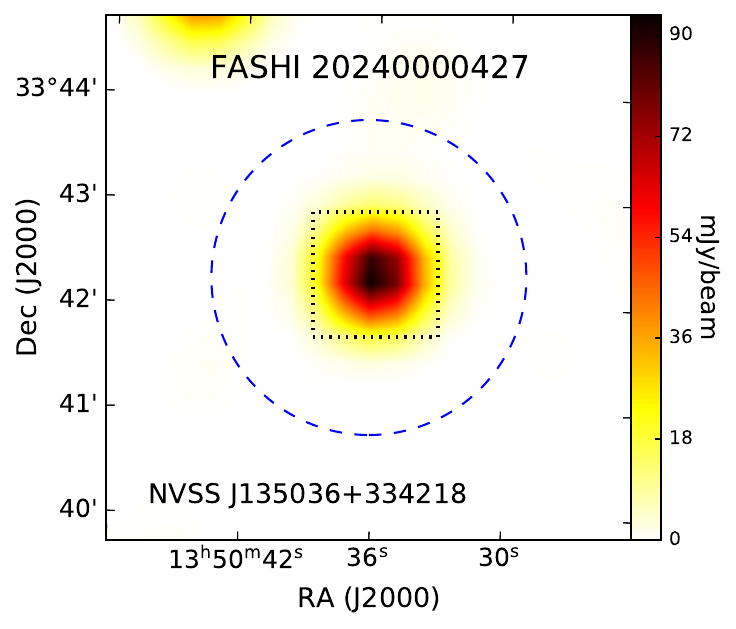}
 \includegraphics[height=0.27\textwidth, angle=0]{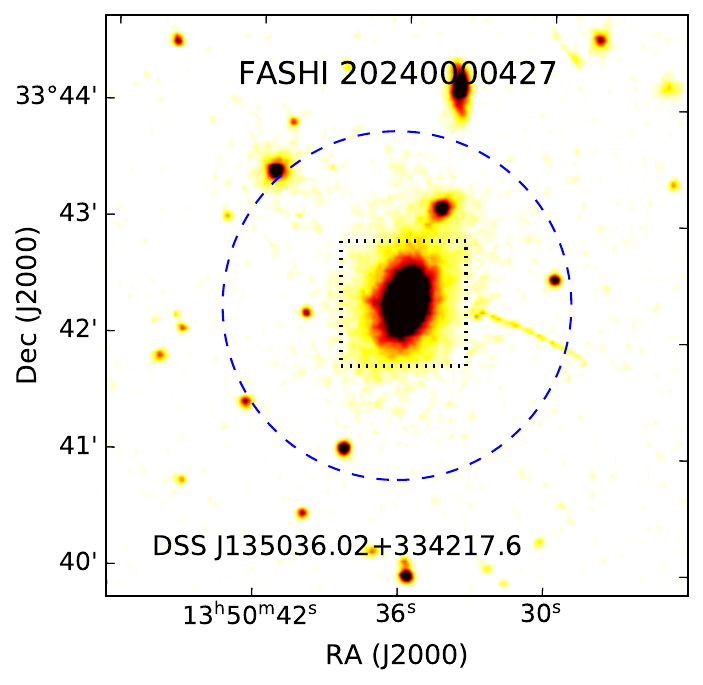}
 \caption{H\,{\scriptsize{I}} absorption galaxy FASHI\,135036.30+334216.5 or ID\,20240000427. \textit{Left}: H\,{\scriptsize{I}} absorption spectrum. The \HI redshift measured by FAST (in red) and an optical spectroscopic redshift (in green) are presented in the panel. The red curve line is the fit to the spectrum, and the green dashed line is the optical spectroscopic redshift. \textit{Middle}: Blue circle indicates the position of \HI source with a beam size of $\sim$2.9$'$. The background shows 1.4\,GHz radio continuum distribution. The bright source within a square is the radio counterpart. \textit{Right}: The background shows DSS image. The bright source within a square is the DSS counterpart. The other 50 H\,{\scriptsize{I}} absorption galaxies are shown in the online version of Figure\,\ref{Fig:FASHI_hi}.}
 \label{Fig:FASHI_hi}
 \end{figure*}

\subsection{Source finding}
\label{sec:extraction}

The process of extracting sources for FASHI is comprised of two steps, involving both automated and interactive procedures. The application utilized for identifying sources within \HI data cubes is designated as the \HI Source Finding Application (\texttt{SoFiA}) in version 2\footnote{\url{https://github.com/SoFiA-Admin/SoFiA-2}}. The \texttt{SoFiA} source finding algorithm involves the smoothing of the data on various spatial and spectral scales specified by the user, along with the measurement of the noise level in each smoothing iteration. Subsequently, a user-defined flux threshold is applied relative to the noise in order to retain all pixels with statistically significant flux density. The fully automated and reliable source finding tool, \texttt{SoFiA}, is crucial for the scientific success of surveys \citep[e.g.,][]{Westmeier2022,Zhang2024fashi}.

In the \texttt{SoFiA} setup, the detection threshold is 4.5$\sigma$ ($\sigma$ means the local noise level within the bounding box of the source in native units of the data cube; see the \texttt{SoFiA} Cookbook); the smoothing kernels are \texttt{kernelsXY = 0, 3, 6} and \texttt{kernelsZ = 0, 3, 7, 15} respectively; in the linking parameters, the minimum size of the sources in the spatial dimension is 5 pixels/channels each in XY and Z space, while the maximum size is 50 pixels in XY space, but not limited in Z space; furthermore, we set the \texttt{reliability.enable} to false, so that artifacts such as RFI and bad baselines occasionally remain in the spectral data. Based on the above setup, we found that the source candidates found by \texttt{SoFiA} contain some false signals. Therefore, a manual interactive source extraction has to be done, judging by the source moment maps, spectral profile, signal-to-noise ratio, and so on.

The \HI absorption source finding procedure is similar to the \HI emission source extraction in FASHI \citep{Zhang2024fashi}. The only difference is to let \texttt{input.invert} be \texttt{true}, so that to search for negative signals, such as absorption lines.

The candidates found by \texttt{SoFiA} include fake sources due to RFI in FAST observation \citep[e.g.,][]{Zhang2022rfi}. A manual source extraction is needed to judge the source moment maps, spectral profile, signal to noise ratio, coordinate, flux density, optical counterparts and radio continuum sources, and then discard fake sources.

As for FASHI ID 20240000003, 20240000024, and 20240000022, the flux densities of the absorption peaks are not below zero, although they exhibit clear absorption characteristics. They thus cannot be extracted using the aforementioned method. In fact, the three sources were extracted from the FASHI emission sources by manual checking \citep{Zhang2024fashi}.

In the end, we discovered and conformed 51 \HI absorber, including 21 previously known and 30 new ones (see Figures\,\ref{Fig:location} and \ref{Fig:FASHI_hi}). Detailed information on these sources is presented in Appendix\,\ref{sec:individual}.

\subsection{Line fitting}
\label{sec:fit}

Gaussian fitting has been widely employed to derive absorption line properties, such as the width of the profile, and to ascertain the presence of multiple components \citep[e.g.,][]{Zhangbo2021,Hu2023}. When multiple peaked profiles occur, as is the case in our absorption sample, Gaussian fitting methods have the disadvantage of necessitating a prior assumption regarding the physical conditions of the gas, namely the number of components to be fitted. The busy-function fitting method, as described in \citet{Westmeier2014}, was employed in this study. This method has been successfully utilized in \citet{Gereb2015}.

In the context of line fitting, if there is only one absorption line present, it is possible to utilise the busy-function fitting method to fit the absorption line directly. In the case that an emission line exists, but the absorption line is not obscured by a nearby emission line, the emission line will be masked first, and then the absorption line will be fitted (e.g., FASHI 20240000196, 20240000264, 20240000051, 20240000013). In the case that the absorption line is inaccessible to the emission line, the absorption line is masked, and then the emission line is fitted. Subsequently, the original line is employed to subtract the emission, after which the absorption line can be fitted (e.g., FASHI 20240000005, 20240000472, 20240000016).

\section{Detection Results and Analysis}
\label{sec:result}

\subsection{Locations of FASHI \HI absorbers}

The FASHI data has already covered over an area of about 10000 square degrees within $0^h\lesssim{\rm RA}\lesssim24^h$,  $-14^\circ\lesssim{\rm Dec}\lesssim66^\circ$ with a typical spectral detection sensitivity of $\sim$1.50\,mJy for a velocity resolution of $\sim$6.4\,$\kms$ at 1.4\,GHz. Because the FASHI project can only operate in schedule-filler mode, parts of the above sky areas are not fully covered. In Figure\,\ref{Fig:location} we show the locations of 51 \HI absorption sources already detected by FAST at $z\lesssim0.09$. This suggests that the current detection rate of the \HI absorbers is $5.1\times10^{-3}$ sources per square degree, while the FASHI \HI emission sources have a detection rate of about 5.5 sources per square degree \citep{Zhang2024fashi}. Based on this detection rate, we estimate that the probability of occurrence for the \HI absorbers in all \HI galaxies is 1/1078\footnote{In this work, 51 \HI absorbers are detected in an area of about 10000 square degrees. The probability of occurrence for the \HI absorbers in all HI galaxies is $51/(5.5\times10000)\approx1/1078$.}. However, considering that the survey area does not have uniform sensitivity\footnote{The sensitivity map is shown in \citet{Zhang2024fashi}}, the measured detection rate is approximate. At the current detection sensitivity, the final number of \HI absorbers detected by FASHI would be about 100 if FAST were able to cover 22000 square degrees at $z\lesssim0.09$.

The ALFALFA survey used the seven-beam Arecibo $L$-band Feed Array to perform a drift-scan survey, covering an area of about 7000 square degrees over a redshift range of $z<0.06$ and cataloging 31502 \HI emission galaxies. \citet{Wu2015} predicted that the whole ALFALFA could eventually produce about 25 \HI absorbers. Their probability of occurrence is about 1/1260 in all \HI galaxies. This number is slightly lower than that found by FASHI.

Additionally, from Figure\,\ref{Fig:location} we can see that the sources are mostly located at $30^\circ\lesssim{\rm Dec}\lesssim60^\circ$. This is because the current FASHI observations focus mainly on $30^\circ\lesssim{\rm Dec}\lesssim60^\circ$, where the data rms is much better than the region of $-15^\circ\lesssim{\rm Dec}\lesssim30^\circ$. In the near future FASHI will spend more time observing the region of $-15^\circ\lesssim{\rm Dec}\lesssim30^\circ$ and it is expected that the number of detections over that area will increase. With increasing detection sensitivity and area, the FASHI \HI absorber sample could be expected to be much larger than the current number.

\begin{table*}[htp]
\caption{Coordinates and counterparts of FASHI \HI absorption galaxies}
\label{tab:name}
\vskip 0pt
\centering \footnotesize 
\renewcommand{\arraystretch}{0.98}
\setlength{\tabcolsep}{1.0mm}{
\begin{tabular}{ccccccc}
\hline \hline
[1]  &  [2]  & [3]   & [4]  & [5] & [6]  & [7]   \\ 
FASHI ID & FASHI & RA & Dec & NVSS & OC & Galaxy \\
& J2000 & deg & deg & J2000 & J2000     \\
\hline
\bf{20240000480} & J030428.71-115315.1 &46.1196 &-11.8875 & J030429-115422 & J030429.63-115422.4 & IRAS03021-1205 \\
\bf{20240000482} & J050451.78-101501.7 &76.2157 &-10.2505 & J050453-101451 & J050453.06-101452.5 & IRAS05025-1018 \\
20240000405 & J141310.09-031140.8 &213.2921 &-3.1947 & J141314-031227 & J141314.84-031227.2 & NGC5506 \\
\bf{20240000476} & J105833.00-023559.1 &164.6375 &-2.5997 & J105835-023552 & J105835.37-023551.9 & PGC156086 \\
20240000399 & J031558.84-022532.1 &48.9952 &-2.4256 & J031600-022539 & J031600.78-022538.2 & NGC1266 \\
\bf{20240000093} & J152121.51+042039.1 &230.3396 &4.3442 & J152122+042033 & J152122.54+042030.1 & PGC054794 \\
20240000483 & J160329.87+171133.9 &240.8745 &17.1927 & J160332+171158 & J160332.09+171155.3 & NGC6034 \\
20240000003 & J081937.14+210636.1 &124.9048 &21.1100 & J081937+210653 & J081937.95+210651.6 & UGC04332 \\
20240000408 & J153455.59+233013.2 &233.7316 &23.5037 & J153457+233011 & J153457.38+233012.0 & IC1127 IC4553 Arp220 \\
20240000409 & J131501.12+243651.4 &198.7546 &24.6143 & J131503+243707 & J131503.52+243707.7 & IC0860 \\
20240000116 & J005922.61+270337.1 &14.8442 &27.0603 & J005924+270332 & J005924.43+270332.3 & IC0064 UGC00613 \\
\bf{20240000122} & J012406.73+295238.5 &21.0281 &29.8774 & J012407+295246 & J012407.50+295244.0 & LEDA1881594 \\
20240060788 & J005747.27+302108.9 &14.4469 &30.3525 & J005748+302114 & J005748.89+302108.8 & NGC0315 \\
20240000140 & J135215.88+312631.6 &208.0662 &31.4421 & J135217+312646 & J135217.82+312646.6 & UGC08782 3C293 \\
\bf{20240000155} & J074735.82+313625.2 &116.8992 &31.6070 & J074742+313757 & J074736.14+313659.6 & PGC1955807 \\
20240000014 & J053117.26+315418.4 &82.8219 &31.9051 & J053118+315412 & J053118.79+315412.3 & NVSSJ053118+315412 \\
20240000159 & J011932.90+321101.9 &19.8871 &32.1839 & J011935+321050 & J011935.00+321050.1 & 4C31.04 \\
\bf{20240000027} & J230222.13+321111.5 &345.5922 &32.1865 & J230221+321125 & J230222.47+321120.6 & PGC070309 \\
\bf{20240000026} & J225442.56+321324.2 &343.6773 &32.2234 & J225445+321248 & J225444.97+321249.2 & PGC069981 \\
\bf{20240000129} & J122512.74+321353.4 &186.3031 &32.2315 & J122513+321401 & J122513.11+321401.6 & PGC1987158 \\
\bf{20240000141} & J090733.90+325729.4 &136.8912 &32.9582 & J090734+325722 & J090734.87+325723.1 & PGC165523 \\
20240000427 & J135036.30+334216.5 &207.6513 &33.7046 & J135036+334218 & J135036.02+334217.6 & NGC5318 \\
\bf{20240000196} & J120547.43+335016.9 &181.4476 &33.8380 & J120547+335024 & J120547.67+335022.1 & PGC083530 \\
20240000174 & J150803.69+342339.2 &227.0154 &34.3942 & J150805+342323 & J150805.08+342315.9 & PGC054033 IRAS15060+3434 \\
\bf{20240000243} & J103327.36+360550.9 &158.3640 &36.0975 & J103328+360606 & J103328.56+360559.5 & PGC2074966 \\
20240000024 & J143235.35+361806.7 &218.1473 &36.3019 & J143239+361808 & J143239.90+361808.2 & NGC5675 \\
20240000037 & J150033.79+364829.3 &225.1408 &36.8081 & J150034+364844 & J150034.58+364845.0 & 2MASXJ150034+364845 \\
\bf{20240000039} & J160542.81+371049.2 &241.4284 &37.1803 & J160543+371046 & J160543.21+371044.7 & PGC084723 \\
\bf{20240000036} & J051419.90+371721.2 &78.5829 &37.2892 & J051421+371714 & J051421.10+371714.0 & NVSSJ051421+371714 \\
\bf{20240000500} & J095055.65+375800.8 &147.7319 &37.9669 & J095058+375758 & J095058.71+375758.7 & PGC090952 \\
\bf{20240000022} & J132849.73+383419.9 &202.2072 &38.5722 & J132852+383438 & J132852.11+383439.5 & UGC08471 \\
\bf{20240000443} & J031646.12+395955.4 &49.1921 &39.9987 & J031646+400011 & J031646.70+400013.0 & UGC02625 IC311 \\
\bf{20240000264} & J123943.52+401459.7 &189.9313 &40.2499 & J123945+401450 & J123945.16+401448.9 & PGC089547 \\
\bf{20240000051} & J071908.19+403355.0 &109.7841 &40.5653 & J071908+403343 & J071908.66+403341.5 & LAMOSTJ071908.70+403341.7 \\
\bf{20240000445} & J032421.15+404714.8 &51.0881 &40.7874 & J032422+404719 & J032422.86+404719.2 & IRAS03211+4036 UGC2715 \\
20240000501 & J031946.81+413034.4 &49.9450 &41.5096 & J031948+413042 & J031948.26+413041.4 & 3C84 NGC1275 \\
\bf{20240000013} & J075134.30+425242.1 &117.8929 &42.8784 & J075135+425249 & J075135.64+425248.3 & UGC04056 \\
\bf{20240000454} & J044226.82+440615.0 &70.6117 &44.1042 & J044228+440638 & J044228.49+440638.5 & NVSSJ044228+440638 \\
\bf{20240000059} & J033337.92+445011.9 &53.4080 &44.8366 & J033337+444941 & J033337.90+444949.0 & NVSSJ033337+444941 \\
\bf{20240000507} & J060801.90+450814.2 &92.0079 &45.1373 & J060803+450825 & J060803.44+450825.0 & LEDA5061018 \\
\bf{20240000078} & J032614.32+492357.3 &51.5597 &49.3993 & J032615+492357 & J032616.00+492359.0 & NVSSJ032615+492357 \\
\bf{20240000512} & J024600.14+511458.1 &41.5006 &51.2495 & J024600+511515 & J024600.54+511514.0 & IRAS02425+5102 \\
\bf{20240000007} & J044433.30+525213.5 &71.1388 &52.8704 & J044434+525224 & J044434.76+525224.8 & 2MASXJ04443470+5252243 \\
\bf{20240000083} & J030630.87+541208.7 &46.6286 &54.2024 & J030632+541158 & J030632.74+541158.5 & NVSSJ030632+541158 \\
\bf{20240000005} & J043636.25+542201.5 &69.1510 &54.3671 & J043638+542159 & J043638.61+542159.3 & 2MASXJ04363857+5421598 \\
20240000023 & J134440.84+555306.2 &206.1702 &55.8850 & J134442+555313 & J134442.09+555312.8 & UGC08696 Mrk273 \\
20240000513 & J125614.58+565222.9 &194.0608 &56.8730 & J125614+565223 & J125614.17+565225.1 & UGC08058 Mrk231 \\
20240000515 & J085519.79+575155.5 &133.8324 &57.8654 & J085521+575143 & J085519.04+575140.7 & SDSSJ085519.04+575140.7 \\
\bf{20240000472} & J162636.67+580849.8 &246.6528 &58.1472 & J162637+580911 & J162637.74+580904.3 & SDSSJ162637.74+580904.2 \\
20240000016 & J112832.57+583323.0 &172.1357 &58.5564 & J112832+583346 & J112831.99+583346.8 & NGC3690 \\
20240000471 & J114852.43+592501.9 &177.2185 &59.4172 & J114850+592457 & J114850.36+592456.4 & NGC3894 \\
\hline
\end{tabular}
}
\begin{flushleft}
\textbf{Notes.} The FASHI IDs in bold indicate newly discovered absorber.  
\\
\end{flushleft}
\end{table*}

\begin{table*}[htp]
\caption{Physical parameters of FASHI \HI absorption galaxies}
\label{tab:Physics}
\vskip 0pt
\centering \footnotesize 
\renewcommand{\arraystretch}{0.98}
\setlength{\tabcolsep}{1.2mm}{
\begin{tabular}{cccccccccc}
\hline \hline
[1]  &  [2]  & [3]   & [4]  & [5] & [6]  & [7]  & [8] & [9] & [10] \\ 
FASHI ID & $V_{\rm heli}$ & $z_{\rm HI}$ & $z_{\rm OC}$   & $W_{50}$ & $S_{\rm HI}$ & $S_{\rm 1.4GHz}$  & ${\rm log}L_{\rm 1.4GHz}$  & $\tau_{\rm HI}$ & $N_{\rm HI}$ \\
& $\kms$      &     &  &   $\kms$  & mJy & mJy & W\,Hz$^{-1}$ & & 10$^{21}$cm$^{-2}$    \\
\hline
\bf{20240000480} &6960.0 &0.02322 &0.03077 &$ 48.7 \pm  4.8 $&$-19.03 \pm  2.56 $&$  29.4 \pm   1.3 $&$ 22.75 $&$ 1.042 \pm 0.260 $&$  9.26 \pm  2.31 $\\
\bf{20240000482} &11837.5 &0.03949 &0.04032 &$176.2 \pm  7.8 $&$-34.20 \pm  1.19 $&$1482.4 \pm  44.5 $&$ 24.69 $&$ 0.023 \pm 0.001 $&$  0.75 \pm  0.03 $\\
20240000405 &1809.4 &0.00604 &0.00607 &$ 72.3 \pm  5.8 $&$-32.97 \pm  1.96 $&$ 338.8 \pm  10.2 $&$ 22.38 $&$ 0.102 \pm 0.007 $&$  1.35 \pm  0.09 $\\
\bf{20240000476} &10888.2 &0.03632 &0.03622 &$115.9 \pm  9.0 $&$-16.25 \pm  1.33 $&$ 205.4 \pm   6.2 $&$ 23.74 $&$ 0.082 \pm 0.007 $&$  1.74 \pm  0.16 $\\
20240000399 &2172.1 &0.00725 &0.00732 &$ 88.6 \pm  8.3 $&$ -8.33 \pm  0.60 $&$ 115.6 \pm   3.5 $&$ 22.08 $&$ 0.075 \pm 0.006 $&$  1.21 \pm  0.10 $\\
\bf{20240000093} &15660.3 &0.05224 &0.05216 &$ 25.5 \pm  3.6 $&$-20.20 \pm  1.88 $&$ 407.3 \pm  13.4 $&$ 24.36 $&$ 0.051 \pm 0.005 $&$  0.24 \pm  0.02 $\\
20240000483 &10224.0 &0.03410 &0.03402 &$ 25.4 \pm  4.3 $&$-28.35 \pm  2.21 $&$ 456.0 \pm  15.2 $&$ 24.03 $&$ 0.064 \pm 0.006 $&$  0.30 \pm  0.03 $\\
20240000003 &5444.8 &0.01816 &0.01842 &$158.1 \pm  9.6 $&$ -5.02 \pm  0.34 $&$  32.9 \pm   1.4 $&$ 22.34 $&$ 0.166 \pm 0.014 $&$  4.77 \pm  0.41 $\\
20240000408 &5435.7 &0.01813 &0.01810 &$334.9 \pm  4.2 $&$-50.51 \pm  0.66 $&$ 326.3 \pm   9.8 $&$ 23.32 $&$ 0.168 \pm 0.006 $&$ 10.27 \pm  0.37 $\\
20240000409 &3903.1 &0.01302 &0.01291 &$116.1 \pm 10.6 $&$-11.67 \pm  0.76 $&$  30.8 \pm   1.0 $&$ 22.00 $&$ 0.476 \pm 0.045 $&$ 10.08 \pm  0.94 $\\
20240000116 &13960.5 &0.04657 &0.04583 &$ 25.8 \pm  1.8 $&$-70.19 \pm  3.33 $&$ 112.6 \pm   4.0 $&$ 23.69 $&$ 0.977 \pm 0.098 $&$  4.59 \pm  0.46 $\\
\bf{20240000122} &16656.6 &0.05556 & &$129.0 \pm 10.0 $&$-11.05 \pm  0.64 $&$  27.1 \pm   0.9 $&$ 23.24 $&$ 0.524 \pm 0.046 $&$ 12.32 \pm  1.09 $\\
20240060788 &5412.4 &0.01805 &0.01746 &$ 10.2 \pm  1.6 $&$-114.01 \pm 24.82 $&$ 772.1 \pm  25.3 $&$ 23.67 $&$ 0.160 \pm 0.038 $&$  0.30 \pm  0.07 $\\
20240000140 &13445.5 &0.04485 &0.04519 &$ 57.8 \pm  2.1 $&$-257.61 \pm  4.20 $&$4085.4 \pm 144.0 $&$ 25.23 $&$ 0.065 \pm 0.003 $&$  0.69 \pm  0.03 $\\
\bf{20240000155} &17767.5 &0.05927 &0.06055 &$381.4 \pm 10.9 $&$ -6.61 \pm  0.27 $&$   7.3 \pm   0.5 $&$ 22.75 $&$ 2.352 \pm 0.761 $&$163.56 \pm 52.93 $\\
20240000014 &19953.8 &0.06656 &0.06120 &$ 87.0 \pm  4.1 $&$-22.58 \pm  1.38 $&$  39.8 \pm   1.3 $&$ 23.50 $&$ 0.838 \pm 0.091 $&$ 13.29 \pm  1.44 $\\
20240000159 &18153.2 &0.06055 &0.05996 &$ 12.9 \pm  1.8 $&$-68.26 \pm  4.13 $&$2635.2 \pm  79.1 $&$ 25.30 $&$ 0.026 \pm 0.002 $&$  0.06 \pm  0.00 $\\
\bf{20240000027} &6411.2 &0.02139 &0.02128 &$145.2 \pm 17.6 $&$ -6.78 \pm  1.63 $&$   2.6 \pm   0.4 $&$ 21.37 $&$ 2.303 \pm 1.000 $&$ 60.94 \pm 26.47 $\\
\bf{20240000026} &6252.9 &0.02086 &0.02169 &$ 58.0 \pm  9.1 $&$ -9.56 \pm  0.98 $&$  32.0 \pm   1.0 $&$ 22.48 $&$ 0.355 \pm 0.046 $&$  3.75 \pm  0.48 $\\
\bf{20240000129} &17951.4 &0.05988 &0.05923 &$ 61.3 \pm  8.7 $&$ -5.50 \pm  0.47 $&$  49.3 \pm   1.5 $&$ 23.56 $&$ 0.118 \pm 0.011 $&$  1.32 \pm  0.13 $\\
\bf{20240000141} &14732.8 &0.04914 &0.04906 &$ 78.3 \pm  4.6 $&$-23.02 \pm  0.92 $&$  46.8 \pm   1.5 $&$ 23.37 $&$ 0.677 \pm 0.050 $&$  9.67 \pm  0.71 $\\
20240000427 &4256.6 &0.01420 &0.01434 &$ 48.6 \pm  4.8 $&$-21.78 \pm  1.16 $&$ 101.1 \pm   3.1 $&$ 22.61 $&$ 0.243 \pm 0.017 $&$  2.15 \pm  0.15 $\\
\bf{20240000196} &15762.2 &0.05258 &0.05388 &$ 86.6 \pm 12.6 $&$ -5.21 \pm  0.49 $&$   8.5 \pm   0.5 $&$ 22.71 $&$ 0.949 \pm 0.175 $&$ 14.98 \pm  2.77 $\\
20240000174 &13484.4 &0.04498 &0.04503 &$117.7 \pm  3.2 $&$-52.47 \pm  1.24 $&$ 133.1 \pm   4.0 $&$ 23.74 $&$ 0.501 \pm 0.025 $&$ 10.76 \pm  0.53 $\\
\bf{20240000243} &17503.2 &0.05838 &0.05876 &$ 18.8 \pm  2.7 $&$-13.73 \pm  1.88 $&$ 105.4 \pm   3.2 $&$ 23.88 $&$ 0.140 \pm 0.021 $&$  0.48 \pm  0.07 $\\
20240000024 &3947.6 &0.01317 &0.01325 &$ 15.9 \pm  0.0 $&$ -9.90 \pm  0.00 $&$ 118.8 \pm   3.6 $&$ 22.61 $&$ 0.087 \pm 0.003 $&$  0.25 \pm  0.01 $\\
20240000037 &19850.7 &0.06621 &0.06605 &$ 61.6 \pm  3.9 $&$-16.20 \pm  2.01 $&$  57.3 \pm   1.8 $&$ 23.72 $&$ 0.332 \pm 0.051 $&$  3.74 \pm  0.57 $\\
\bf{20240000039} &19906.3 &0.06640 &0.06649 &$ 39.6 \pm  0.0 $&$ -8.68 \pm  1.47 $&$  28.4 \pm   1.2 $&$ 23.42 $&$ 0.365 \pm 0.077 $&$  2.64 \pm  0.56 $\\
\bf{20240000036} &22626.6 &0.07547 & &$ 19.2 \pm  2.2 $&$-17.23 \pm  1.80 $&$ 143.3 \pm   4.3 $&$ 24.24 $&$ 0.128 \pm 0.015 $&$  0.45 \pm  0.05 $\\
\bf{20240000500} &12199.6 &0.04069 &0.04053 &$ 31.8 \pm  1.8 $&$-50.84 \pm  1.99 $&$  66.4 \pm   2.0 $&$ 23.35 $&$ 1.451 \pm 0.161 $&$  8.41 \pm  0.94 $\\
\bf{20240000022} &7837.8 &0.02614 &0.02649 &$ 25.7 \pm  6.0 $&$-11.34 \pm  1.74 $&$  11.1 \pm   1.3 $&$ 22.19 $&$ 2.303 \pm 1.000 $&$ 10.78 \pm  4.68 $\\
\bf{20240000443} &4276.0 &0.01426 &0.01419 &$ 25.4 \pm  4.4 $&$-11.05 \pm  1.73 $&$  35.2 \pm   1.4 $&$ 22.14 $&$ 0.377 \pm 0.074 $&$  1.74 \pm  0.34 $\\
\bf{20240000264} &15751.0 &0.05254 &0.05386 &$294.1 \pm  0.0 $&$ -1.32 \pm  0.09 $&$  15.2 \pm   0.6 $&$ 22.96 $&$ 0.091 \pm 0.007 $&$  4.87 \pm  0.40 $\\
\bf{20240000051} &19848.0 &0.06621 &0.06593 &$101.6 \pm  3.7 $&$ -8.03 \pm  0.61 $&$  16.9 \pm   0.6 $&$ 23.19 $&$ 0.645 \pm 0.076 $&$ 11.95 \pm  1.40 $\\
\bf{20240000445} &3842.8 &0.01282 &0.01299 &$ 37.3 \pm  8.3 $&$-12.28 \pm  1.66 $&$  55.5 \pm   1.7 $&$ 22.26 $&$ 0.250 \pm 0.039 $&$  1.70 \pm  0.27 $\\
20240000501 &8115.0 &0.02707 &0.01754 &$ 10.9 \pm  2.4 $&$-2196.30 \pm 10.16 $&$22829.2 \pm 684.9 $&$ 25.14 $&$ 0.101 \pm 0.003 $&$  0.20 \pm  0.01 $\\
\bf{20240000013} &9223.3 &0.03077 &0.03203 &$127.9 \pm 17.1 $&$ -5.27 \pm  0.53 $&$   2.7 \pm   0.4 $&$ 21.75 $&$ 2.303 \pm 1.000 $&$ 53.67 \pm 23.31 $\\
\bf{20240000454} &5362.5 &0.01789 & &$179.3 \pm  8.7 $&$-23.78 \pm  0.92 $&$  54.9 \pm   1.7 $&$ 22.54 $&$ 0.568 \pm 0.038 $&$ 18.56 \pm  1.24 $\\
\bf{20240000059} &19841.4 &0.06618 & &$ 26.6 \pm  6.7 $&$ -6.61 \pm  1.29 $&$  10.3 \pm   0.5 $&$ 22.98 $&$ 1.027 \pm 0.359 $&$  4.98 \pm  1.74 $\\
\bf{20240000507} &10214.9 &0.03407 &0.03566 &$ 36.6 \pm  2.3 $&$-25.54 \pm  1.11 $&$  56.2 \pm   1.7 $&$ 23.16 $&$ 0.606 \pm 0.044 $&$  4.04 \pm  0.29 $\\
\bf{20240000078} &23584.3 &0.07867 & &$167.3 \pm  2.7 $&$-31.48 \pm  2.20 $&$  64.0 \pm   2.0 $&$ 23.93 $&$ 0.677 \pm 0.074 $&$ 20.64 \pm  2.26 $\\
\bf{20240000512} &10463.7 &0.03490 &0.03476 &$ 73.9 \pm  1.5 $&$-157.22 \pm  2.33 $&$ 223.5 \pm   6.7 $&$ 23.74 $&$ 1.216 \pm 0.079 $&$ 16.37 \pm  1.07 $\\
\bf{20240000007} &9943.1 &0.03317 &0.03273 &$ 95.5 \pm  6.5 $&$-12.66 \pm  1.09 $&$  77.2 \pm   2.4 $&$ 23.22 $&$ 0.179 \pm 0.018 $&$  3.12 \pm  0.31 $\\
\bf{20240000083} &19527.9 &0.06514 & &$ 35.8 \pm  4.1 $&$ -9.26 \pm  0.80 $&$  55.3 \pm   1.7 $&$ 23.70 $&$ 0.183 \pm 0.018 $&$  1.19 \pm  0.12 $\\
\bf{20240000005} &5647.5 &0.01884 & &$ 37.9 \pm  3.1 $&$-10.56 \pm  1.20 $&$  58.7 \pm   1.8 $&$ 22.61 $&$ 0.198 \pm 0.026 $&$  1.37 \pm  0.18 $\\
20240000023 &11288.1 &0.03765 &0.03732 &$521.8 \pm 11.9 $&$-12.57 \pm  0.58 $&$ 144.7 \pm   5.1 $&$ 23.61 $&$ 0.091 \pm 0.006 $&$  8.64 \pm  0.52 $\\
20240000513 &12650.1 &0.04220 &0.04176 &$188.0 \pm 11.5 $&$-16.77 \pm  0.73 $&$ 308.9 \pm  12.1 $&$ 24.04 $&$ 0.056 \pm 0.003 $&$  1.91 \pm  0.11 $\\
20240000515 &7737.9 &0.02581 &0.02601 &$ 10.4 \pm  0.0 $&$-101.22 \pm 11.53 $&$ 651.6 \pm  19.6 $&$ 23.94 $&$ 0.169 \pm 0.022 $&$  0.32 \pm  0.04 $\\
\bf{20240000472} &4836.8 &0.01613 & &$ 26.0 \pm  2.5 $&$-60.37 \pm  3.88 $&$ 532.9 \pm  18.3 $&$ 23.44 $&$ 0.120 \pm 0.009 $&$  0.57 \pm  0.04 $\\
20240000016 &3140.3 &0.01047 &0.01044 &$214.6 \pm  1.4 $&$-69.90 \pm  0.78 $&$ 677.1 \pm  25.4 $&$ 23.16 $&$ 0.109 \pm 0.005 $&$  4.26 \pm  0.18 $\\
20240000471 &3292.4 &0.01098 &0.01081 &$158.0 \pm  8.5 $&$-24.73 \pm  1.19 $&$ 481.4 \pm  14.4 $&$ 23.04 $&$ 0.053 \pm 0.003 $&$  1.52 \pm  0.09 $\\
\hline
\end{tabular}
}
\begin{flushleft}
\textbf{Notes.} The FASHI IDs in bold are first discovered in this work. For FASHI 20240000013, 20240000022, and 20240000027, $\mid S_{\rm HI}\mid$ is larger than $S_{\rm 1.4GHz}$, we assume $S_{\rm HI}/S_{\rm 1.4GHz}=-0.9$ for $\tau_{\rm HI}$ estimation. Perhaps the line intensity measurements have a large uncertainty due to the mixing between the emission and absorption lines. Further confirmation, such as high spatial resolution observations, is needed to explain this.
\\
\end{flushleft}
\end{table*}

\begin{table*}
\caption{\textbf{Cross-matched sources between the absorbers and GSWLC.}}
\label{tab:GSWLC}
\centering 
\setlength{\tabcolsep}{1.3mm}{
\begin{tabular}{cccccccccccccc}
\hline \hline
[1]  &  [2]  & [3]   & [4]  & [5] & [6]  & [7]  & [8] & [9] & [10] & [11]  \\   
FASHI ID & OBJID & RA & Dec & $z_{\odot}$ & log($M_{\star})$ & log($\rm SFR_{SED})$ & $A_{\rm V}$  & $g$ & $r$ & $z$   \\
B1950 &    & deg & deg &   & $\rm M_{\odot}$ & $\rm M_{\odot}\,yr^{-1}$ & mag & mag &  mag &  mag \\
\hline
20240000093 &1237662266464600297 &230.344 &4.342 &0.0522 &11.55 $\pm$ 0.00 &-0.52 $\pm$0.14 &0.15 $\pm$ 0.03 &14.65 &13.72 &13.01 \\
20240000483 &1237665440998031492 &240.884 &17.199 &0.0340 &11.20 $\pm$ 0.01 &0.33 $\pm$0.05 &0.49 $\pm$ 0.01 &13.95 &13.08 &12.40 \\
20240000003 &1237664836462182621 &124.908 &21.114 &0.0185 &10.82 $\pm$ 0.04 &0.65 $\pm$0.03 &0.76 $\pm$ 0.01 &14.27 &13.44 &12.76 \\
20240000408 &1237665537075511390 &233.738 &23.504 &0.0184 &10.87 $\pm$ 0.02 &0.37 $\pm$0.08 &0.69 $\pm$ 0.07 &13.59 &12.93 &12.40 \\
20240000140 &1237665331471515679 &208.075 &31.446 &0.0452 &11.33 $\pm$ 0.01 &0.85 $\pm$0.07 &0.68 $\pm$ 0.08 &14.62 &13.74 &13.04 \\
20240000155 &1237657595683406347 &116.910 &31.633 &0.0600 &9.54 $\pm$ 0.07 &-0.09 $\pm$0.08 &0.16 $\pm$ 0.05 &16.16 &15.50 &15.01 \\
20240000129 &1237665329852842117 &186.305 &32.234 &0.0592 &11.59 $\pm$ 0.01 &0.17 $\pm$0.27 &0.43 $\pm$ 0.10 &14.85 &13.97 &13.26 \\
20240000141 &1237660763766849674 &136.895 &32.956 &0.0491 &11.05 $\pm$ 0.03 &0.67 $\pm$0.11 &0.63 $\pm$ 0.08 &15.47 &14.61 &14.21 \\
20240000427 &1237665025986986033 &207.650 &33.705 &0.0143 &10.90 $\pm$ 0.01 &-0.42 $\pm$0.23 &0.49 $\pm$ 0.03 &13.04 &12.17 &11.89 \\
20240000174 &1237662306730967257 &227.039 &34.422 &0.0453 &9.19 $\pm$ 0.08 &-0.38 $\pm$0.08 &0.17 $\pm$ 0.05 &16.35 &15.55 &16.00 \\
20240000243 &1237662224058220698 &158.369 &36.100 &0.0588 &11.17 $\pm$ 0.01 &0.33 $\pm$0.12 &0.12 $\pm$ 0.04 &15.76 &14.96 &14.42 \\
20240000024 &1237662305117143073 &218.166 &36.302 &0.0132 &10.98 $\pm$ 0.02 &-0.18 $\pm$0.31 &0.56 $\pm$ 0.10 &13.02 &12.22 &11.77 \\
20240000037 &1237661850409304180 &225.144 &36.813 &0.0661 &10.97 $\pm$ 0.01 &-0.27 $\pm$0.26 &0.23 $\pm$ 0.07 & & & \\
20240000039 &1237662303518457905 &241.430 &37.179 &0.0665 &10.58 $\pm$ 0.08 &0.91 $\pm$0.12 &0.39 $\pm$ 0.12 &16.42 &15.78 &15.41 \\
20240000022 &1237664294759366699 &202.234 &38.574 &0.0265 &10.15 $\pm$ 0.04 &-0.08 $\pm$0.48 &0.50 $\pm$ 0.10 &14.24 &13.47 &13.03 \\
20240000023 &1237661387602853939 &206.176 &55.887 &0.0373 &10.77 $\pm$ 0.05 &0.35 $\pm$0.33 &0.39 $\pm$ 0.14 &14.74 &14.00 &13.62 \\
20240000515 &1237651274037329931 &133.829 &57.861 &0.0260 &9.15 $\pm$ 0.05 &-0.78 $\pm$0.04 &0.13 $\pm$ 0.00 &17.34 &16.77 &16.78 \\
20240000016 &1237655107301277787 &172.140 &58.563 &0.0105 &10.40 $\pm$ 0.04 &0.85 $\pm$0.08 &0.86 $\pm$ 0.09 &12.47 &11.90 &11.98 \\
20240000471 &1237655107839393797 &177.267 &59.433 &0.0107 &10.45 $\pm$ 0.03 &-0.32 $\pm$0.26 &0.46 $\pm$ 0.09 &12.11 &11.29 &10.97 \\
\hline
\end{tabular}}
\begin{flushleft}
\textbf{Notes.} In this catalog, the listed Columns [2]-[8] are taken from GSWLC \citep{Salim2016} and the Columns [9]-[11] are taken from the SGA catalog. \\
\end{flushleft}
\end{table*}

\subsection{Coordinate catalog}
\label{sec:coord_cata}

Table\,\ref{tab:name} presents the source coordinates and counterparts of FASHI \HI absorption galaxies. Each column is introduced as follows:

\begin{itemize} 

\item {Column\,1: Index number for each FASHI \HI source. This index number is unique to each FASHI source.}

\item {Column\,2: Centroid coordinate (J2000) for each FASHI \HI source in the format of \texttt{Jhhmmss.ss$\pm$ddmmss.s}.}

\item {Columns\,3-4: Right ascension (RA) and declination (Dec) in units of deg from the FASHI source centroid (J2000).}

\item {Column\,5: Centroid coordinate (J2000) or name for each NVSS radio  source in the format of \texttt{Jhhmmss$\pm$ddmmss} \citep{Condon1998}.}

\item {Column\,6: Centroid coordinate (J2000) for each optical counterpart in the format of \texttt{Jhhmmss.ss$\pm$ddmmss.s}.}

\item {Column\,7: Representatives of the corresponding galaxy counterparts.}

\end{itemize}

\subsection{Physical parameter catalog}
\label{sec:Physics}

Table\,\ref{tab:Physics} presents the physical parameters of FASHI \HI absorption galaxies. Each column is introduced as follows:

\begin{itemize} 

\item {Column\,1: Index number for each FASHI \HI source. This index number is unique to each FASHI source.}

\item {Column\,2: Heliocentric velocity of the \HI source in units of $\kms$. All results are presented in this work using the optical heliocentric definition of velocity.}

\item {Column\,3: Redshift, $z_{\rm HI}$ of the 21\,cm \HI line, corresponding to the heliocentric velocity in Column\,2.}

\item {Column\,4: Redshift, $z_{\rm OC}$ of the optical counterparts collected from the literature.}

\item {Column\,5: Velocity width of the \HI absorption line profile, $W_{50}$ in $\kms$, measured at the level of 50\% of the peak using the busy-function fitting method \citep{Westmeier2014}.} 

\item {Column\,6: The depth of the \HI absorption line, $S_{\rm HI}$ in mJy in negative value.}

\item {Column\,7: Integrated 1.4\,GHz continuum flux density of radio source, $S_{\rm 1.4GHz}$ in mJy as provided by NVSS survey \citep{Condon1998}.}

\item {Column\,8: Assuming isotropic emission for a source, the radio luminosity (power) is calculated with the following relation \citep{Yun2001}:
\begin{equation}
{\rm log}\frac{L_{\rm{1.4GHz}}}{\rm W\,Hz^{-1}} = 17.08+2{\rm log}\frac{D_{\rm L}}{\rm Mpc}+{\rm log}\frac{S_{\rm 1.4GHz}}{\rm mJy},
\label{eq:lumeq}
\end{equation} 
where $D_{\rm L}$ is the cosmological luminosity distance in Mpc and was calculated with a calculator from \citet{Wright2006}.}

\item {Column\,9: Optical depth $\tau_{\rm HI}$, estimated with equation \citep[e.g.,][]{Wolfe1975}: 
\begin{equation}
 \tau_{\rm HI} \approx -{\rm ln}(1+\frac{S_{\rm HI}}{c_{\rm f}S_{\rm 1.4GHz}}),
\end{equation}
where we assume covering factor $c_{\rm f}=1$ \citep{Morganti2018} indicating the fraction of background radio continuum intercepted by the foreground \HI cloud, $S_{\rm HI}$ and $S_{\rm 1.4GHz}$ are listed in the above columns.}

\item {Column\,10: \HI column density, $N_{\rm HI}$ can be estimated with equation \citep[e.g.,][]{Wolfe1975}: 
\begin{equation}
 N_{\rm HI} = 1.823\times 10^{18} T_{\rm s}\int\tau_{\rm HI} {\rm d}V,
\end{equation}
where we assume spin temperature $T_{\rm s}=100\,$K for the \HI gas \citep{Morganti2018}.}

\end{itemize}

 \subsection{Absorbers cross-matching with GSWLC}
 \label{sec:catalog_GSWLC}

 Table\,\ref{tab:GSWLC} shows the 19 sources recovered when the FASHI observed absorbers are cross-matched with the GALEX-SDSS-WISE Legacy Catalog (GSWLC) catalog \citep{Salim2016} under the condition $\rm\delta_{RA} \leq 3'$, $\rm\delta_{Dec} \leq 3'$ and $\rm\delta_{HI\,velocity} \leq 300\,\kms$. Furthermore, the cross-matched sources could be classified as associated absorption, since the redshift difference between the \HI absorption and the background is much smaller than $3000\,\kms$ \citep[e.g.,][]{Aditya2024}. Each column in Table\,\ref{tab:GSWLC} is introduced as follows:
 
 \begin{itemize} 
 
 \item {Column\,1: Index number for each FASHI \HI source. This index number is unique to each FASHI source.}
 
 \item {Column\,2: OBJID, SDSS photometric identification number, from GSWLC.}
 
 \item {Columns\,3-4: RA and Dec with centroid coordinate (J2000) in units deg, from GSWLC.}
 
 \item {Column\,5: Redshift, $z_{\odot}$, from GSWLC.}
 
 \item {Column\,6: Stellar mass, log($M_{\star})$ with its error, from GSWLC.}
 
 \item {Column\,7: UV/optical (SED) star formation rate, log($\rm SFR_{SED})$ with its error, from GSWLC.}
 
 \item {Column\,8: Dust attenuation, $A_{\rm V}$, in rest-frame $V$, from GSWLC.}

  \item {Columns\,9-11: $g$, $r$, and $z$ band photometry, in AB mag, from the local galaxy sample Siena Galaxy Atlas (SGA).}

 \end{itemize}

\section{Discussion}
\label{sec:discu}

\subsection{Comparison with known detections}

 \begin{figure}[htp]
 \centering
 \includegraphics[height=0.35\textwidth, angle=0]{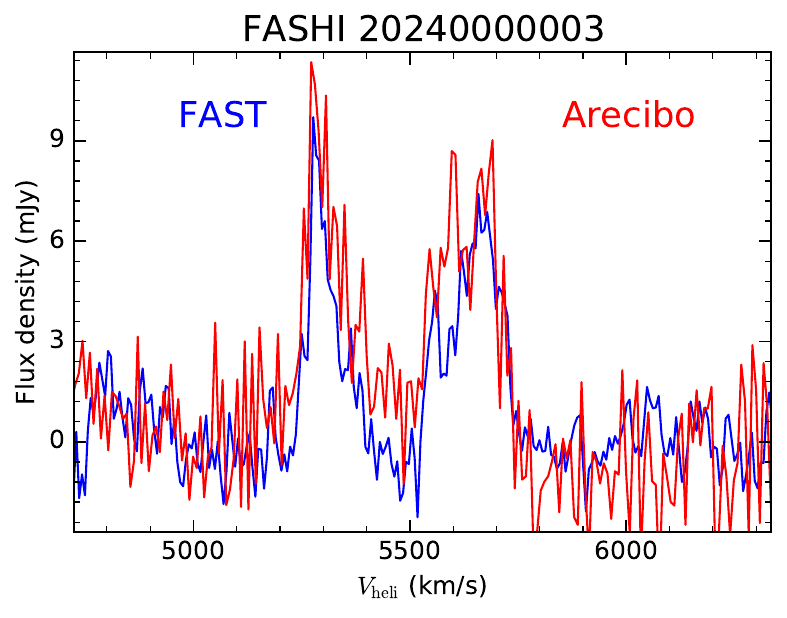}
 \includegraphics[height=0.35\textwidth, angle=0]{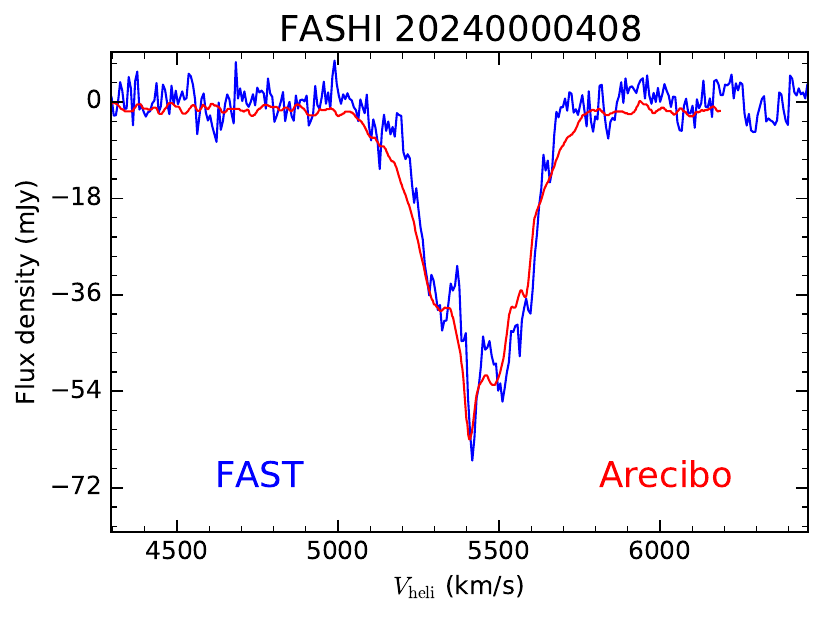}
 \caption{Two examples of FAST spectra superimposed on Arecibo spectra \citep{Mirabel1988,Springob2005}. The both show good agreement in velocity and flux. }
 \label{Fig:compare}
 \end{figure}

In Figure\,\ref{Fig:compare}, two examples of FAST spectra are superimposed on Arecibo spectra \citep{Mirabel1988,Springob2005}. The both show good agreement in velocity and flux. We also checked other previously observed \HI absorbers in Appendix\,\ref{sec:individual}. We found that most of the absorbers have good agreement in flux, e.g., 20240000405, 20240000116, and 20240000014. In fact, we have made a detailed comparison between FASHI and ALFALFA survey data in \citet{Zhang2024fashi}. Comparisons of the measured rms, SNR, and integrated flux parameters between FASHI and ALFALFA sources suggest that FAST provides highly reliable data sets. However, there are also some absorbers that have large flux discrepancies between FAST and the other instruments. A likely reason is that the low spatial resolution observations may introduce large flux pollution from nearby objects or unknown RFI, and the high spatial resolution VLBI observations may miss some flux due to lack of large scale structure coverage. 

\subsection{Luminosity across redshift}

 \begin{figure}[htp]
 \centering
 \includegraphics[width=0.48\textwidth, angle=0]{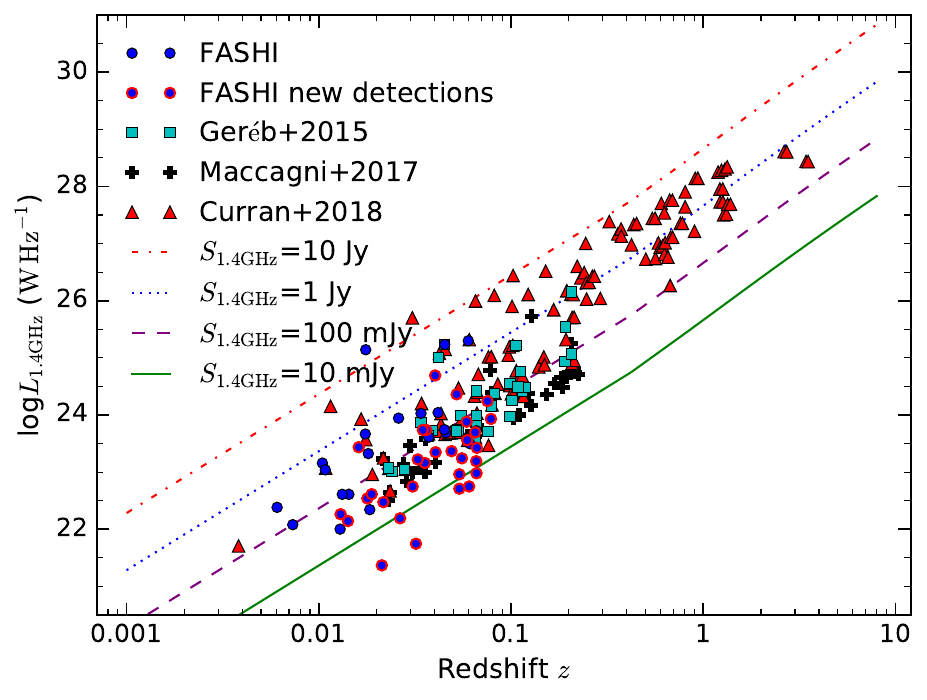}
 \includegraphics[width=0.48\textwidth, angle=0]{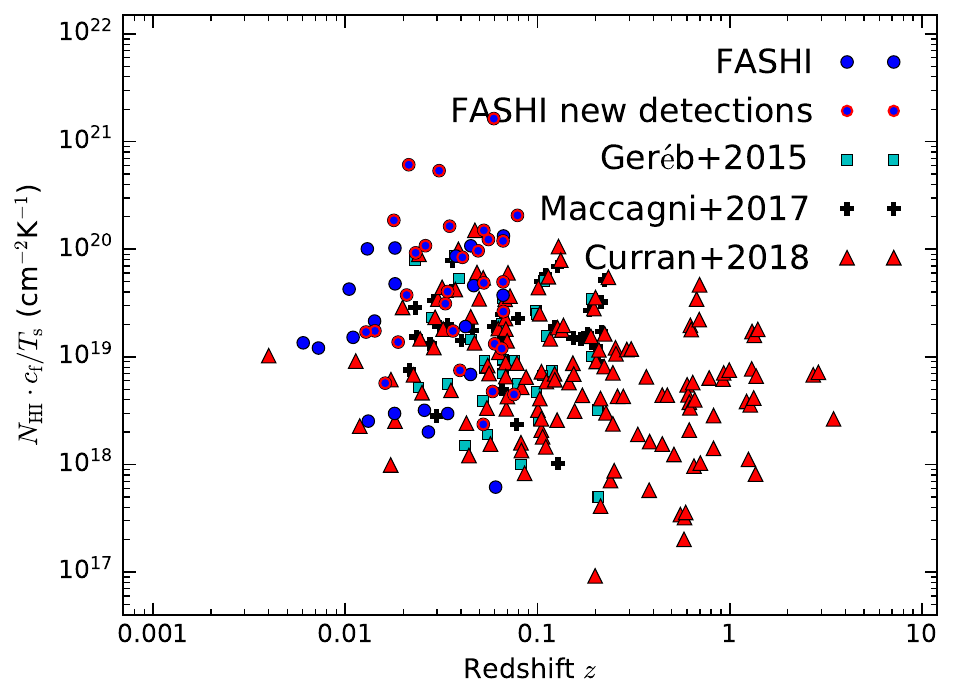}
 \includegraphics[width=0.48\textwidth, angle=0]{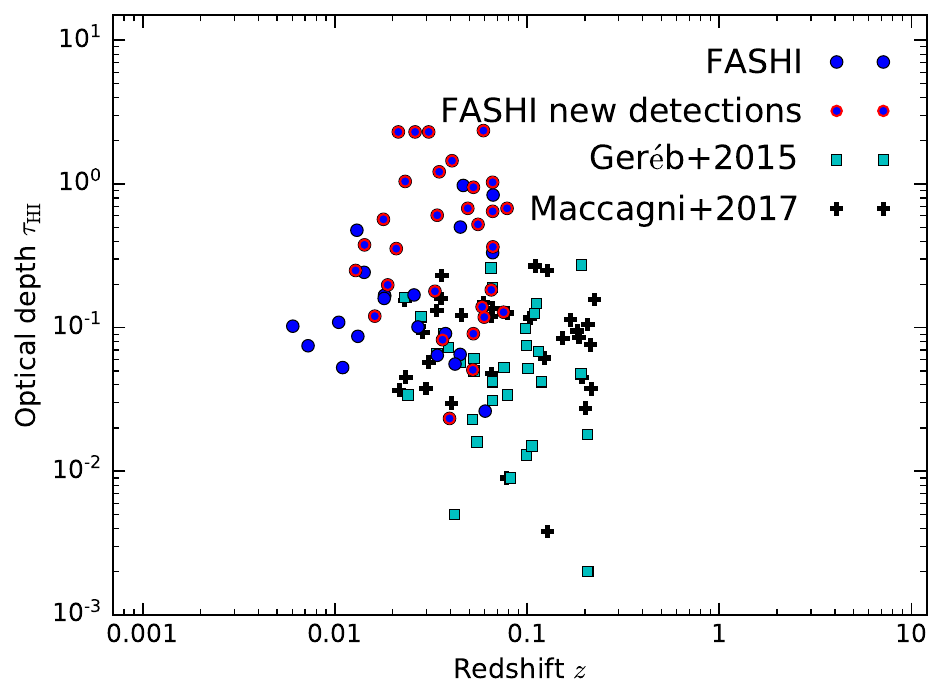}
 \caption{The rest-frame 1.4\,GHz continuum luminosity ($upper$) of the background sources, the \HI column density ($middle$), and the \HI optical depth versus redshift ($lower$) for the 21-cm \HI absorbers, including data points from \citet{Gereb2015}, \citet{Maccagni2017}, \citet{Curran2018}, and FASHI. In the upper panel, the different lines indicate different 1.4\,GHz continuum flux density.}
 \label{Fig:para-z}
 \end{figure}

 \begin{figure}[htp]
 \centering
 \includegraphics[width=0.48\textwidth, angle=0]{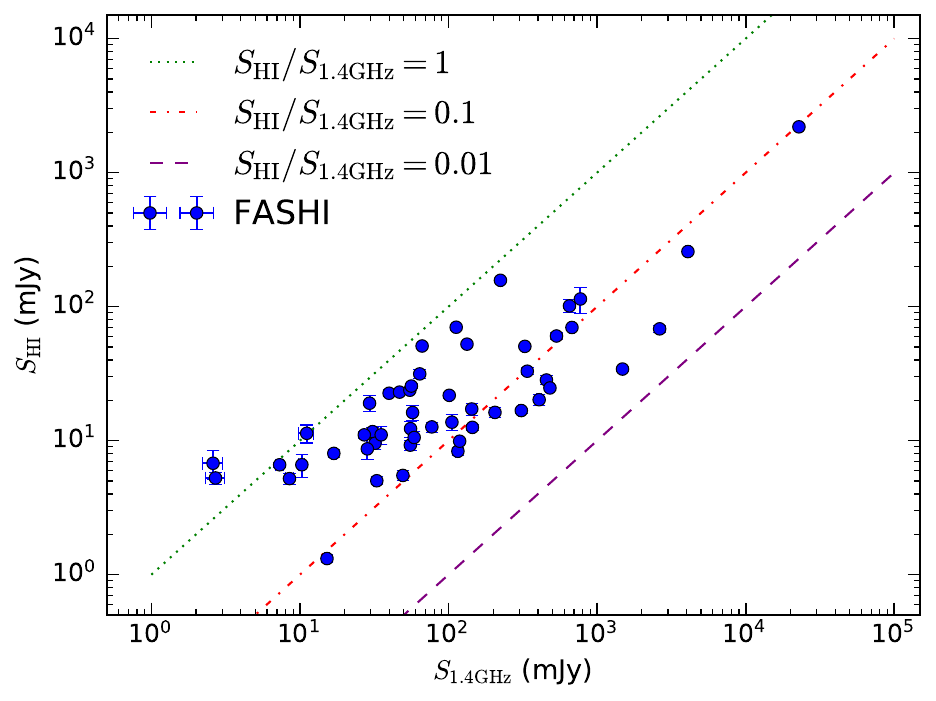}
 \caption{The distribution between the depth of the \HI absorption line ($S_{\rm HI}$) and the integrated 1.4\,GHz continuum flux density ($S_{\rm 1.4GHz}$) for the detected absorbers by FASHI.}
 \label{Fig:s-s}
 \end{figure}

 Figure\,\ref{Fig:para-z} shows the luminosity-redshift distribution, which includes 51 \HI absorbers observed by FASHI, some sample compiled by \citet{Gereb2015}, \citet{Maccagni2017}, and \citet{Curran2018}. A notable difference is that the measured 1.4\,GHz continuum luminosity of the background sources towards the FASHI absorbers is lower than that for the sample in \citet{Gereb2015}, \citet{Maccagni2017}, and \citet{Curran2018}. Figure\,\ref{Fig:para-z} shows that the radio flux densities of the FASHI absorbers are mainly concentrated in the flux density range of $S_{\rm 1.4GHz}=10\sim100$ mJy, but even as low as $2.6\pm0.4$\,mJy for FASHI\,20240000027. The absorbers in the \citet{Gereb2015}, \citet{Maccagni2017}, and \citet{Curran2018} sample are located at $S_{\rm 1.4GHz}=100\sim10000$\,mJy. The discrepancy is probably due to the selection effect. For example, the sample in \citet{Gereb2015}, \citet{Maccagni2017}, and \citet{Curran2018} come from the flux-selected sources, whereas the FASHI absorbers represent a completely untargeted and unbiased survey.

Figure\,\ref{Fig:s-s} shows the distribution between the depth of the \HI absorption line ($S_{\rm HI}$) and the integrated 1.4\,GHz continuum flux density ($S_{\rm 1.4GHz}$) for the absorbers detected by FASHI. We find that many detected absorbers have a relatively lower 1.4\,GHz continuum flux density ($S_{\rm 1.4GHz}=10$\,mJy) than previous flux-selected surveys, which usually have a selection threshold of $S_{\rm 1.4GHz}>50$\,mJy \citep[e.g.,][]{Yun2001,Gereb2014,Gereb2015,Curran2018}.

\subsection{Column density and optical depth across redshift}

Figure\,\ref{Fig:para-z} also shows the distribution between the \HI column density and redshift for the 51 \HI absorbers detected by FASHI and some other sample compiled by \citet{Gereb2015}, \citet{Maccagni2017}, and \citet{Curran2018}. The column densities vary from $6.2\times10^{19}$ to $1.6\times10^{23}\rm\,cm^{-2}$, and the median is $3.7\times10^{21}\rm\,cm^{-2}$ for the FASHI data. It also shows that there is no apparent number variation across redshift, which is consistent with the previous result measured by \citet{Gupta2006}. However, we found weak evidence of a dependence on redshift among all data points, including FASHI, \citet{Gereb2015}, \citet{Maccagni2017}, and \citet{Curran2018}. Figure\,\ref{Fig:para-z} also shows the distribution between \HI optical depth and redshift for the 51 \HI absorbers detected by FASHI and some other samples compiled by \citet{Gereb2015} and \citet{Maccagni2017}. It also shows that there is no apparent number variation across redshift. We also found weak evidence for a redshift dependence among all data points, including FASHI, \citet{Gereb2015} and \citet{Maccagni2017}. The negative correlation in column density and optical depth across redshift may be due to different detection limits.

 \subsection{Distribution of velocity}

 \begin{figure}[htp]
 \centering
 \includegraphics[width=0.48\textwidth, angle=0]{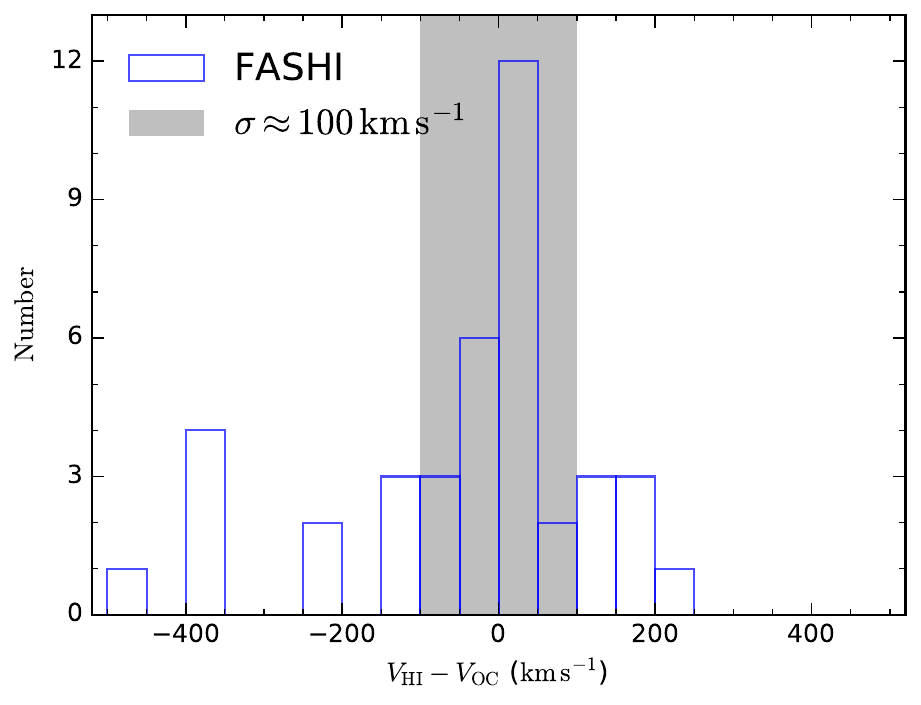}
 \caption{The histogram of velocities for the principal \HI absorption components with respect to the systemic velocity estimated from optical counterparts. The derived velocity error is about $\sigma=100\,\kms$ shown with a shaded bar.}
 \label{Fig:hist_vel}
 \end{figure}

The morphology and kinematics of \HI gas in radio galaxies are found to be very complex. \HI can trace rotating disks, offset clouds, and complex morphological structures of unsettled gas, such as infall and outflow \citep{Gereb2015}. \HI gas associated with the circumnuclear disk/torus and rotating around the nucleus may appear both blue- and redshifted relative to the systemic velocity \citep{Chandola2013}. Figure\,\ref{Fig:hist_vel} shows the velocity distribution for the main \HI absorption components with respect to the systemic velocity estimated from the optical counterparts. We have a total of 43 samples in this figure, since other samples have no known spectroscopic redshift parameters. We find that the number of blue-shifted and redshifted absorbers is 20 and 23, respectively. It seems that in this untargeted survey the redshifted number is slightly higher than the blueshifted number. However, if we consider the uncertainties of the systemic velocity ($\sigma\approx100\,\kms$), the number of blue- and redshifted absorbers should be almost equal (see Figure\,\ref{Fig:hist_vel}). We also found that there are four blueshifted absorbers at $<-300\,\kms$, but none at the high redshifted position ($>300\,\kms$). Therefore, blueshifted absorption is relatively more common, which is consistent with the previous conclusion in, e.g., \citet{Vermeulen2003}. The relative velocity of the absorbing gas could be significantly blueshifted if it is due to gas which has been accelerated by interaction with the radio jet, while infalling material fuelling the central engine would appear redshifted. More recent studies \citep[e.g.,][]{Gereb2015,Maccagni2017} also show that if the line profiles show \HI gas deviating from regular rotation, this \HI gas is mostly seen as blueshifted absorption. It is possible that the blueshifted absorption could be from AGN-driven outflows or driven by a radio jet. In addition, a very narrow and high redshifted absorption is detected and likely tracing an infalling cloud at larger distance from the nucleus.

\subsection{Distributions of line width and optical depth}

 \begin{figure}[htp]
 \centering
 \includegraphics[width=0.48\textwidth, angle=0]{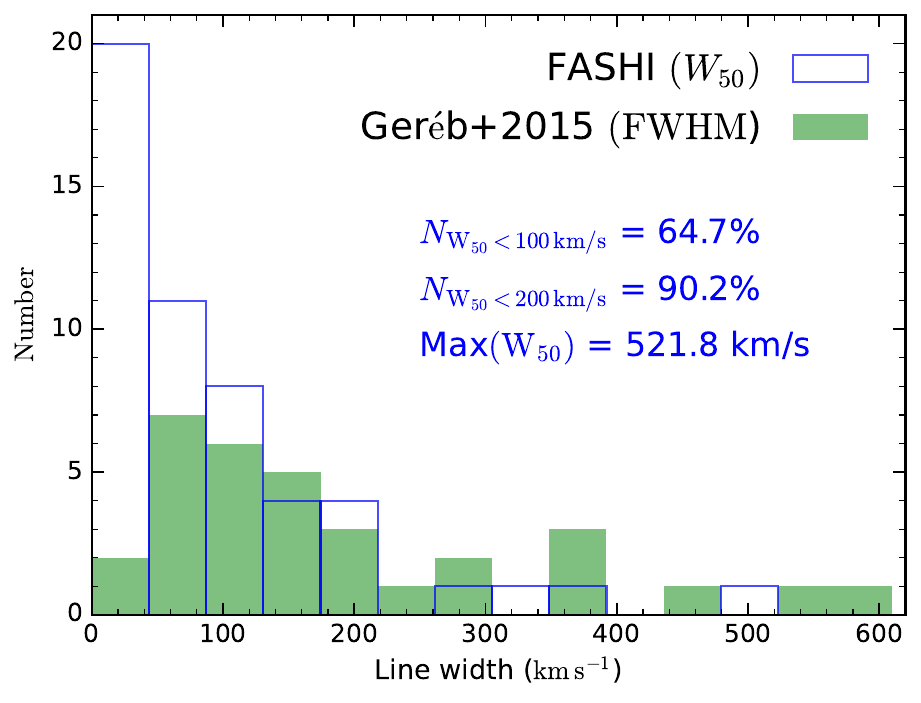}
  \includegraphics[width=0.48\textwidth, angle=0]{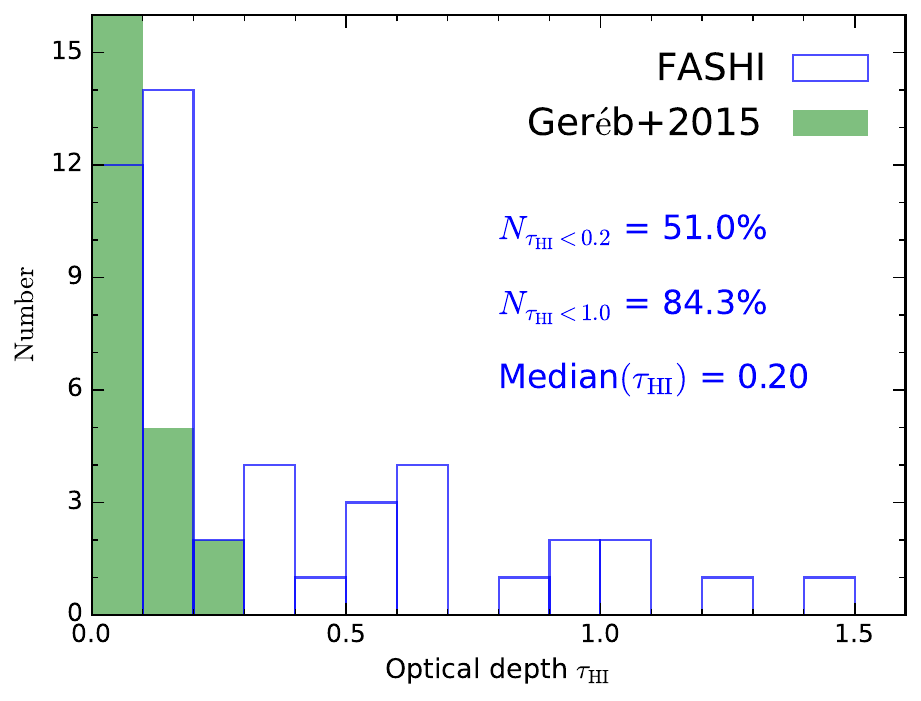}
\caption{The histograms of line width ($upper$) and optical depth ($lower$) for the 21\,cm \HI absorbers in FASHI and \citet{Gereb2015}.}
 \label{Fig:hist_para}
 \end{figure}

Figure\,\ref{Fig:hist_para} shows the distribution of line width for the 21\,cm \HI absorber measured by FASHI and \citet{Gereb2015}. We find that 64.7\% of the FASHI sample has a relatively narrow line width with $W_{50}<100\,\kms$, while only 9.8\% of the FASHI sample has a wide line width with $W_{50}>200\,\kms$. FASHI\,20240000023 has the widest line width with $W_{50}=521.8\pm11.9\,\kms$. The sample in \citet{Gereb2015} has a similar trend in the distribution of the line width as the FASHI. \citet{Taylor1999} utilized VLBI to demonstrate that the narrow component could be observed against both jets and the nucleus, while the broad component is visible exclusively against the nucleus. They proposed that the broad component represents a circumnuclear torus, while the narrow component is indicative of inward-flowing cold gas. \citet{Beswick2004} also found that the narrow \HI absorption lines are often attributed to gas at a distance from the center of the galaxy, as they are often indicative of absorption by ambient gas. Furthermore, major merger systems typically exhibit broad \HI absorption profiles, which trace perturbed disk-like structures \citep[e.g.,][]{Gereb2015,Maccagni2017}.

Figure\,\ref{Fig:hist_para} shows the optical depth distribution for the 21\,cm \HI absorbers measured by FASHI and \citet{Gereb2015}. In the FASHI sample, we find that most \HI absorbers are optically thin with $N_{\tau_{\rm HI}<0.2}=51.0\%$ and $N_{\tau_{\rm HI}<1.0}=84.3\%$, and the median is $\tau_{\rm HI}=0.2$. We note that our measured optical depths are mostly higher than previous measurements \citep[e.g.,][]{Gereb2015}. This may be because our untargeted survey is not biased towards the strong radio continuum sources. 

\subsection{Several special \HI absorbers}

\citet{Gereb2015} presented an analysis of the \HI absorption in 32 objects observed with the Westerbork Synthesis Radio Telescope (WSRT). The 32 \HI absorption sample show a broad variety of widths, shapes, and kinematical properties, but they do not include a kind of special line profile mentioned below. A common feature of such absorbers is the coexistence of an absorption line with an associated emission line. In the case of FASHI 20240000003, 20240000024, 20240000022, 20240000443, 20240000005, 20240000472, and 20240000016, the absorption line is situated in close proximity to the line centre of the emission line. In the case of FASHI 20240000196, 20240000264, and 20240000013, the absorption line is situated at the blueshifted wing of the emission profile. For FASHI 20240000027 and 20240000051, the absorption line is located at the redshifted wing of the emission line. The emission component may come from the nearby interacting galaxy or from the same galaxy as the absorbers. The coexistence of absorption and emission lines is very rare and requires high spatial resolution observations, such as VLBI, to resolve the host galaxy in order to understand their formation.

 \subsection{Host galaxy properties}
 
 \begin{figure}[htp]
 \centering
 \includegraphics[width=0.47\textwidth, angle=0]{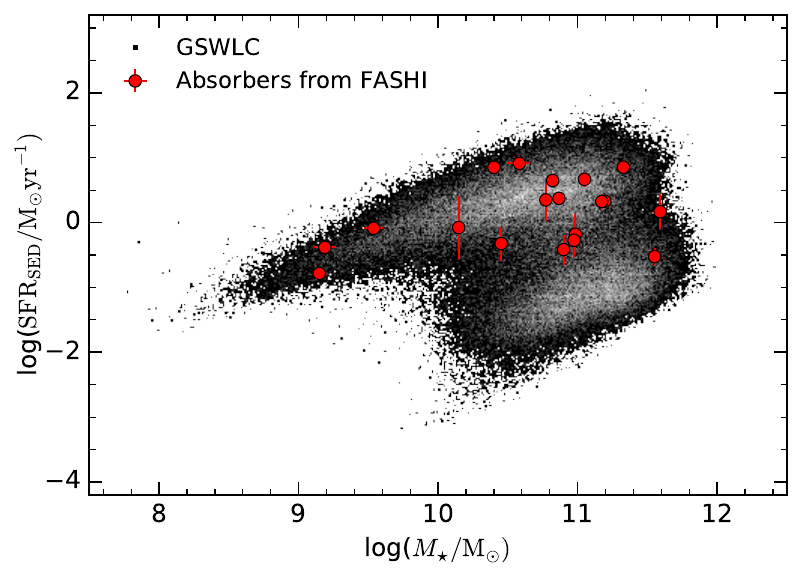}
 \includegraphics[width=0.47\textwidth, angle=0]{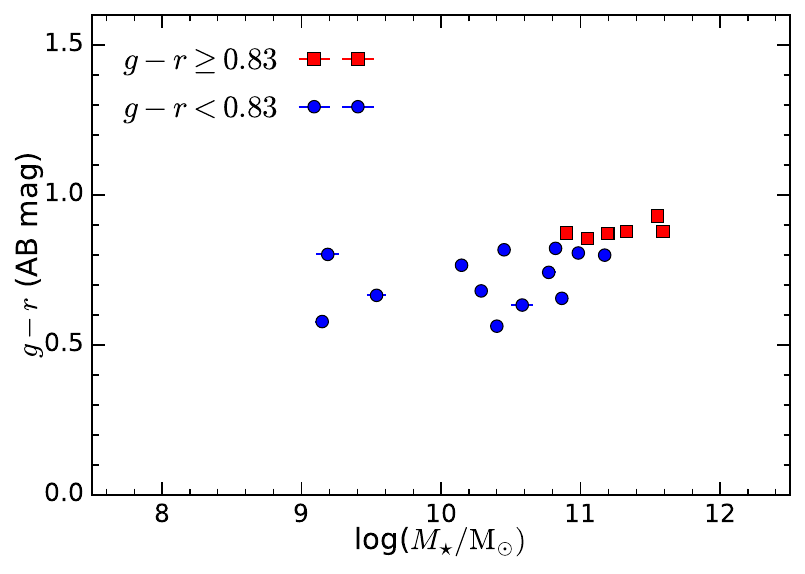}
 \caption{$Upper$: Two-dimensional distributions of star formation rate log($\rm SFR_{SED}$) and stellar mass log($M_{\star}$). The background shows all the data in the GSWLC. The red dots indicate cross-matched and associated \HI absorbers detected by FASHI. $Lower$: Color-stellar mass diagram of the FASHI absorbers. The blue and red galaxies are separated by $g-r=0.83$ \citep{Yang2006,Weinmann2006}.}
 \label{Fig:GSWLC}
 \end{figure}
 
 \begin{figure}[htp]
 \centering
 \includegraphics[width=0.47\textwidth, angle=0]{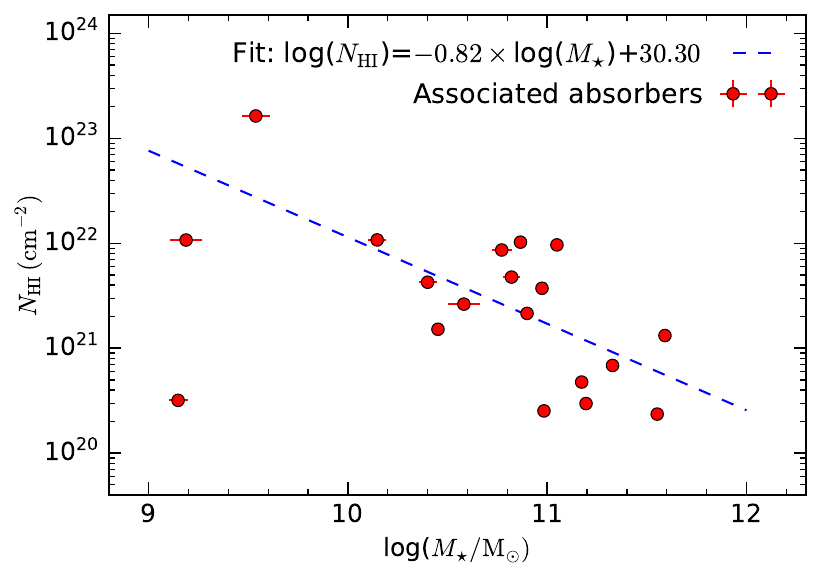}
 \caption{The relationship between the H\,I column density and the stellar mass in the host galaxy. The red dots indicate cross-matched and associated \HI absorbers detected by FASHI.}
 \label{Fig:N_Mstar}
 \end{figure}

The cross-matched 19 absorbers in the upper panel of Figure\,\ref{Fig:GSWLC} and Table\,\ref{tab:GSWLC} belong to the associated \HI absorption and are also thought to originate from within the AGN host galaxy. The multi-wavelength SED fitting results in Figure\,\ref{Fig:GSWLC} show that the associated \HI absorbers host galaxies have relatively high star formation rates with $-0. 78\pm0.04<\rm log(SFR_{SED})<0.91\pm0.12\,M_{\odot}\,yr^{-1}$, and the stellar masses are in the range of $\rm 9.15\pm0.05<log(M_{\star})<11.59\pm0.01\,M_{\odot}$.

The lower panel of Figure\,\ref{Fig:GSWLC} shows the color-stellar mass diagram of the FASHI absorbers. The color magnitudes are taken from the SGA catalog. The blue and red galaxies are separated by $g-r=0.83$ \citep{Yang2006,Weinmann2006}. We find that most of the sources are blue galaxies, and the red galaxies have colors close to the threshold of $g-r=0.83$. This proves that the \HI absorbers are more common in star-forming galaxies and less common in elliptical galaxies. This result is consistent with the inferred high star formation rates for the \HI absorbers in the upper panel of Figure\,\ref{Fig:GSWLC}.

In Figure\,\ref{Fig:N_Mstar}, we also plot the distribution between the H\,I column density and stellar mass log($M_{\star}$) for the cross-matched and associated \HI absorbers detected by FAST. It presents a negative correlation between the \HI column density and log($M_{\star}$). We then fit a relationship that is log($N_{\rm HI}$)=$-0.82\times$log($M_{\star}$)+30.30. This may indicate that when the \HI column density for the \HI absorbers is high, the host galaxy has a low stellar mass. It is likely that the high-stellar-mass galaxies are in their old stage for \HI absorbers, so gas in the AGN envelope is scarce. We also check the relationship between \HI line width, star formation rates, and \HI column density. However, we do not get a clear linear relationship.

\subsection{Completeness and Reliability}

In the current reduced FASHI data, the sensitivities for detection are uneven in different regions \citep[see details in ][]{Zhang2024fashi}. This is because FASHI could only use the schedule-filler time. As a result, the completeness becomes relatively low at the regions with a low detection sensitivity or a relatively high redshift. In addition, the strange gaps of the FASHI targets at $\sim$190, $\sim$210, and $\sim$266\,Mpc \citep[see details in ][]{Zhang2024fashi} also result in a relatively low completeness. A full test for completeness will be done with the \HI emission sources and presented in a forthcoming paper, since the \HI absorption sources have a much smaller sample for statistics.

The currently released \HI absorbers were extracted using \texttt{SoFiA} with a threshold of 4.5$\sigma$. The sources were checked for reliability by interactive manual source extraction based on 0th, 1st and 2nd moments, integrated spectral profile and SNR ($\rm SNR > 5.0$). Human intervention was used to optimize measurement accuracy and improve sample reliability by rejecting spurious detections corresponding to low-level RFI, poorly sampled data, and residual baseline fluctuations. The optical and radio counterparts of each source were identified and listed in Table\,\ref{tab:name}. Out of 51 \HI obsorbers, only 8 sources have no corresponding spectroscopic redshift data, but they have obvious \HI absorption line profiles and corresponding NVSS counterparts. As a result, all of the \HI absorbers that have been released are of relatively high confidence.

\section{Summary and Future work}
\label{sec:summary}

FASHI is designed to cover the entire sky observable by the Five-hundred-meter Aperture Spherical radio Telescope (FAST). Based on the FASHI data, in the first \HI absorption data release we perform an untargeted survey of 21\,cm \HI absorption galaxies at redshift $z\lesssim0.09$ over an area of about 10000 square degrees. With \texttt{SoFiA} source finding algorithm, we extracted 51 \HI absorbers, including 21 previously known and 30 new ones, with 8 sources having no optical spectroscopic redshift. The current detection rate of the \HI absorbers is $5.1\times10^{-3}$ sources per square degree, while the FASHI \HI emission sources have a detection rate of about 5.5 sources per square degree. We can further estimate that the probability of occurrence for the \HI absorbers in all \HI galaxies is 1/1078. With the current detection sensitivity, the final number of \HI absorbers detected by FASHI would be about 100 if FAST were able to cover 22000 square degrees at $z\lesssim0.09$. In the near future, the redshift will be extended to $z\gtrsim0.09$, and many more absorbers would be detected and released.

The radio flux densities of the FASHI absorbers are mainly concentrated in the range of $S_{\rm 1.4GHz}=10\sim100$ mJy, but also as low as $2.6\pm0.4$\,mJy. We also find that 64.7\% of the sample has a relatively narrow line width with $W_{50}<100\,\kms$, while only 9.8\% of the sample has a wide line width with $W_{50}>200\,\kms$. FASHI\,20240000023 has the widest line width with $W_{50}=521.8\pm11.9\,\kms$. Such results would provide some important clues for future flux-selected \HI absorber surveys. We also detected some special \HI absorbers. A common feature is the coexistence of an absorption line with an associated emission line, for example FASHI 20240000003, 20240000024, 20240000022, 20240000443, 20240000005, 20240000472, 20240000016, 20240000196, 20240000264, 20240000013, 20240000027 and 20240000051. The emission component may originate from the nearby interacting galaxy or the same galaxy as the absorbers. High spatial resolution observations, such as VLBI, are necessary to conduct such studies.

We present the multi-wavelength SED fitting results for the associated \HI absorption. We find that the host galaxies of the associated \HI absorbers have relatively high star formation rates with $-0. 78\pm0.04<\rm log(SFR_{SED})<0.91\pm0.12\,M_{\odot}\,yr^{-1}$. We find a negative correlation between the H\,I column density and log($M_{\star}$) with a fit of log($N_{\rm HI}$)=$-0.82\times$log($M_{\star}$)+30.30. This may indicate that when the \HI column density for the \HI absorbers is high, the host galaxy has a small stellar mass. We check the relationship between \HI line width, star formation rates, and \HI column density, but we do not get a clear linear relationship.

FAST has significantly improved the capabilities and performance for \HI absorption observations, providing a true untargeted survey of 21\,cm \HI absorption galaxies for such studies over an area of about 10000 square degrees. However, the currently released data cover only a low redshift range with $z\lesssim0.09$. With the accumulation of FAST observational data, higher redshifts up to $z\approx0.42$ and larger area up to 18000 square degrees would be covered in 4-5 years. In addition, FAST with UWB receiver could cover much higher redshifts up to $z\approx1.84$ \citep{Zhang2023uwb}. The detection sensitivity would also be greatly improved. This will lead to the detection of more \HI absorbers. We also plan to observe each \HI absorber with single point observing mode to obtain its high sensitivity and high velocity resolution spectra, which could help us to clearly see the hyperfine velocity profile.

\section*{Acknowledgements}
\addcontentsline{toc}{section}{Acknowledgements}

This work is supported by the National Key R\&D Program of China (No.\,2022YFA1602901), the West Light Foundation of the Chinese Academy of Sciences, the National Natural Science Foundation of China (Nos.\,11803044, 11933003, 12173045, 12373001, 12373011, and 11933011), the science research grants from the China Manned Space Project (No.\,CMS-CSST-2021-A05), and the Guizhou Provincial Science and Technology Projects (QKHFQ[2023]003, QKHPTRC-ZDSYS[2023]003, QKHFQ[2024]001-1). This work is also sponsored by the Chinese Academy of Sciences (CAS), through a grant to the CAS South America Center for Astronomy (CASSACA). FAST is a Chinese national mega-science facility, operated by the National Astronomical Observatories of Chinese Academy of Sciences (NAOC). We also wish to thank the anonymous referee for comments that improved the clarity of the paper.

This research has made use of the SIMBAD database, operated at CDS, Strasbourg, France. This research has made use of the NASA/IPAC Extragalactic Database (NED), which is funded by the National Aeronautics and Space Administration and operated by the California Institute of Technology.

\bibliography{references}{}

\begin{thebibliography}{}
\expandafter\ifx\csname natexlab\endcsname\relax\def\natexlab#1{#1}\fi
\providecommand{\url}[1]{\href{#1}{#1}}
\providecommand{\dodoi}[1]{doi:~\href{http://doi.org/#1}{\nolinkurl{#1}}}
\providecommand{\doeprint}[1]{\href{http://ascl.net/#1}{\nolinkurl{http://ascl.net/#1}}}
\providecommand{\doarXiv}[1]{\href{https://arxiv.org/abs/#1}{\nolinkurl{https://arxiv.org/abs/#1}}}

\bibitem[{{Aditya} \& {Kanekar}(2018)}]{Aditya2018}
{Aditya}, J.~N.~H.~S., \& {Kanekar}, N. 2018, \mnras, 481, 1578,
  \dodoi{10.1093/mnras/sty2184}

\bibitem[{{Aditya} {et~al.}(2024){Aditya}, {Yoon}, {Allison}, {An}, {Chhetri},
  {Curran}, {Darling}, {Emig}, {Glowacki}, {Kerrison}, {Koribalski}, {Mahony},
  {Moss}, {Morgan}, {Sadler}, {Soria}, {Su}, {Weng}, \& {Whiting}}]{Aditya2024}
{Aditya}, J.~N.~H.~S., {Yoon}, H., {Allison}, J.~R., {et~al.} 2024, \mnras,
  527, 8511, \dodoi{10.1093/mnras/stad3722}

\bibitem[{{Alatalo} {et~al.}(2011){Alatalo}, {Blitz}, {Young}, {Davis},
  {Bureau}, {Lopez}, {Cappellari}, {Scott}, {Shapiro}, {Crocker},
  {Mart{\'\i}n}, {Bois}, {Bournaud}, {Davies}, {de Zeeuw}, {Duc}, {Emsellem},
  {Falc{\'o}n-Barroso}, {Khochfar}, {Krajnovi{\'c}}, {Kuntschner}, {Lablanche},
  {McDermid}, {Morganti}, {Naab}, {Oosterloo}, {Sarzi}, {Serra}, \&
  {Weijmans}}]{Alatalo2011}
{Alatalo}, K., {Blitz}, L., {Young}, L.~M., {et~al.} 2011, \apj, 735, 88,
  \dodoi{10.1088/0004-637X/735/2/88}

\bibitem[{{Allison} {et~al.}(2022){Allison}, {Sadler}, {Amaral}, {An},
  {Curran}, {Darling}, {Edge}, {Ellison}, {Emig}, {Gaensler},
  {Garratt-Smithson}, {Glowacki}, {Grasha}, {Koribalski}, {Lagos}, {Lah},
  {Mahony}, {Mao}, {Morganti}, {Moss}, {Pettini}, {Pimbblet}, {Power}, {Salas},
  {Staveley-Smith}, {Whiting}, {Wong}, {Yoon}, {Zheng}, \&
  {Zwaan}}]{Allison2022}
{Allison}, J.~R., {Sadler}, E.~M., {Amaral}, A.~D., {et~al.} 2022, \pasa, 39,
  e010, \dodoi{10.1017/pasa.2022.3}

\bibitem[{{Baan} \& {Haschick}(1990)}]{Baan1990}
{Baan}, W.~A., \& {Haschick}, A. 1990, \apj, 364, 65, \dodoi{10.1086/169385}

\bibitem[{{Baan} \& {Haschick}(1981)}]{Baan1981}
{Baan}, W.~A., \& {Haschick}, A.~D. 1981, \apjl, 243, L143,
  \dodoi{10.1086/183461}

\bibitem[{{Baan} {et~al.}(1978){Baan}, {Haschick}, \& {Greenfield}}]{Baan1978}
{Baan}, W.~A., {Haschick}, A.~D., \& {Greenfield}, P.~E. 1978, \apjl, 222, L7,
  \dodoi{10.1086/182680}

\bibitem[{{Baan} {et~al.}(1982){Baan}, {Wood}, \& {Haschick}}]{Baan1982}
{Baan}, W.~A., {Wood}, P.~A.~D., \& {Haschick}, A.~D. 1982, \apjl, 260, L49,
  \dodoi{10.1086/183868}

\bibitem[{Baek {et~al.}(2015)Baek, Park, Ahn, \& Choo}]{Baek2015}
Baek, S.-J., Park, A., Ahn, Y.-J., \& Choo, J. 2015, The Analyst, 140 1, 250

\bibitem[{{Beswick} {et~al.}(2004){Beswick}, {Peck}, {Taylor}, \&
  {Giovannini}}]{Beswick2004}
{Beswick}, R.~J., {Peck}, A.~B., {Taylor}, G.~B., \& {Giovannini}, G. 2004,
  \mnras, 352, 49, \dodoi{10.1111/j.1365-2966.2004.07892.x}

\bibitem[{{Bicay} \& {Giovanelli}(1986)}]{Bicay1986}
{Bicay}, M.~D., \& {Giovanelli}, R. 1986, \aj, 91, 732, \dodoi{10.1086/114054}

\bibitem[{{Chandola} {et~al.}(2013){Chandola}, {Gupta}, \&
  {Saikia}}]{Chandola2013}
{Chandola}, Y., {Gupta}, N., \& {Saikia}, D.~J. 2013, \mnras, 429, 2380,
  \dodoi{10.1093/mnras/sts499}

\bibitem[{{Condon} {et~al.}(1998){Condon}, {Cotton}, {Greisen}, {Yin},
  {Perley}, {Taylor}, \& {Broderick}}]{Condon1998}
{Condon}, J.~J., {Cotton}, W.~D., {Greisen}, E.~W., {et~al.} 1998, \aj, 115,
  1693, \dodoi{10.1086/300337}

\bibitem[{{Conway}(1996)}]{Conway1996}
{Conway}, J.~E. 1996, in Extragalactic Radio Sources, ed. R.~D. {Ekers},
  C.~{Fanti}, \& L.~{Padrielli}, Vol. 175, 92

\bibitem[{{Courtois} \& {Tully}(2015)}]{Courtois2015}
{Courtois}, H.~M., \& {Tully}, R.~B. 2015, \mnras, 447, 1531,
  \dodoi{10.1093/mnras/stu2405}

\bibitem[{{Curran} \& {Duchesne}(2018)}]{Curran2018}
{Curran}, S.~J., \& {Duchesne}, S.~W. 2018, \mnras, 476, 3580,
  \dodoi{10.1093/mnras/sty443}

\bibitem[{{Curran} {et~al.}(2016){Curran}, {Reeves}, {Allison}, \&
  {Sadler}}]{Curran2016}
{Curran}, S.~J., {Reeves}, S.~N., {Allison}, J.~R., \& {Sadler}, E.~M. 2016,
  \mnras, 459, 4136, \dodoi{10.1093/mnras/stw943}

\bibitem[{{Darling} {et~al.}(2011){Darling}, {Macdonald}, {Haynes}, \&
  {Giovanelli}}]{Darling2011}
{Darling}, J., {Macdonald}, E.~P., {Haynes}, M.~P., \& {Giovanelli}, R. 2011,
  \apj, 742, 60, \dodoi{10.1088/0004-637X/742/1/60}

\bibitem[{{de Vaucouleurs} {et~al.}(1976){de Vaucouleurs}, {de Vaucouleurs}, \&
  {Corwin}}]{Vaucouleurs1976}
{de Vaucouleurs}, G., {de Vaucouleurs}, A., \& {Corwin}, J.~R. 1976, Second
  reference catalogue of bright galaxies, 1976, 0

\bibitem[{{De Young} {et~al.}(1973){De Young}, {Roberts}, \&
  {Saslaw}}]{Young1973}
{De Young}, D.~S., {Roberts}, M.~S., \& {Saslaw}, W.~C. 1973, \apj, 185, 809,
  \dodoi{10.1086/152456}

\bibitem[{{Dickey}(1986)}]{Dickey1986}
{Dickey}, J.~M. 1986, \apj, 300, 190, \dodoi{10.1086/163793}

\bibitem[{{Dickey}(1997)}]{Dickey1997}
---. 1997, \aj, 113, 1939, \dodoi{10.1086/118408}

\bibitem[{{Ferruit} {et~al.}(1997){Ferruit}, {Adam}, {Binette}, \&
  {P{\'e}contal}}]{Ferruit1997}
{Ferruit}, P., {Adam}, G., {Binette}, L., \& {P{\'e}contal}, E. 1997, \na, 2,
  345, \dodoi{10.1016/S1384-1076(97)00023-7}

\bibitem[{{Gallimore} {et~al.}(1999){Gallimore}, {Baum}, {O'Dea}, {Pedlar}, \&
  {Brinks}}]{Gallimore1999}
{Gallimore}, J.~F., {Baum}, S.~A., {O'Dea}, C.~P., {Pedlar}, A., \& {Brinks},
  E. 1999, \apj, 524, 684, \dodoi{10.1086/307853}

\bibitem[{{Ger{\'e}b} {et~al.}(2015){Ger{\'e}b}, {Maccagni}, {Morganti}, \&
  {Oosterloo}}]{Gereb2015}
{Ger{\'e}b}, K., {Maccagni}, F.~M., {Morganti}, R., \& {Oosterloo}, T.~A. 2015,
  \aap, 575, A44, \dodoi{10.1051/0004-6361/201424655}

\bibitem[{{Ger{\'e}b} {et~al.}(2014){Ger{\'e}b}, {Morganti}, \&
  {Oosterloo}}]{Gereb2014}
{Ger{\'e}b}, K., {Morganti}, R., \& {Oosterloo}, T.~A. 2014, \aap, 569, A35,
  \dodoi{10.1051/0004-6361/201423999}

\bibitem[{{Giovanelli} \& {Haynes}(2015)}]{Giovanelli2015}
{Giovanelli}, R., \& {Haynes}, M.~P. 2015, \aapr, 24, 1,
  \dodoi{10.1007/s00159-015-0085-3}

\bibitem[{{Gupta} {et~al.}(2006){Gupta}, {Salter}, {Saikia}, {Ghosh}, \&
  {Jeyakumar}}]{Gupta2006}
{Gupta}, N., {Salter}, C.~J., {Saikia}, D.~J., {Ghosh}, T., \& {Jeyakumar}, S.
  2006, \mnras, 373, 972, \dodoi{10.1111/j.1365-2966.2006.11064.x}

\bibitem[{{Gupta} {et~al.}(2016){Gupta}, {Srianand}, {Baan}, {Baker},
  {Beswick}, {Bhatnagar}, {Bhattacharya}, {Bosma}, {Carilli}, {Cluver},
  {Combes}, {Cress}, {Dutta}, {Fynbo}, {Heald}, {Hilton}, {Hussain}, {Jarvis},
  {Jozsa}, {Kamphuis}, {Kembhavi}, {Kerp}, {Kloeckner}, {Krogager}, {Kulkarni},
  {Ledoux}, {Mahabal}, {Mauch}, {Moodley}, {Momjian}, {Morganti}, {Noterdaeme},
  {Oosterloo}, {Petitjean}, {Schroeder}, {Serra}, {Sievers}, {Spekkens},
  {Vaisanen}, {van der Hulst}, {Vivek}, {Wang}, {Wong}, \& {Zungu}}]{Gupta2016}
{Gupta}, N., {Srianand}, R., {Baan}, W., {et~al.} 2016, in MeerKAT Science: On
  the Pathway to the SKA, 14, \dodoi{10.22323/1.277.0014}

\bibitem[{{Haynes} {et~al.}(2011){Haynes}, {Giovanelli}, {Martin}, {Hess},
  {Saintonge}, {Adams}, {Hallenbeck}, {Hoffman}, {Huang}, {Kent}, {Koopmann},
  {Papastergis}, {Stierwalt}, {Balonek}, {Craig}, {Higdon}, {Kornreich},
  {Miller}, {O'Donoghue}, {Olowin}, {Rosenberg}, {Spekkens}, {Troischt}, \&
  {Wilcots}}]{Haynes2011}
{Haynes}, M.~P., {Giovanelli}, R., {Martin}, A.~M., {et~al.} 2011, \aj, 142,
  170, \dodoi{10.1088/0004-6256/142/5/170}

\bibitem[{{Haynes} {et~al.}(2018){Haynes}, {Giovanelli}, {Kent}, {Adams},
  {Balonek}, {Craig}, {Fertig}, {Finn}, {Giovanardi}, {Hallenbeck}, {Hess},
  {Hoffman}, {Huang}, {Jones}, {Koopmann}, {Kornreich}, {Leisman}, {Miller},
  {Moorman}, {O'Connor}, {O'Donoghue}, {Papastergis}, {Troischt}, {Stark}, \&
  {Xiao}}]{Haynes2018}
{Haynes}, M.~P., {Giovanelli}, R., {Kent}, B.~R., {et~al.} 2018, \apj, 861, 49,
  \dodoi{10.3847/1538-4357/aac956}

\bibitem[{{Heckman} {et~al.}(1978){Heckman}, {Balick}, \&
  {Sullivan}}]{Heckman1978}
{Heckman}, T.~M., {Balick}, B., \& {Sullivan}, W.~T., I. 1978, \apj, 224, 745,
  \dodoi{10.1086/156423}

\bibitem[{{Helou} {et~al.}(1991){Helou}, {Madore}, {Schmitz}, {Bicay}, {Wu}, \&
  {Bennett}}]{Helou1991}
{Helou}, G., {Madore}, B.~F., {Schmitz}, M., {et~al.} 1991, in Astrophysics and
  Space Science Library, Vol. 171, Databases and On-line Data in Astronomy, ed.
  M.~A. {Albrecht} \& D.~{Egret}, 89--106, \dodoi{10.1007/978-94-011-3250-3_10}

\bibitem[{{Hu} {et~al.}(2023){Hu}, {Wang}, {Li}, {Xu}, {Yang}, {Lagache},
  {Pen}, {Zheng}, {Shu}, {Zheng}, {Li}, {Ching}, \& {Chen}}]{Hu2023}
{Hu}, W., {Wang}, Y., {Li}, Y., {et~al.} 2023, \aap, 675, A40,
  \dodoi{10.1051/0004-6361/202245549}

\bibitem[{{Iwasawa} {et~al.}(2011){Iwasawa}, {Mazzarella}, {Surace}, {Sanders},
  {Armus}, {Evans}, {Howell}, {Komossa}, {Petric}, {Teng}, {U}, \&
  {Veilleux}}]{Iwasawa2011}
{Iwasawa}, K., {Mazzarella}, J.~M., {Surace}, J.~A., {et~al.} 2011, \aap, 528,
  A137, \dodoi{10.1051/0004-6361/201015872}

\bibitem[{{Jiang} {et~al.}(2019){Jiang}, {Yue}, {Gan}, {Yao}, {Li}, {Pan},
  {Sun}, {Yu}, {Liu}, {Tang}, {Qian}, {Lu}, {Yan}, {Peng}, {Zhang}, {Wang},
  {Li}, \& {Li}}]{Jiang2019}
{Jiang}, P., {Yue}, Y., {Gan}, H., {et~al.} 2019, Science China Physics,
  Mechanics, and Astronomy, 62, 959502, \dodoi{10.1007/s11433-018-9376-1}

\bibitem[{{Jiang} {et~al.}(2020){Jiang}, {Tang}, {Hou}, {Liu}, {Kr{\v{c}}o},
  {Qian}, {Sun}, {Ching}, {Liu}, {Duan}, {Yue}, {Gan}, {Yao}, {Li}, {Pan},
  {Yu}, {Liu}, {Li}, {Peng}, {Yan}, \& {FAST Collaboration}}]{Jiang2020}
{Jiang}, P., {Tang}, N.-Y., {Hou}, L.-G., {et~al.} 2020, Research in Astronomy
  and Astrophysics, 20, 064, \dodoi{10.1088/1674-4527/20/5/64}

\bibitem[{{Jing} {et~al.}(2024){Jing}, {Wang}, {Xu}, {Liu}, {Chen}, {Liang},
  {Xu}, {Cao}, {Wang}, {Hu}, {Zhang}, {Guo}, {Gao}, {Ai}, {Gan}, {Gao}, {Han},
  {Hou}, {Hou}, {Jiang}, {Kong}, {Li}, {Liu}, {Shao}, {Pan}, {Pan}, {Qian},
  {Sun}, {Tang}, {Yang}, {Zhang}, {Zhang}, \& {Zhu}}]{Jing2024}
{Jing}, Y., {Wang}, J., {Xu}, C., {et~al.} 2024, Science China Physics,
  Mechanics, and Astronomy, 67, 259514, \dodoi{10.1007/s11433-023-2333-8}

\bibitem[{{Lavaux} \& {Hudson}(2011)}]{Lavaux2011}
{Lavaux}, G., \& {Hudson}, M.~J. 2011, \mnras, 416, 2840,
  \dodoi{10.1111/j.1365-2966.2011.19233.x}

\bibitem[{{Maccagni} {et~al.}(2017){Maccagni}, {Morganti}, {Oosterloo},
  {Ger{\'e}b}, \& {Maddox}}]{Maccagni2017}
{Maccagni}, F.~M., {Morganti}, R., {Oosterloo}, T.~A., {Ger{\'e}b}, K., \&
  {Maddox}, N. 2017, \aap, 604, A43, \dodoi{10.1051/0004-6361/201730563}

\bibitem[{{Mahony} {et~al.}(2013){Mahony}, {Morganti}, {Emonts}, {Oosterloo},
  \& {Tadhunter}}]{Mahony2013}
{Mahony}, E.~K., {Morganti}, R., {Emonts}, B.~H.~C., {Oosterloo}, T.~A., \&
  {Tadhunter}, C. 2013, \mnras, 435, L58, \dodoi{10.1093/mnrasl/slt094}

\bibitem[{{Mirabel}(1983)}]{Mirabel1983}
{Mirabel}, I.~F. 1983, \apjl, 270, L35, \dodoi{10.1086/184065}

\bibitem[{{Mirabel} \& {Sanders}(1988)}]{Mirabel1988}
{Mirabel}, I.~F., \& {Sanders}, D.~B. 1988, \apj, 335, 104,
  \dodoi{10.1086/166909}

\bibitem[{{Morganti} \& {Oosterloo}(2018)}]{Morganti2018}
{Morganti}, R., \& {Oosterloo}, T. 2018, \aapr, 26, 4,
  \dodoi{10.1007/s00159-018-0109-x}

\bibitem[{{Nagar} {et~al.}(2002){Nagar}, {Oliva}, {Marconi}, \&
  {Maiolino}}]{Nagar2002}
{Nagar}, N.~M., {Oliva}, E., {Marconi}, A., \& {Maiolino}, R. 2002, \aap, 391,
  L21, \dodoi{10.1051/0004-6361:20021039}

\bibitem[{{Nan} {et~al.}(2011){Nan}, {Li}, {Jin}, {Wang}, {Zhu}, {Zhu},
  {Zhang}, {Yue}, \& {Qian}}]{Nan2011}
{Nan}, R., {Li}, D., {Jin}, C., {et~al.} 2011, International Journal of Modern
  Physics D, 20, 989, \dodoi{10.1142/S0218271811019335}

\bibitem[{{Nilson}(1973)}]{Nilson1973}
{Nilson}, P. 1973, {Uppsala general catalogue of galaxies}

\bibitem[{{Peck} \& {Taylor}(2000)}]{Peck2000}
{Peck}, A., \& {Taylor}, G.~B. 2000, in EVN Symposium 2000, Proceedings of the
  5th european VLBI Network Symposium, ed. J.~E. {Conway}, A.~G. {Polatidis},
  R.~S. {Booth}, \& Y.~M. {Pihlstr{\"o}m}, 119,
  \dodoi{10.48550/arXiv.astro-ph/0009373}

\bibitem[{{Reich} \& {Reich}(1986)}]{Reich1986}
{Reich}, P., \& {Reich}, W. 1986, \aaps, 63, 205

\bibitem[{{Reich} {et~al.}(2001){Reich}, {Testori}, \& {Reich}}]{Reich2001}
{Reich}, P., {Testori}, J.~C., \& {Reich}, W. 2001, \aap, 376, 861,
  \dodoi{10.1051/0004-6361:20011000}

\bibitem[{{Reich}(1982)}]{Reich1982}
{Reich}, W. 1982, \aaps, 48, 219

\bibitem[{{Sadler} {et~al.}(2007){Sadler}, {Cannon}, {Mauch}, {Hancock},
  {Wake}, {Ross}, {Croom}, {Drinkwater}, {Edge}, {Eisenstein}, {Hopkins},
  {Johnston}, {Nichol}, {Pimbblet}, {De Propris}, {Roseboom}, {Schneider}, \&
  {Shanks}}]{Sadler2007}
{Sadler}, E.~M., {Cannon}, R.~D., {Mauch}, T., {et~al.} 2007, \mnras, 381, 211,
  \dodoi{10.1111/j.1365-2966.2007.12231.x}

\bibitem[{{Salim} {et~al.}(2016){Salim}, {Lee}, {Janowiecki}, {da Cunha},
  {Dickinson}, {Boquien}, {Burgarella}, {Salzer}, \& {Charlot}}]{Salim2016}
{Salim}, S., {Lee}, J.~C., {Janowiecki}, S., {et~al.} 2016, \apjs, 227, 2,
  \dodoi{10.3847/0067-0049/227/1/2}

\bibitem[{{Springob} {et~al.}(2005){Springob}, {Haynes}, {Giovanelli}, \&
  {Kent}}]{Springob2005}
{Springob}, C.~M., {Haynes}, M.~P., {Giovanelli}, R., \& {Kent}, B.~R. 2005,
  \apjs, 160, 149, \dodoi{10.1086/431550}

\bibitem[{{Struve} \& {Conway}(2012)}]{Struve2012}
{Struve}, C., \& {Conway}, J.~E. 2012, \aap, 546, A22,
  \dodoi{10.1051/0004-6361/201218768}

\bibitem[{{Su} {et~al.}(2022){Su}, {Sadler}, {Allison}, {Mahony}, {Moss},
  {Whiting}, {Yoon}, {Aditya}, {Bellstedt}, {Robotham}, {Garratt-Smithson},
  {Gu}, {Koribalski}, {Soria}, \& {Weng}}]{Su2022}
{Su}, R., {Sadler}, E.~M., {Allison}, J.~R., {et~al.} 2022, \mnras, 516, 2947,
  \dodoi{10.1093/mnras/stac2257}

\bibitem[{{Taylor} {et~al.}(1999){Taylor}, {O'Dea}, {Peck}, \&
  {Koekemoer}}]{Taylor1999}
{Taylor}, G.~B., {O'Dea}, C.~P., {Peck}, A.~B., \& {Koekemoer}, A.~M. 1999,
  \apjl, 512, L27, \dodoi{10.1086/311873}

\bibitem[{{Toba} {et~al.}(2014){Toba}, {Oyabu}, {Matsuhara}, {Malkan},
  {Gandhi}, {Nakagawa}, {Isobe}, {Shirahata}, {Oi}, {Ohyama}, {Takita},
  {Yamauchi}, \& {Yano}}]{Toba2014}
{Toba}, Y., {Oyabu}, S., {Matsuhara}, H., {et~al.} 2014, \apj, 788, 45,
  \dodoi{10.1088/0004-637X/788/1/45}

\bibitem[{{van Gorkom} {et~al.}(1989){van Gorkom}, {Knapp}, {Ekers}, {Ekers},
  {Laing}, \& {Polk}}]{Gorkom1989}
{van Gorkom}, J.~H., {Knapp}, G.~R., {Ekers}, R.~D., {et~al.} 1989, \aj, 97,
  708, \dodoi{10.1086/115016}

\bibitem[{{Vermeulen} {et~al.}(2003){Vermeulen}, {Pihlstr{\"o}m}, {Tschager},
  {de Vries}, {Conway}, {Barthel}, {Baum}, {Braun}, {Bremer}, {Miley}, {O'Dea},
  {R{\"o}ttgering}, {Schilizzi}, {Snellen}, \& {Taylor}}]{Vermeulen2003}
{Vermeulen}, R.~C., {Pihlstr{\"o}m}, Y.~M., {Tschager}, W., {et~al.} 2003,
  \aap, 404, 861, \dodoi{10.1051/0004-6361:20030468}

\bibitem[{{Wang} {et~al.}(2018){Wang}, {Luo}, {Shen}, {Hou}, {Kong}, {Song},
  {Zhang}, {Wu}, {Cao}, {Hou}, {Wang}, {Zhang}, \& {Zhao}}]{Wang2018}
{Wang}, L.-L., {Luo}, A.~L., {Shen}, S.-Y., {et~al.} 2018, \mnras, 474, 1873,
  \dodoi{10.1093/mnras/stx2798}

\bibitem[{{Weinmann} {et~al.}(2006){Weinmann}, {van den Bosch}, {Yang}, \&
  {Mo}}]{Weinmann2006}
{Weinmann}, S.~M., {van den Bosch}, F.~C., {Yang}, X., \& {Mo}, H.~J. 2006,
  \mnras, 366, 2, \dodoi{10.1111/j.1365-2966.2005.09865.x}

\bibitem[{{Wenger} {et~al.}(2000){Wenger}, {Ochsenbein}, {Egret}, {Dubois},
  {Bonnarel}, {Borde}, {Genova}, {Jasniewicz}, {Lalo{\"e}}, {Lesteven}, \&
  {Monier}}]{Wenger2000}
{Wenger}, M., {Ochsenbein}, F., {Egret}, D., {et~al.} 2000, \aaps, 143, 9,
  \dodoi{10.1051/aas:2000332}

\bibitem[{{Westmeier} {et~al.}(2014){Westmeier}, {Jurek}, {Obreschkow},
  {Koribalski}, \& {Staveley-Smith}}]{Westmeier2014}
{Westmeier}, T., {Jurek}, R., {Obreschkow}, D., {Koribalski}, B.~S., \&
  {Staveley-Smith}, L. 2014, \mnras, 438, 1176, \dodoi{10.1093/mnras/stt2266}

\bibitem[{{Westmeier} {et~al.}(2022){Westmeier}, {Deg}, {Spekkens}, {Reynolds},
  {Shen}, {Gaudet}, {Goliath}, {Huynh}, {Venkataraman}, {Lin}, {O'Beirne},
  {Catinella}, {Cortese}, {D{\'e}nes}, {Elagali}, {For}, {J{\'o}zsa},
  {Howlett}, {van der Hulst}, {Jurek}, {Kamphuis}, {Kilborn}, {Kleiner},
  {Koribalski}, {Lee-Waddell}, {Murugeshan}, {Rhee}, {Serra}, {Shao},
  {Staveley-Smith}, {Wang}, {Wong}, {Zwaan}, {Allison}, {Anderson}, {Ball},
  {Bock}, {Brodrick}, {Bunton}, {Cooray}, {Gupta}, {Hayman}, {Mahony}, {Moss},
  {Ng}, {Pearce}, {Raja}, {Roxby}, {Voronkov}, {Warhurst}, {Courtois}, \&
  {Said}}]{Westmeier2022}
{Westmeier}, T., {Deg}, N., {Spekkens}, K., {et~al.} 2022, \pasa, 39, e058,
  \dodoi{10.1017/pasa.2022.50}

\bibitem[{{Wolfe} \& {Burbidge}(1975)}]{Wolfe1975}
{Wolfe}, A.~M., \& {Burbidge}, G.~R. 1975, \apj, 200, 548,
  \dodoi{10.1086/153821}

\bibitem[{{Wright}(2006)}]{Wright2006}
{Wright}, E.~L. 2006, \pasp, 118, 1711, \dodoi{10.1086/510102}

\bibitem[{{Wu} {et~al.}(2015){Wu}, {Haynes}, {Giovanelli}, {Zhu}, \&
  {Chen}}]{Wu2015}
{Wu}, Z.~Z., {Haynes}, M.~P., {Giovanelli}, R., {Zhu}, M., \& {Chen}, R.~R.
  2015, Acta Astronomica Sinica, 56, 112

\bibitem[{{Yang} {et~al.}(2006){Yang}, {van den Bosch}, {Mo}, {Mao}, {Kang},
  {Weinmann}, {Guo}, \& {Jing}}]{Yang2006}
{Yang}, X., {van den Bosch}, F.~C., {Mo}, H.~J., {et~al.} 2006, \mnras, 369,
  1293, \dodoi{10.1111/j.1365-2966.2006.10373.x}

\bibitem[{{Yoon} {et~al.}(2024){Yoon}, {Sadler}, {Mahony}, {Aditya}, {Allison},
  {Glowacki}, {Kerrison}, {Moss}, {Su}, {Weng}, {Whiting}, {Wong},
  {Callingham}, {Curran}, {Darling}, {Edge}, {Ellison}, {Emig},
  {Garratt-Smithson}, {German}, {Grasha}, {Koribalski}, {Morganti},
  {Oosterloo}, {P{\'e}roux}, {Pettini}, {Pimbblet}, {Zheng}, {Zwaan}, {Ball},
  {Bock}, {Brodrick}, {Bunton}, {Cooray}, {Edwards}, {Hayman}, {Hotan},
  {Lee-Waddell}, {McClure-Griffiths}, {Ng}, {Phillips}, {Raja}, {Voronkov}, \&
  {Westmeier}}]{Yoon2024}
{Yoon}, H., {Sadler}, E.~M., {Mahony}, E.~K., {et~al.} 2024, arXiv e-prints,
  arXiv:2408.06626, \dodoi{10.48550/arXiv.2408.06626}

\bibitem[{{Yu} {et~al.}(2022){Yu}, {Ho}, {Wang}, \& {Li}}]{Yu2022}
{Yu}, N., {Ho}, L.~C., {Wang}, J., \& {Li}, H. 2022, \apjs, 261, 21,
  \dodoi{10.3847/1538-4365/ac626b}

\bibitem[{{Yu} {et~al.}(2023){Yu}, {Fang}, {Wang}, \& {Wu}}]{Yu2023}
{Yu}, Q., {Fang}, T., {Wang}, J., \& {Wu}, J. 2023, \apj, 952, 144,
  \dodoi{10.3847/1538-4357/acdb76}

\bibitem[{{Yun} {et~al.}(2001){Yun}, {Reddy}, \& {Condon}}]{Yun2001}
{Yun}, M.~S., {Reddy}, N.~A., \& {Condon}, J.~J. 2001, \apj, 554, 803,
  \dodoi{10.1086/323145}

\bibitem[{{Zhang} {et~al.}(2021){Zhang}, {Zhu}, {Wu}, {Yu}, {Jiang}, {Yue},
  {Huang}, \& {Hao}}]{Zhangbo2021}
{Zhang}, B., {Zhu}, M., {Wu}, Z.-Z., {et~al.} 2021, \mnras, 503, 5385,
  \dodoi{10.1093/mnras/stab754}

\bibitem[{{Zhang} {et~al.}(2022){Zhang}, {Xu}, {Wang}, {Jing}, {Liu}, {Zhu}, \&
  {Jiang}}]{Zhang2022rfi}
{Zhang}, C.-P., {Xu}, J.-L., {Wang}, J., {et~al.} 2022, Research in Astronomy
  and Astrophysics, 22, 025015, \dodoi{10.1088/1674-4527/ac3f2d}

\bibitem[{{Zhang} {et~al.}(2023){Zhang}, {Jiang}, {Zhu}, {Pan}, {Cheng}, {Liu},
  {Zhu}, {Sun}, \& {FAST Collaboration}}]{Zhang2023uwb}
{Zhang}, C.-P., {Jiang}, P., {Zhu}, M., {et~al.} 2023, Research in Astronomy
  and Astrophysics, 23, 075016, \dodoi{10.1088/1674-4527/acd58e}

\bibitem[{Zhang {et~al.}(2024)Zhang, Zhu, Jiang, Cheng, Wang, Wang, Xu, Liu,
  Yu, Qian, Yu, Ai, Jing, Xu, Liu, Guan, Sun, Yang, Huang, Hao, \&
  Collaboration}]{Zhang2024fashi}
Zhang, C.-P., Zhu, M., Jiang, P., {et~al.} 2024, SCPMA, 67, 219511,
  \dodoi{10.1007/s11433-023-2219-7}

\bibitem[{{Zwaan} {et~al.}(2015){Zwaan}, {Liske}, {P{\'e}roux}, {Murphy},
  {Bouch{\'e}}, {Curran}, \& {Biggs}}]{Zwaan2015}
{Zwaan}, M.~A., {Liske}, J., {P{\'e}roux}, C., {et~al.} 2015, \mnras, 453,
  1268, \dodoi{10.1093/mnras/stv1717}

\end{thebibliography}
\bibliographystyle{aasjournal}

 \appendix
  

\section{Notes on individual sources}
\label{sec:individual}

 In this work, the optical spectroscopic redshift parameters are mostly obtained from the local galaxy sample Siena Galaxy Atlas\footnote{\url{https://www.legacysurvey.org/sga/sga2020/}} (SGA). Some other sample information comes from the SIMBAD Astronomical Database \citep{Wenger2000} and the NASA/IPAC Extragalactic Database \citep{Helou1991}.
 
FASHI\,J030428.71-115315.1 or (ID\,20240000480): This \HI absorber seems to have been first discovered by FASHI. The optical redshift \citep[$z_{\rm OC}=0.03077$;][]{Lavaux2011} is $\sim2266\,\kms$ larger than the measured \HI redshift ($z_{\rm HI}=0.02322$).

FASHI\,J050451.78-101501.7 or (ID\,20240000482): This \HI absorber seems to have been first detected by FASHI. The radio source is the Seyfert 2 galaxy, IRAS\,05025-1018 (Simbad).

FASHI\,J141310.09-031140.8 or (ID\,20240000405): This absorber was reported e.g. by \citet{Dickey1986}. There are two absorption lines, located at $V_{\rm heli} \approx 1800\,\kms$ and $V_{\rm heli} \approx 2000\,\kms$. The measured spectral profile is in good agreement with that in \citet{Gallimore1999}. The radio source belongs to the Seyfert galaxy \citep{Gallimore1999,Nagar2002}.

FASHI\,J105833.00-023559.1 or (ID\,20240000476): This \HI absorber seems to have been first detected by FASHI. The source is a radio galaxy.

FASHI\,J031558.84-022532.1 or (ID\,20240000399): This absorber was reported e.g. by \citet{Alatalo2011}. The measured flux density agrees well with that of \citet{Alatalo2011}. The radio source is a lenticular galaxy NGC\,1266 in the constellation Eridanus. The galaxy hosts an obscured active galactic nucleus \citep{Alatalo2011}.

FASHI\,J152121.51+042039.1 or (ID\,20240000093): This \HI absorber seems to have been first detected by FASHI. The radio source belongs to the Seyfert 2 galaxy.

FASHI\,J160329.87+171133.9 or (ID\,20240000483): This absorber was first detected in absorption by VLA observations \citep{Dickey1997}. The radio source is hosted by an S0 optical galaxy in cluster A2151 of the Hercules Supercluster \citep{Gereb2015}.

FASHI\,J081937.14+210636.1 or (ID\,20240000003): The \HI observations were presented by \citet{Bicay1986,Springob2005}. The radio source belongs to the Seyfert 2 galaxy.

FASHI\,J153455.59+233013.2 or (ID\,20240000408): This absorber has been widely detected, e.g. by \citet{Dickey1986}. The radio source is a well-known starburst galaxy, Arp\,220 or IC\,4553 \citep{Baan1982}.

FASHI\,J131501.12+243651.4 or (ID\,20240000409): The \HI observations were presented by \citet{Mirabel1988,Courtois2015}. The radio source belongs to a LINER-type Active Galaxy Nucleus.

FASHI\,J005922.61+270337.1 or (ID\,20240000116): This absorber was first reported with FAST by \citet{Zhangbo2021}, but also detected by \citet{Hu2023}. Our measured flux density is in good agreement with that of \citet{Hu2023}.

FASHI\,J012406.73+295238.5 or (ID\,20240000122): This \HI absorber appears to have been first discovered by FASHI. There is no optical spectroscopic redshift in the previous literature.

FASHI\,J005747.27+302108.9 or (ID\,20240060788): The \HI observations were presented by \citet{Haynes2011}. The radio source belongs to a LINER-type active galactic nucleus.

FASHI\,J135215.88+312631.6 or (ID\,20240000140): This absorber was reported with FAST by \citet{Hu2023}. This object is the famous disk galaxy 3C\,0293. The \HI outflow driven by the radio jet was reported by \citet{Mahony2013}. The extremely broad and multi-component \HI absorption feature may be indicative of a rotating disk \citep[e.g.,][]{Baan1981,Beswick2004}.

FASHI\,J074735.82+313625.2 or (ID\,20240000155): This \HI absorber seems to have been first detected by FASHI. The radio source is weak with $S_{\rm 1.4GHz}=7.3\pm0.5{\rm\,mJy}$, while the \HI linewidth is broad with $W_{50}=381.4\pm10.9\,\kms$. There are three optical counterparts near the radio source.

FASHI\,J053117.26+315418.4 or (ID\,20240000014): This absorber was first detected with FAST by \citet{Hu2023}, whose measured flux density is in good agreement with our detection in this paper. \citet{Hu2023} suggested unsettled gas structures or gas accretion onto the supermassive black hole (SMBH).

FASHI\,J011932.90+321101.9 or (ID\,20240000159): This absorber was reported by \citet{Gorkom1989,Conway1996,Gupta2006,Struve2012,Morganti2018}. It has two \HI absorption velocity components. The low velocity is slightly blueshifted ($\sim$40\,$\kms$), while the high velocity is redshifted ($\sim$200\,$\kms$), relative to the systemic velocity of the galaxy. The low-velocity component may trace the rotating \HI disk, and the high-velocity component may be due to small clouds evaporating from the inner edge of the disk \citep{Gupta2006}.

FASHI\,J230222.13+321111.5 or (ID\,20240000027): The \HI line was presented by \citet{Courtois2015,Yu2022}, but its \HI absorption feature was first discovered by FASHI. The radio source belongs to the Low Surface Brightness Galaxy.

FASHI\,J225442.56+321324.2 or (ID\,20240000026): This \HI absorber seems to have been first detected by FASHI. The galaxy is in a group of galaxies. In the foreground is the Milky Way, which may interfere with detecting the optical counterpart.

FASHI\,J122512.74+321353.4 or (ID\,20240000129): This \HI absorber seems to have been first detected by FASHI. The radio source belongs to the Seyfert 2 galaxy.

FASHI\,J090733.90+325729.4 or (ID\,20240000141): This \HI absorber seems to have been first detected by FASHI. There is no further information about this object.

FASHI\,J135036.30+334216.5 or (ID\,20240000427): This absorber was reported by \citet{Mirabel1983}. The radio source is a LINER-type active galactic nucleus. That the observed \HI absorption occurs in clouds close to the nucleus \citep{Mirabel1983}.

FASHI\,J120547.43+335016.9 or (ID\,20240000196): This \HI absorber seems to have been first discovered by FASHI. The \HI line shows not only an absorption feature but also an emission line profile that agrees well with the redshift of the optical counterpart.

FASHI\,J150803.69+342339.2 or (ID\,20240000174): This absorber was reported by \citet{Baan1978}. The optical counterpart is a binary galaxy system that is connected to each other by a visible bridge (see Figure\,\ref{Fig:FASHI_hi}).

FASHI\,J103327.36+360550.9 or (ID\,20240000243): This \HI absorber appears to have been first discovered by FASHI. There is no further information on this object.

FASHI\,J143235.35+361806.7 or (ID\,20240000024): The radio source is a known Seyfert 2 galaxy, NGC\,5675. This source was observed by \citet{Dickey1986}, but they did not detect an \HI absorption feature. With FAST we detected a faint and narrow absorption line at the center of the \HI emission line.

FASHI\,J150033.79+364829.3 or (ID\,20240000037): This absorber was reported by \citet{Gereb2015}. The \HI line is close to the systemic velocity of the host galaxy and may trace a regularly rotating \HI disk \citep{Gereb2015}.

FASHI\,J160542.81+371049.2 or (ID\,20240000039): This \HI absorber seems to have been first detected by FASHI. The radio source is a LINER-type active galactic nucleus.

FASHI\,J051419.90+371721.2 or (ID\,20240000036): This \HI absorber seems to have been first detected by FASHI. There is no further information about this object. In the foreground is the Milky Way, which may interfere with detecting the optical counterpart.

FASHI\,J095055.65+375800.8 or (ID\,20240000500): This \HI absorber seems to have been first detected by FASHI. The radio source belongs to the Seyfert galaxy \citep{Toba2014}.

FASHI\,J132849.73+383419.9 or (ID\,20240000022): This \HI absorber seems to have been first discovered by FASHI. The \HI line shows not only an absorption feature but also an emission line profile. The \HI line shows a narrow absorption feature at the center of the broad emission line.

FASHI\,J031646.12+395955.4 or (ID\,20240000443): This \HI absorber seems to have been first detected by FASHI. The \HI line shows not only an absorption feature but also an emission line profile. The \HI line shows a prominent absorption feature that nearly coincides with the mean velocity of the emission signal. The \HI line profile of this source is almost identical to that of NGC\,984 reported by \citet{Mirabel1983}. NGC\,984 has a compact bright core with an extended halo. It is a low surface brightness galaxy.

FASHI\,J123943.52+401459.7 or (ID\,20240000264): This \HI absorber seems to have been first detected by FASHI. The \HI line shows not only an absorption feature but also an emission line profile. There is also an \HI emission line that agrees well with the redshift of the optical counterpart.

FASHI\,J071908.19+403355.0 or (ID\,20240000051): This \HI absorber seems to have been first detected by FASHI. The \HI line shows not only an absorption feature but also an emission line profile. There is also an \HI emission line at a low velocity position. In the foreground is the Milky Way, which may interfere with detecting the optical counterpart. The radio source belongs to the Seyfert galaxy \citep{Wang2018}.

FASHI\,J032421.15+404714.8 or (ID\,20240000445): This \HI absorber seems to have been first discovered by FASHI. In the foreground is the Milky Way, which may interfere with detecting the optical counterpart.

FASHI\,J031946.81+413034.4 or (ID\,20240000501): This absorber has been widely reported e.g. by \citet{Young1973,Ferruit1997}. The radio source is the well-known Seyfert 2 galaxy, 3C\,84. The absorbing systems have two very different velocities. The high velocity system is a relatively narrow \HI absorption line redshifted $\sim$3000\,$\kms$ from the systemic velocity of 3C\,84 \citep{Young1973}.

FASHI\,J075134.30+425242.1 or (ID\,20240000013): This \HI absorber seems to have been first discovered by FASHI. The \HI line shows not only an absorption feature but also an emission line profile. The \HI absorption line is located at the blueshift wing of the double peak \HI emission profile. The radio source belongs to a LINER-type active galactic nucleus.

FASHI\,J044226.82+440615.0 or (ID\,20240000454): This \HI absorber seems to have been first detected by FASHI. There is no further information about this object. In the foreground is the Milky Way, which may interfere with detecting the optical counterpart.

FASHI\,J033337.92+445011.9 or (ID\,20240000059): This \HI absorber seems to have been first discovered by FASHI. There is no optical spectroscopic redshift in the previous literature. In the foreground is the Milky Way, which may interfere with detecting the optical counterpart.

FASHI\,J060801.90+450814.2 or (ID\,20240000507): This \HI absorber seems to have been first detected by FASHI. There is no further information about this object. In the foreground is the Milky Way, which may interfere with detecting the optical counterpart.

FASHI\,J032614.32+492357.3 or (ID\,20240000078): : This \HI absorber appears to have been first detected by FASHI. There is no optical spectroscopic redshift in the previous literature. In the foreground is the Milky Way, which may interfere with detecting the optical counterpart.

FASHI\,J024600.14+511458.1 or (ID\,20240000512): This \HI absorber seems to have been first discovered by FASHI. There is no further information about this object. In the foreground is the Milky Way, which may interfere with detecting the optical counterpart.

FASHI\,J044433.30+525213.5 or (ID\,20240000007): This \HI absorber seems to have been first detected by FASHI. There is no further information about this object. In the foreground is the Milky Way, which may interfere with detecting the optical counterpart.

FASHI\,J030630.87+541208.7 or (ID\,20240000083): This \HI absorber seems to have been first discovered by FASHI. There is no optical spectroscopic redshift in the previous literature. In the foreground is the Milky Way, which may interfere with detecting the optical counterpart.

FASHI\,J043636.25+542201.5 or (ID\,20240000005): This \HI absorber seems to have been first detected by FASHI. The \HI line shows not only an absorption feature but also an emission line profile. The \HI line shows a narrow absorption feature at the center of the broad emission line. There is no optical spectroscopic redshift in the previous literature.

FASHI\,J134440.84+555306.2 or (ID\,20240000023): This \HI absorber has been reported in detail by \citet{Gereb2015}. In Figure\,\ref{Fig:FASHI_hi} we can see that the line width is extremely broad with $W_{50}=521.8\pm11.9\,\kms$ and the optical morphology shows a long tidal tail extending 40\,kpc to the south \citep{Iwasawa2011}. This object is the host of an ongoing merger.

FASHI\,J125614.58+565222.9 or (ID\,20240000513): \HI absorption was reported by \citet{Heckman1978,Gallimore1999}. The radio source belongs to the Seyfert 2 galaxy.

FASHI\,J085519.79+575155.5 or (ID\,20240000515): This absorber has been reported in detail by \citet{Zwaan2015,Curran2016}. The \HI absorption line is very narrow and strong.

FASHI\,J162636.67+580849.8 or (ID\,20240000472): This \HI absorber seems to have been first discovered by FASHI. The \HI line shows not only an absorption feature but also an emission line profile. The \HI line shows a narrow but strong absorption feature at the broad emission line center. There is no optical spectroscopic redshift in the previous literature.

FASHI\,J112832.57+583323.0 or (ID\,20240000016): This absorber was reported by \citet{Dickey1986,Baan1990}. The \HI line shows not only an absorption feature but also an emission line profile. The interaction system IC\,694/NGC\,3690 (Arp\,299-Mrk\,171) has been studied in \HI absorption and OH emission with the VLA in the A configuration \citep{Baan1990}. The \HI line shows not only an absorption feature but also an emission profile (see Figure\,\ref{Fig:FASHI_hi}). The emission component may come from the nearby interacting galaxy.

FASHI\,J114852.43+592501.9 or (ID\,20240000471): This absorber was reported by \citet{Gorkom1989,Dickey1986,Peck2000}. This galaxy is classified as elliptical by \citet{Vaucouleurs1976} and as SO by \citet{Nilson1973}. The galaxy has a close binary SO companion, NGC 3895 \citep{Gorkom1989}. The \HI line shows not only an absorption feature but also an emission line profile. The \HI emission component at $\sim3500\,\kms$ may originate from the nearby galaxy (see Figure\,\ref{Fig:FASHI_hi}).

\clearpage

 \section{Figures on individual sources}
\label{sec:figs}
 
 \begin{figure*}[htp]
 \centering
 \renewcommand{\thefigure}{\arabic{figure} (Continued)}
 \setcounter{figure}{1}
 \includegraphics[height=0.22\textwidth, angle=0]{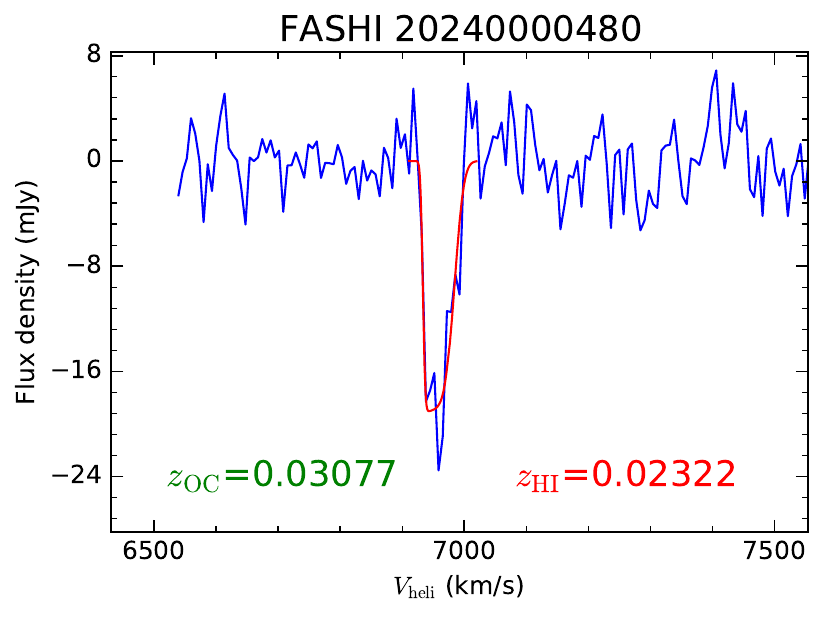}
 \includegraphics[height=0.27\textwidth, angle=0]{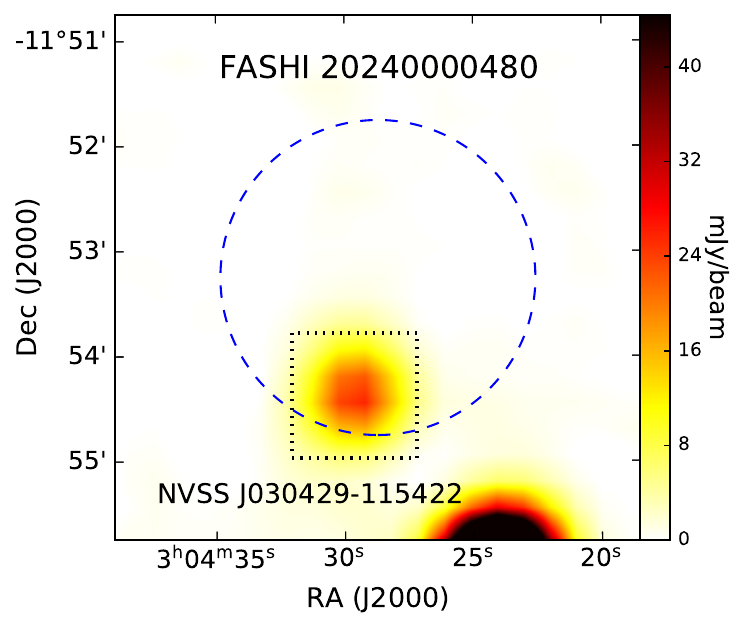}
 \includegraphics[height=0.27\textwidth, angle=0]{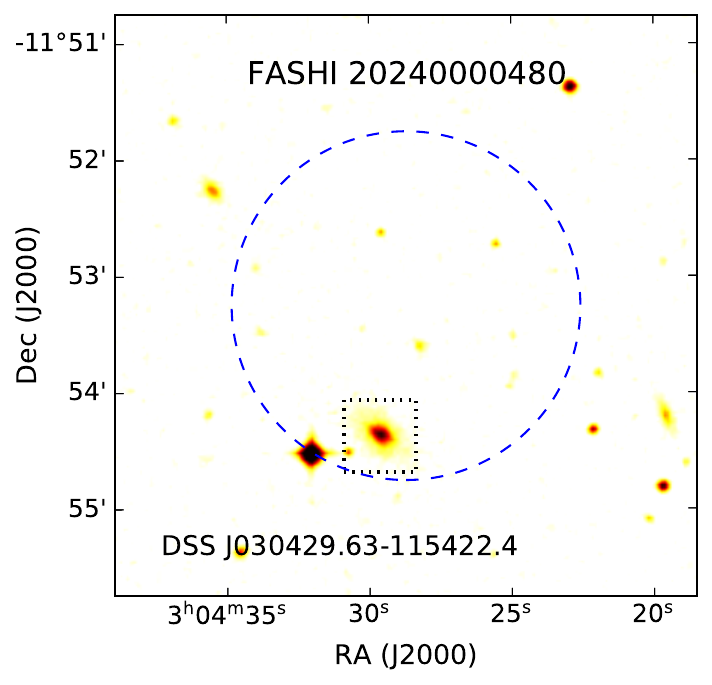}
 \caption{21\,cm H\,{\scriptsize{I}} absorption galaxy FASHI\,030428.71-115315.1 or ID\,20240000480.}
 \end{figure*} 

 \begin{figure*}[htp]
 \centering
 \renewcommand{\thefigure}{\arabic{figure} (Continued)}
 \addtocounter{figure}{-1}
 \includegraphics[height=0.22\textwidth, angle=0]{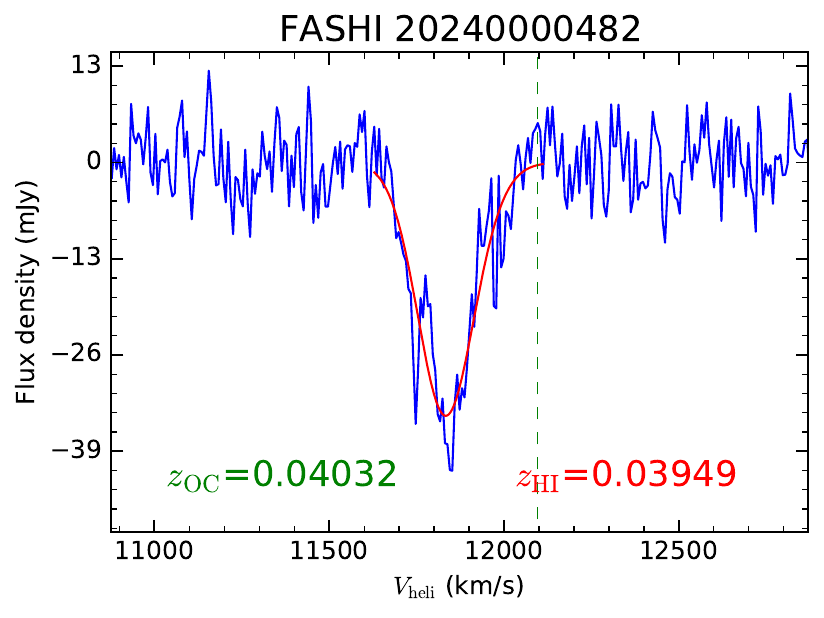}
 \includegraphics[height=0.27\textwidth, angle=0]{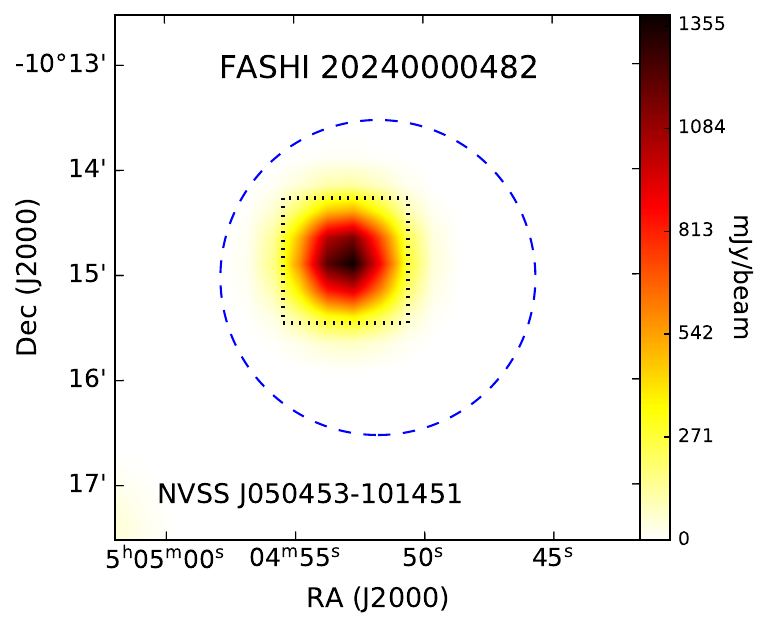}
 \includegraphics[height=0.27\textwidth, angle=0]{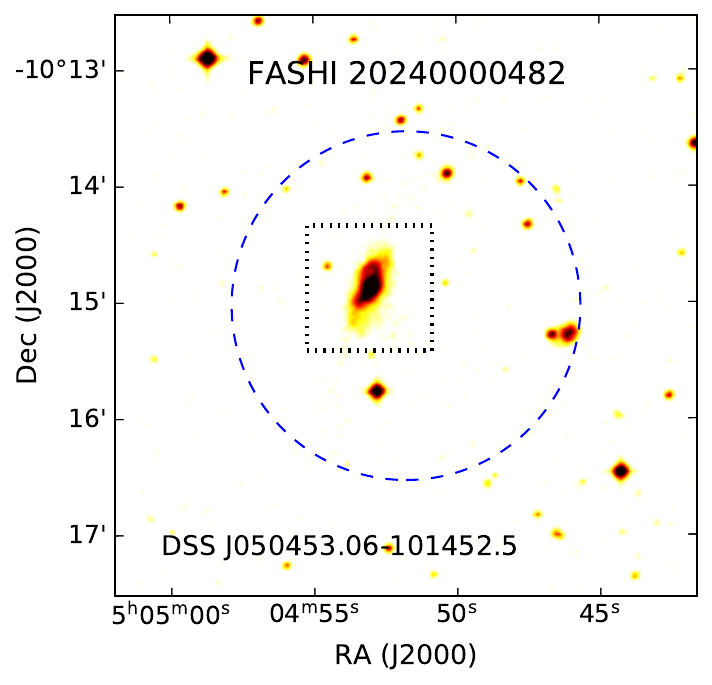}
 \caption{21\,cm H\,{\scriptsize{I}} absorption galaxy FASHI\,050451.78-101501.7 or ID\,20240000482.}
 \end{figure*} 

 \begin{figure*}[htp]
 \centering
 \renewcommand{\thefigure}{\arabic{figure} (Continued)}
 \addtocounter{figure}{-1}
 \includegraphics[height=0.22\textwidth, angle=0]{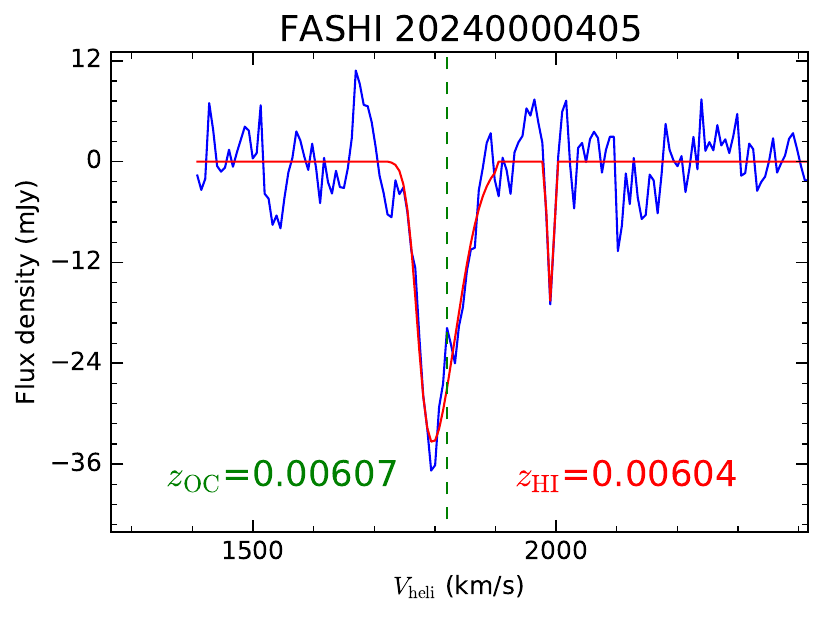}
 \includegraphics[height=0.27\textwidth, angle=0]{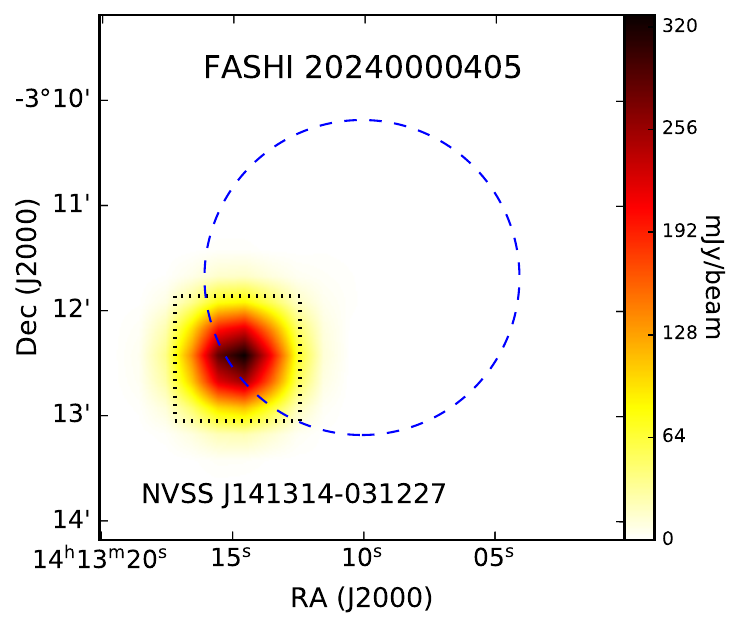}
 \includegraphics[height=0.27\textwidth, angle=0]{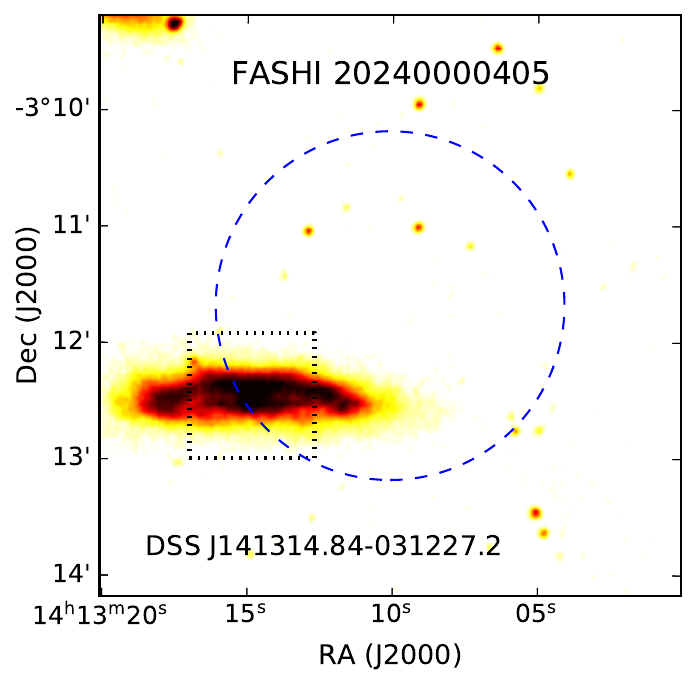}
 \caption{21\,cm H\,{\scriptsize{I}} absorption galaxy FASHI\,141310.09-031140.8 or ID\,20240000405.}
 \end{figure*} 

 \begin{figure*}[htp]
 \centering
 \renewcommand{\thefigure}{\arabic{figure} (Continued)}
 \addtocounter{figure}{-1}
 \includegraphics[height=0.22\textwidth, angle=0]{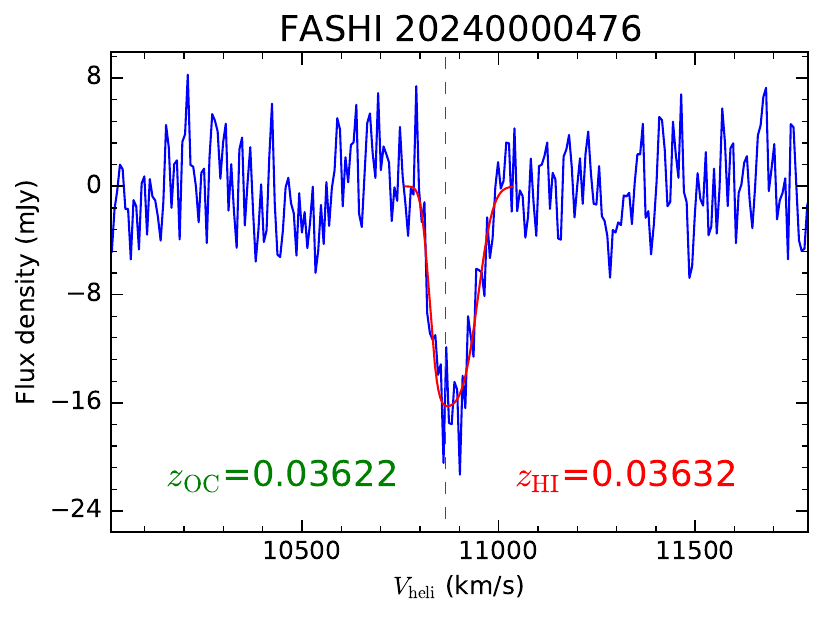}
 \includegraphics[height=0.27\textwidth, angle=0]{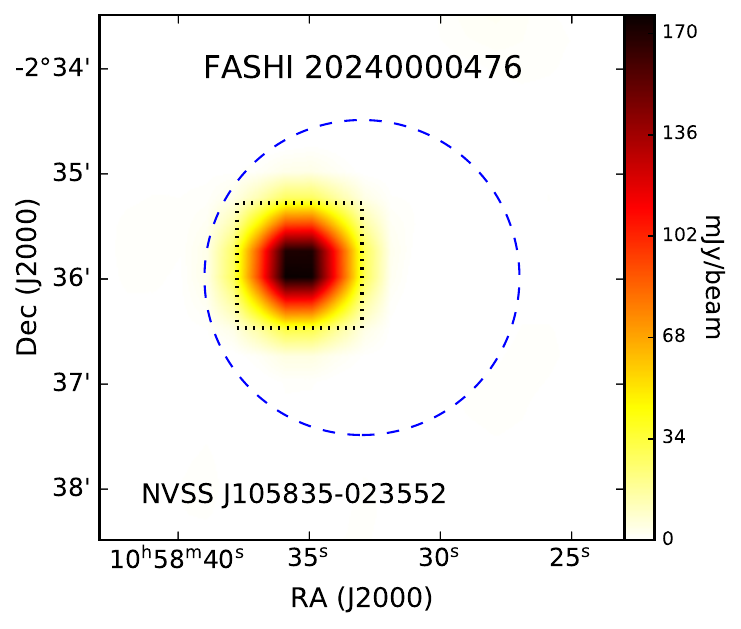}
 \includegraphics[height=0.27\textwidth, angle=0]{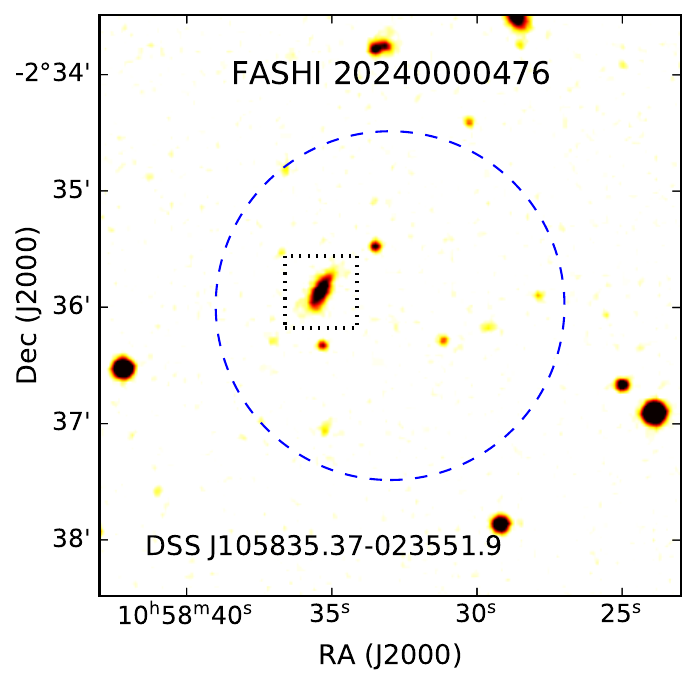}
 \caption{21\,cm H\,{\scriptsize{I}} absorption galaxy FASHI\,105833.00-023559.1 or ID\,20240000476.}
 \end{figure*} 

 \begin{figure*}[htp]
 \centering
 \renewcommand{\thefigure}{\arabic{figure} (Continued)}
 \addtocounter{figure}{-1}
 \includegraphics[height=0.22\textwidth, angle=0]{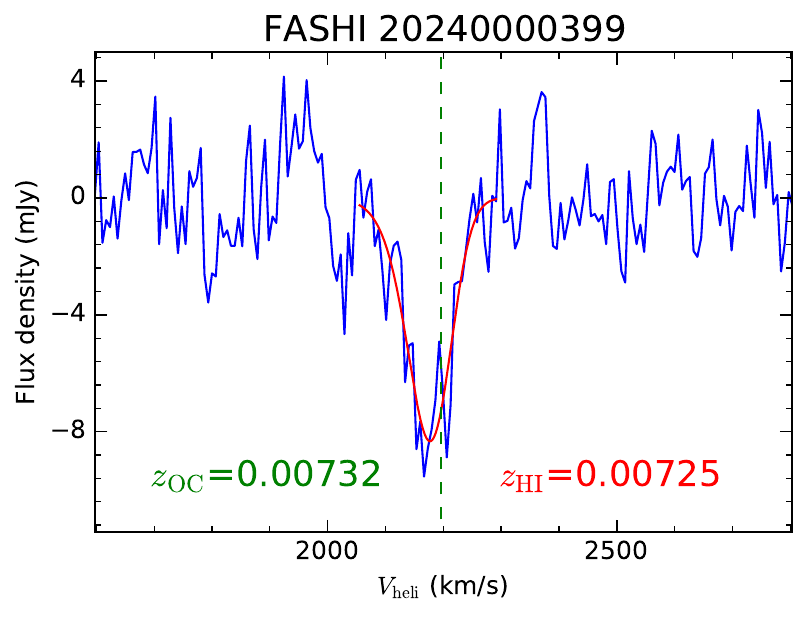}
 \includegraphics[height=0.27\textwidth, angle=0]{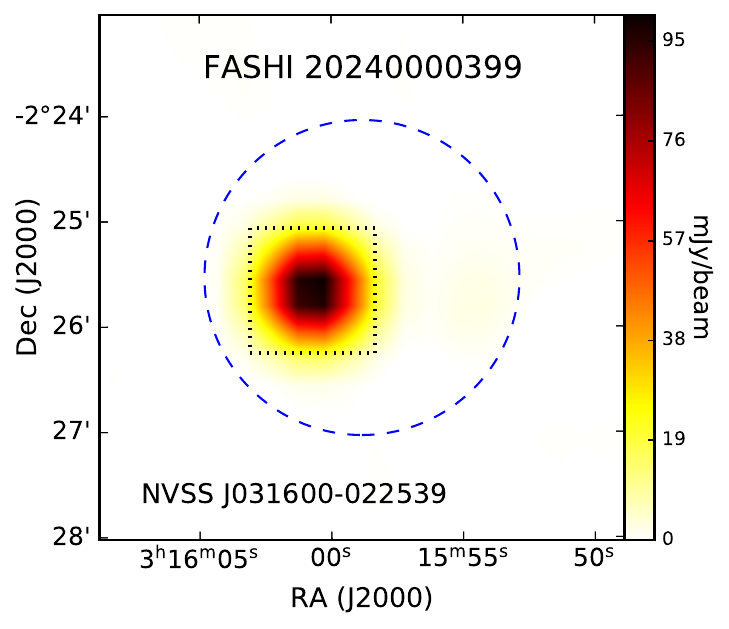}
 \includegraphics[height=0.27\textwidth, angle=0]{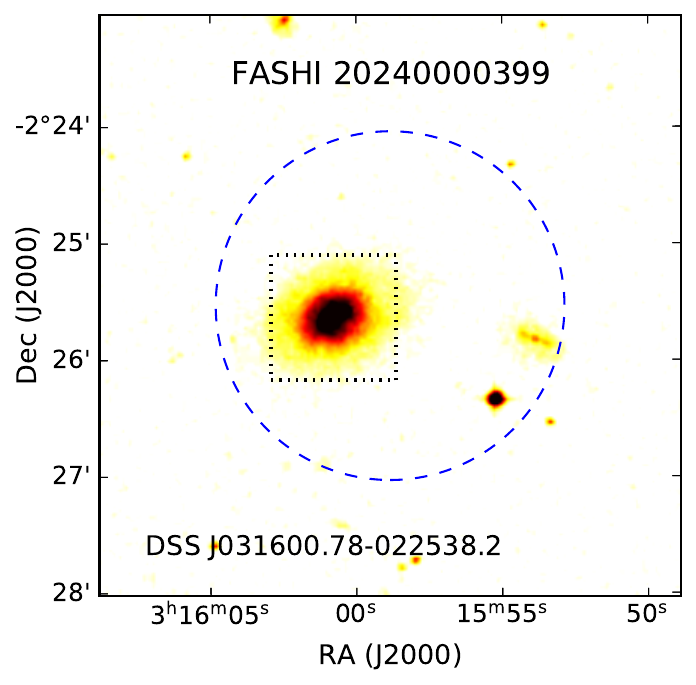}
 \caption{21\,cm H\,{\scriptsize{I}} absorption galaxy FASHI\,031558.84-022532.1 or ID\,20240000399.}
 \end{figure*} 

 \begin{figure*}[htp]
 \centering
 \renewcommand{\thefigure}{\arabic{figure} (Continued)}
 \addtocounter{figure}{-1}
 \includegraphics[height=0.22\textwidth, angle=0]{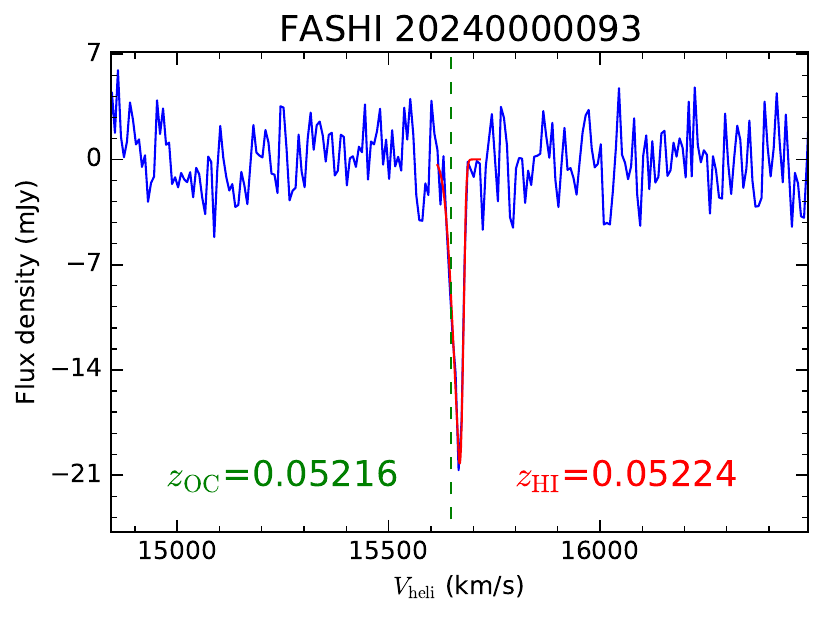}
 \includegraphics[height=0.27\textwidth, angle=0]{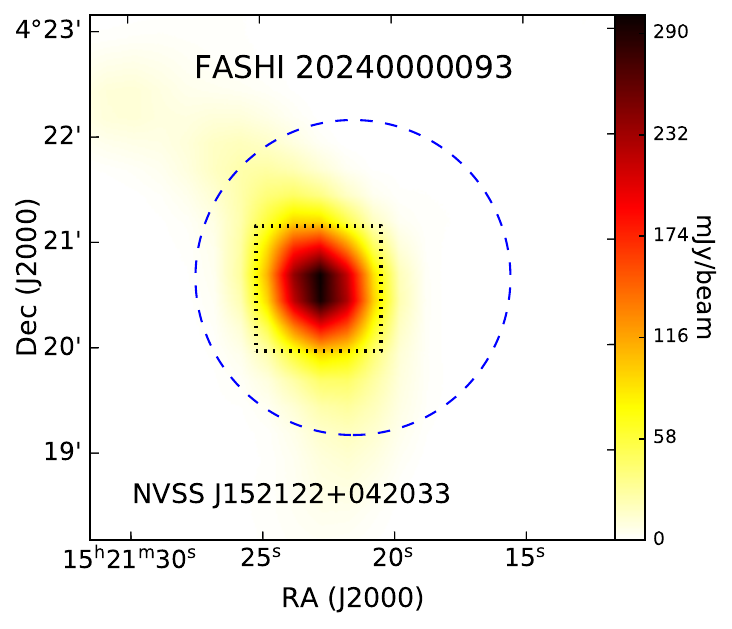}
 \includegraphics[height=0.27\textwidth, angle=0]{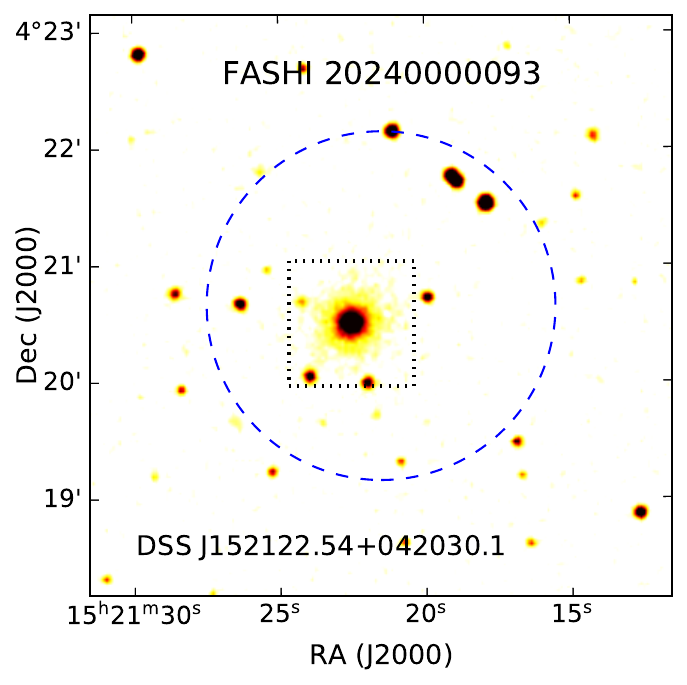}
 \caption{21\,cm H\,{\scriptsize{I}} absorption galaxy FASHI\,152121.51+042039.1 or ID\,20240000093.}
 \end{figure*} 

 \begin{figure*}[htp]
 \centering
 \renewcommand{\thefigure}{\arabic{figure} (Continued)}
 \addtocounter{figure}{-1}
 \includegraphics[height=0.22\textwidth, angle=0]{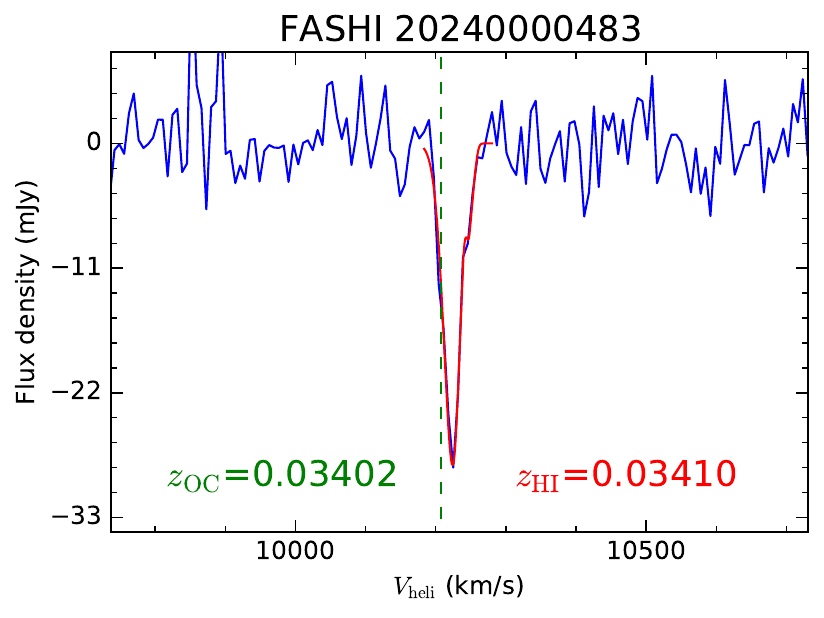}
 \includegraphics[height=0.27\textwidth, angle=0]{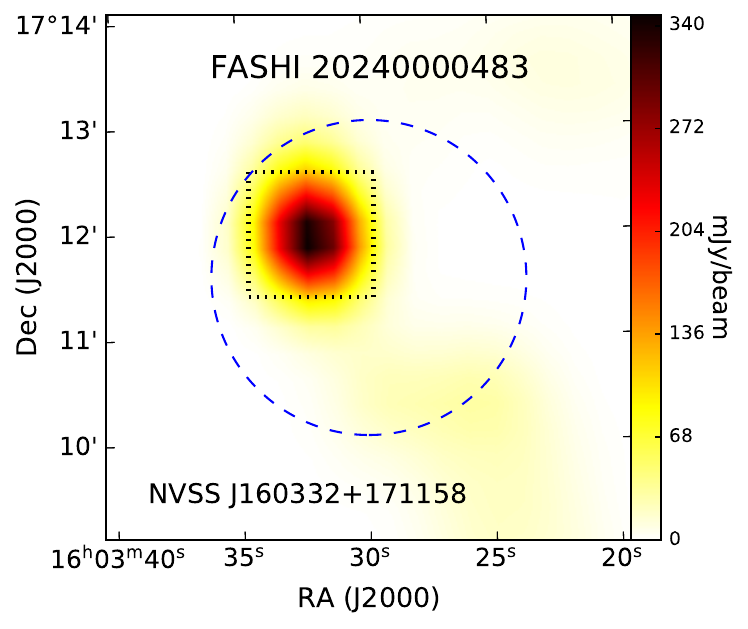}
 \includegraphics[height=0.27\textwidth, angle=0]{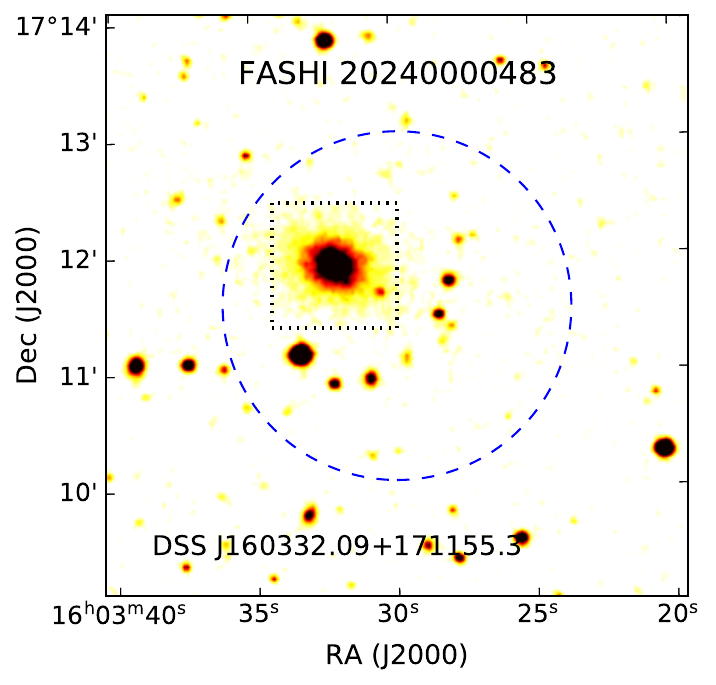}
 \caption{21\,cm H\,{\scriptsize{I}} absorption galaxy FASHI\,160329.87+171133.9 or ID\,20240000483.}
 \end{figure*} 

 \begin{figure*}[htp]
 \centering
 \renewcommand{\thefigure}{\arabic{figure} (Continued)}
 \addtocounter{figure}{-1}
 \includegraphics[height=0.22\textwidth, angle=0]{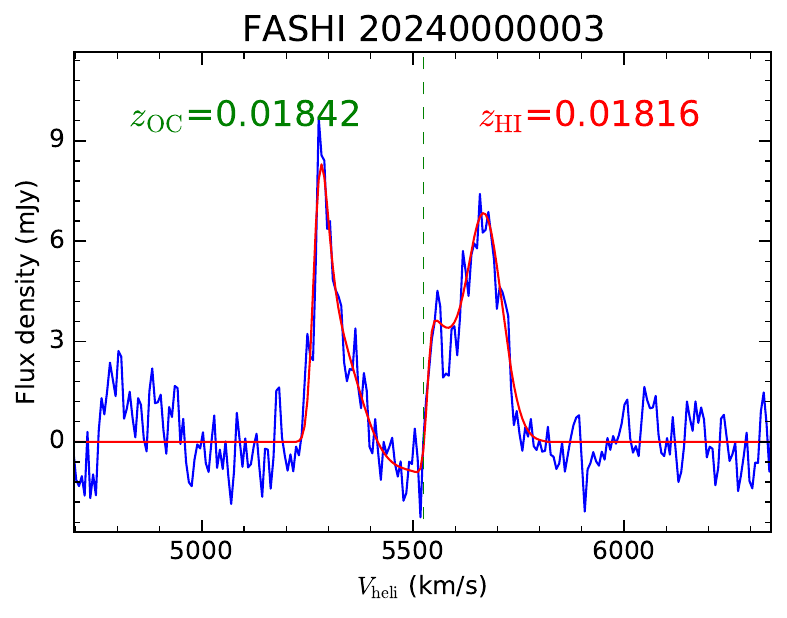}
 \includegraphics[height=0.27\textwidth, angle=0]{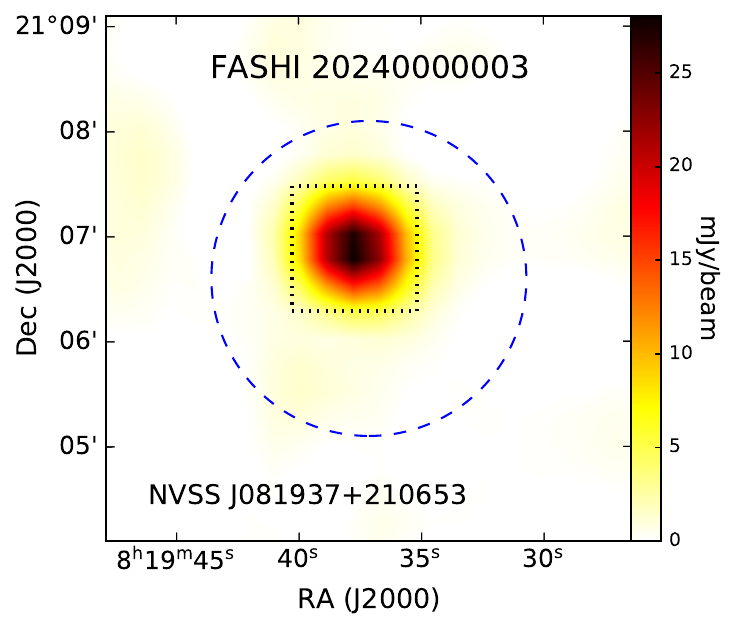}
 \includegraphics[height=0.27\textwidth, angle=0]{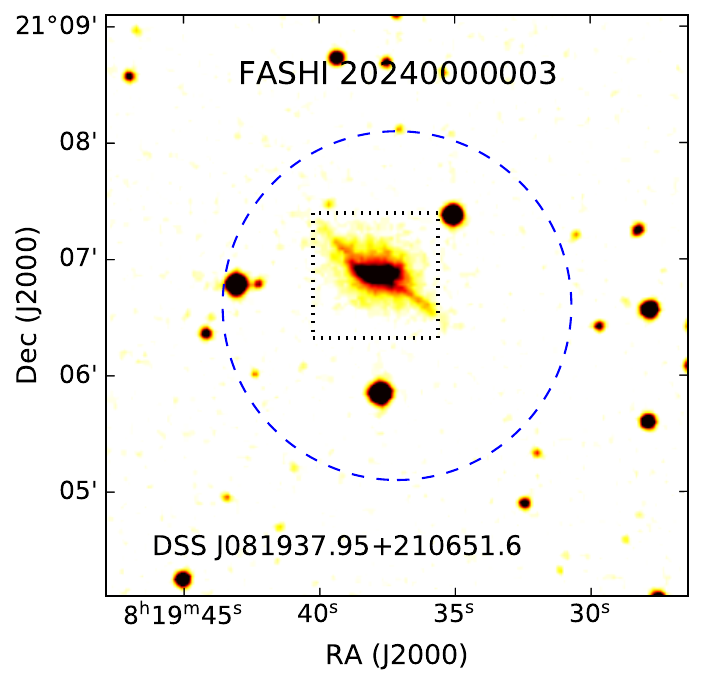}
 \caption{21\,cm H\,{\scriptsize{I}} absorption galaxy FASHI\,081937.14+210636.1 or ID\,20240000003.}
 \end{figure*} 

 \begin{figure*}[htp]
 \centering
 \renewcommand{\thefigure}{\arabic{figure} (Continued)}
 \addtocounter{figure}{-1}
 \includegraphics[height=0.22\textwidth, angle=0]{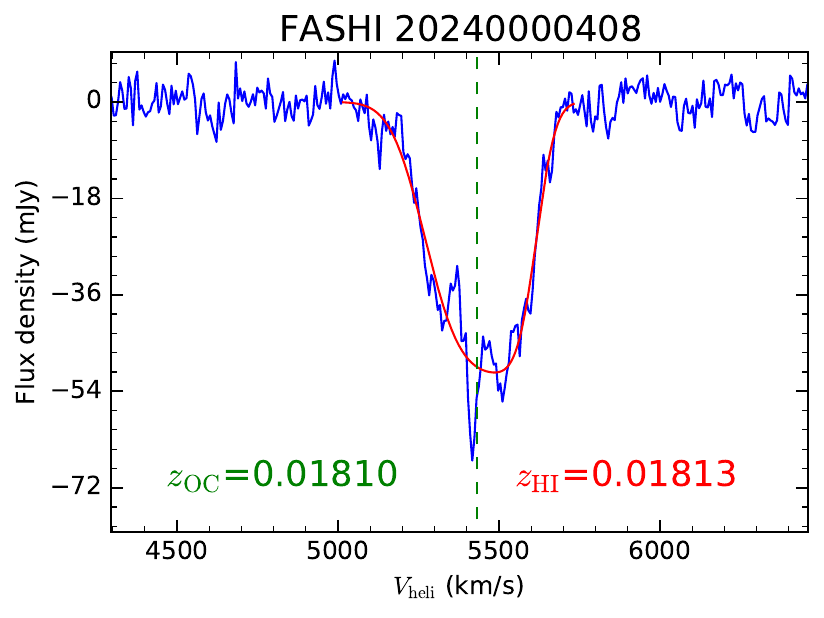}
 \includegraphics[height=0.27\textwidth, angle=0]{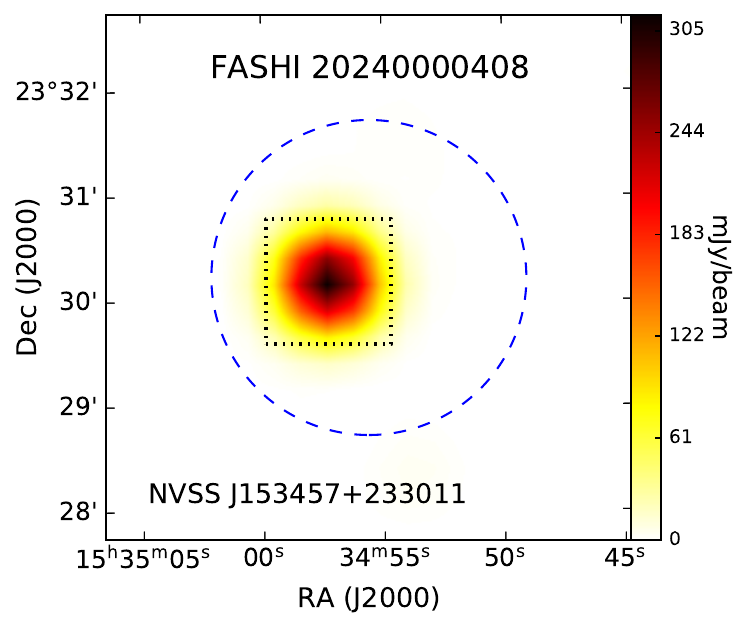}
 \includegraphics[height=0.27\textwidth, angle=0]{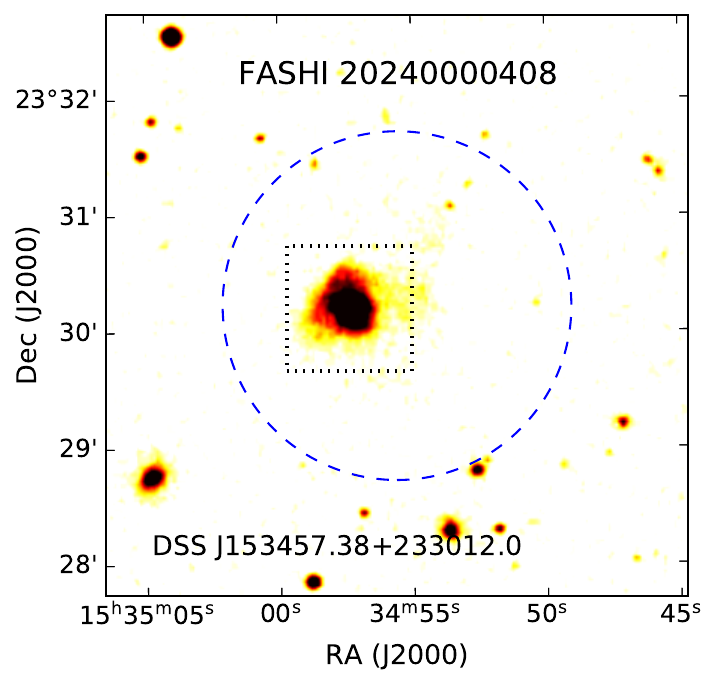}
 \caption{21\,cm H\,{\scriptsize{I}} absorption galaxy FASHI\,153455.59+233013.2 or ID\,20240000408.}
 \end{figure*} 

  \clearpage

 \begin{figure*}[htp]
 \centering
 \renewcommand{\thefigure}{\arabic{figure} (Continued)}
 \addtocounter{figure}{-1}
 \includegraphics[height=0.22\textwidth, angle=0]{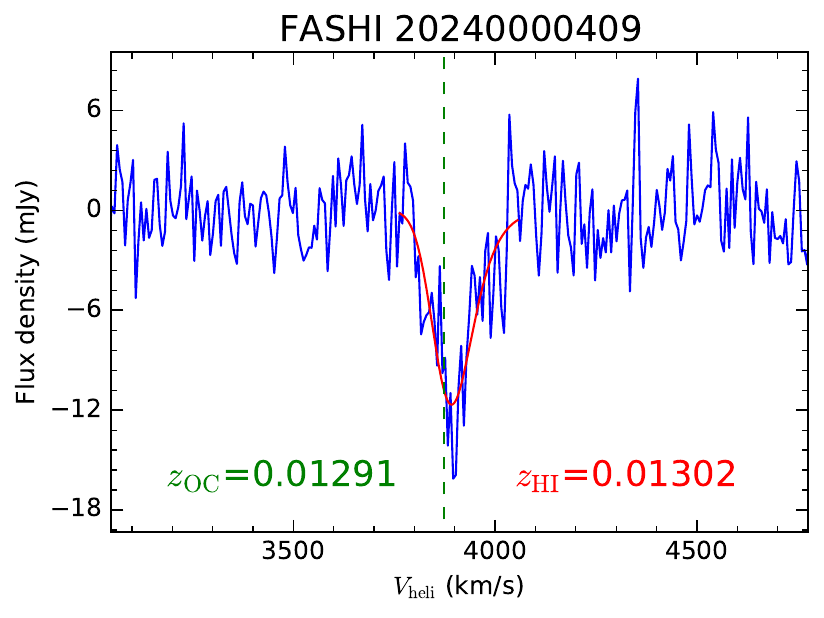}
 \includegraphics[height=0.27\textwidth, angle=0]{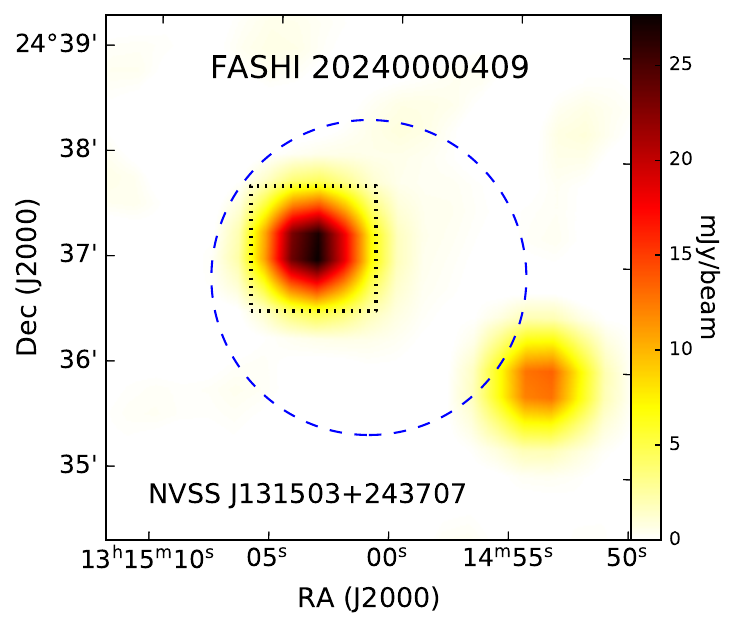}
 \includegraphics[height=0.27\textwidth, angle=0]{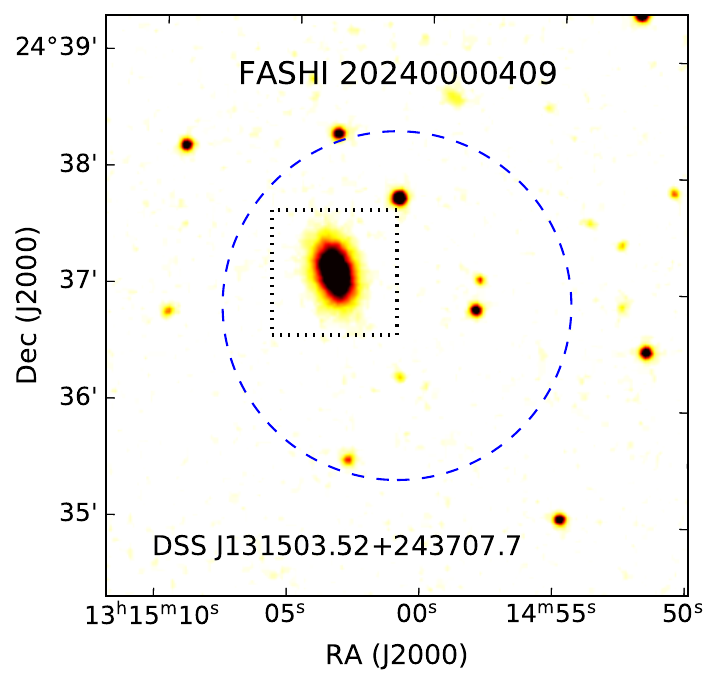}
 \caption{21\,cm H\,{\scriptsize{I}} absorption galaxy FASHI\,131501.12+243651.4 or ID\,20240000409.}
 \end{figure*} 

 \begin{figure*}[htp]
 \centering
 \renewcommand{\thefigure}{\arabic{figure} (Continued)}
 \addtocounter{figure}{-1}
 \includegraphics[height=0.22\textwidth, angle=0]{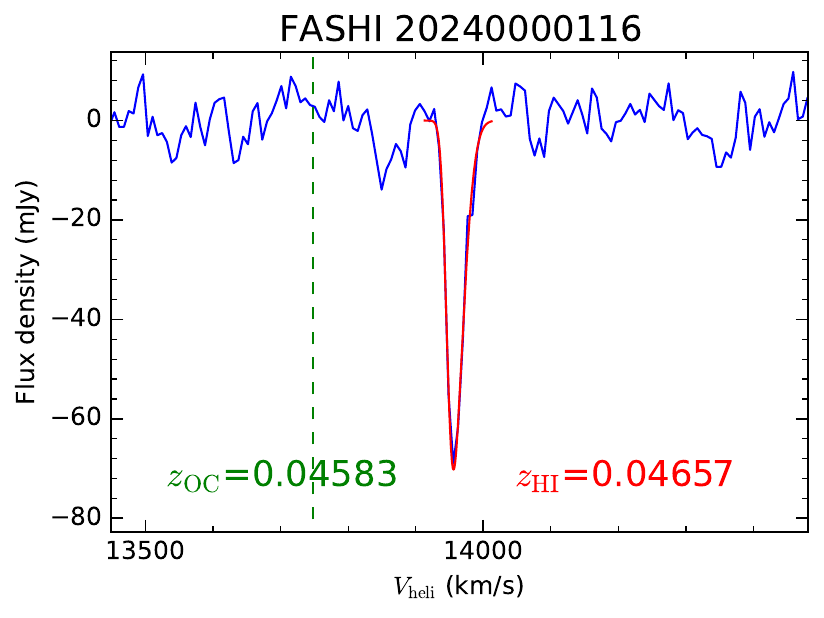}
 \includegraphics[height=0.27\textwidth, angle=0]{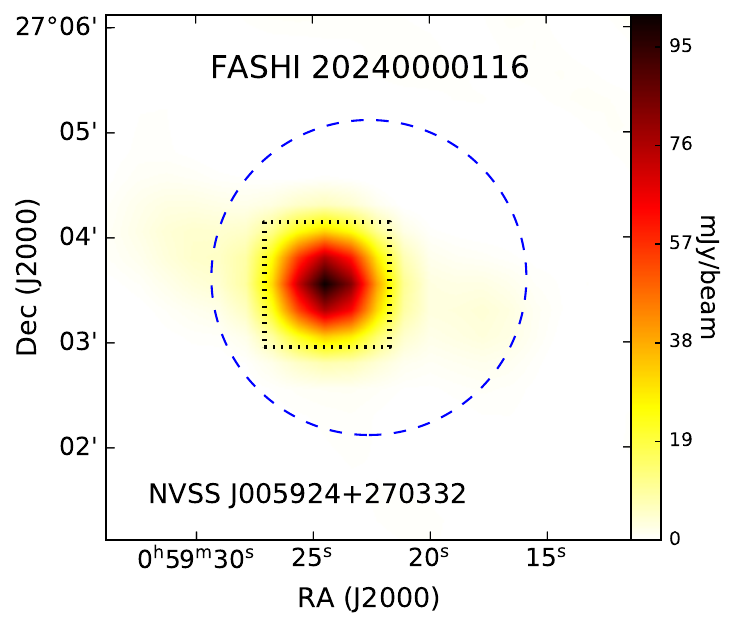}
 \includegraphics[height=0.27\textwidth, angle=0]{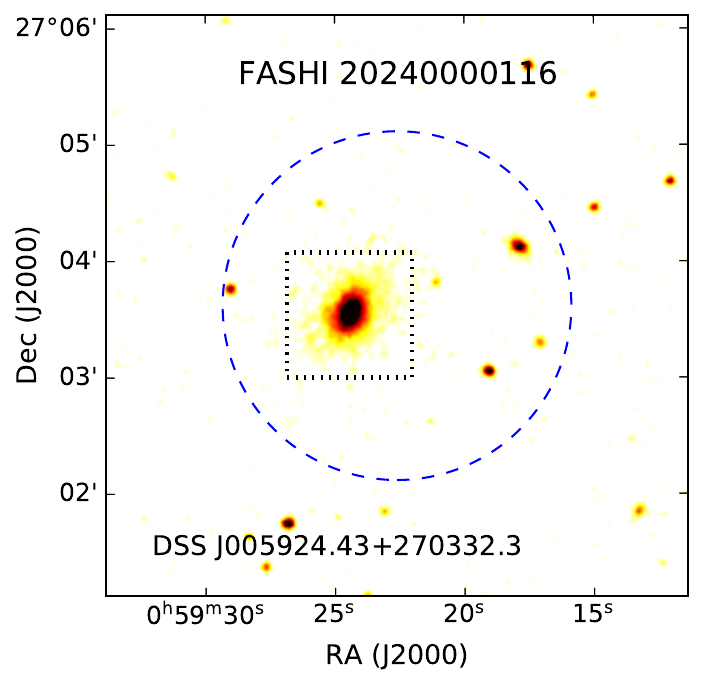}
 \caption{21\,cm H\,{\scriptsize{I}} absorption galaxy FASHI\,005922.61+270337.1 or ID\,20240000116.}
 \end{figure*} 

 \begin{figure*}[htp]
 \centering
 \renewcommand{\thefigure}{\arabic{figure} (Continued)}
 \addtocounter{figure}{-1}
 \includegraphics[height=0.22\textwidth, angle=0]{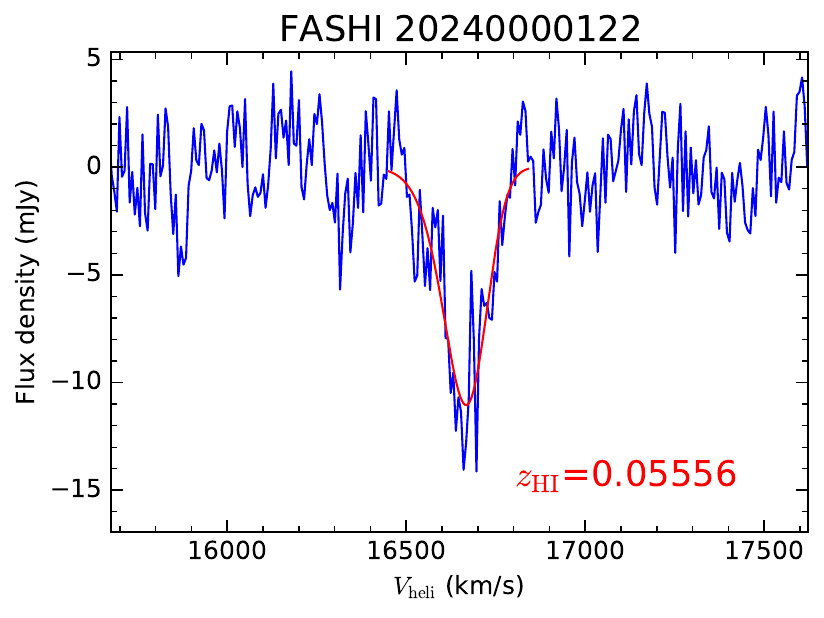}
 \includegraphics[height=0.27\textwidth, angle=0]{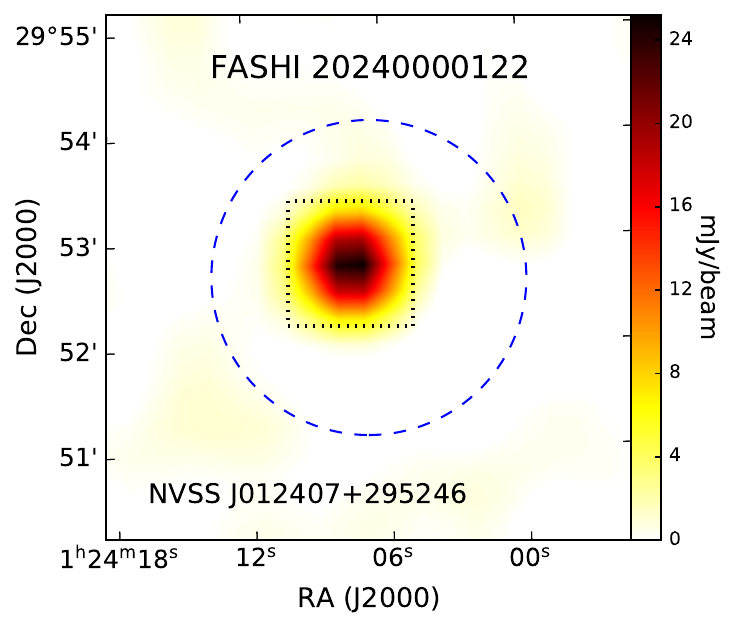}
 \includegraphics[height=0.27\textwidth, angle=0]{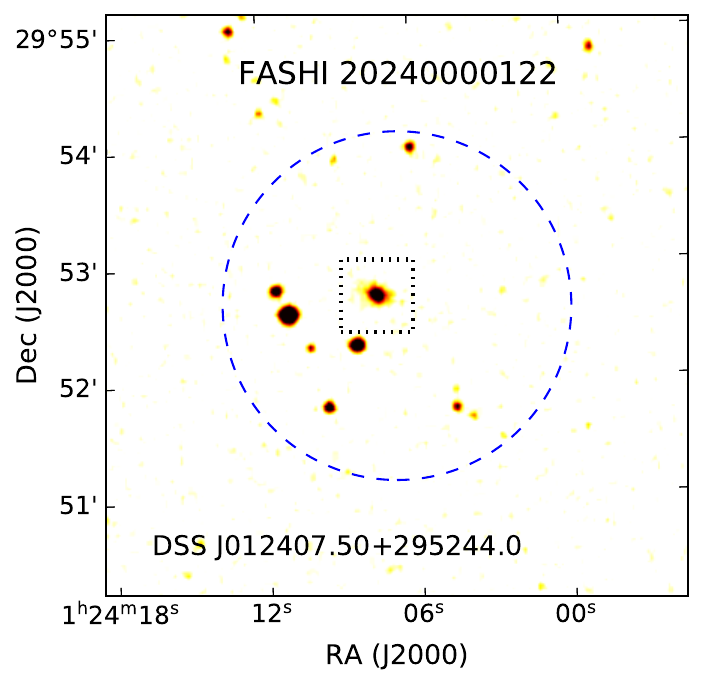}
 \caption{21\,cm H\,{\scriptsize{I}} absorption galaxy FASHI\,012406.73+295238.5 or ID\,20240000122.}
 \end{figure*} 

 \begin{figure*}[htp]
 \centering
 \renewcommand{\thefigure}{\arabic{figure} (Continued)}
 \addtocounter{figure}{-1}
 \includegraphics[height=0.22\textwidth, angle=0]{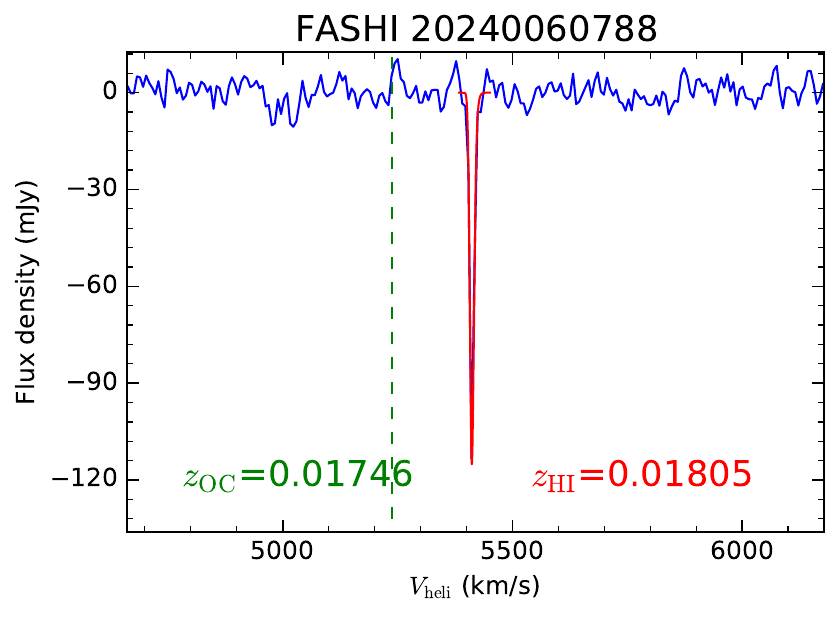}
 \includegraphics[height=0.27\textwidth, angle=0]{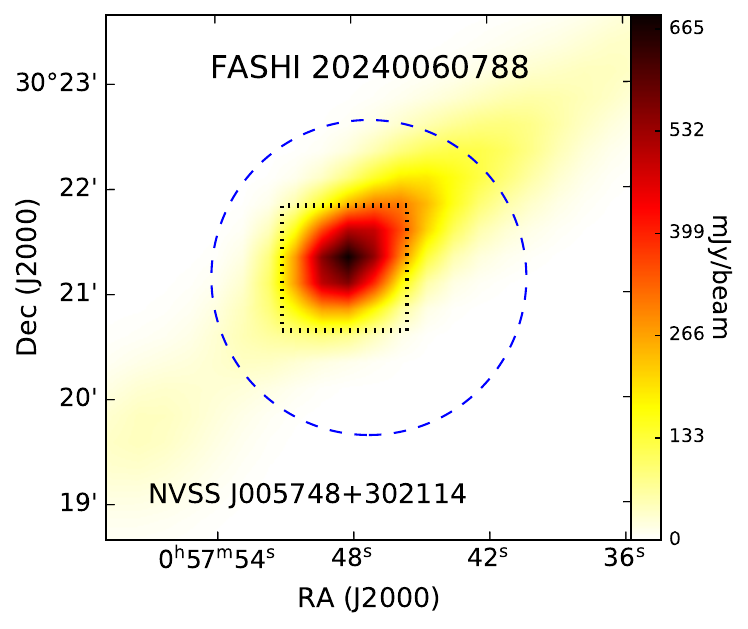}
 \includegraphics[height=0.27\textwidth, angle=0]{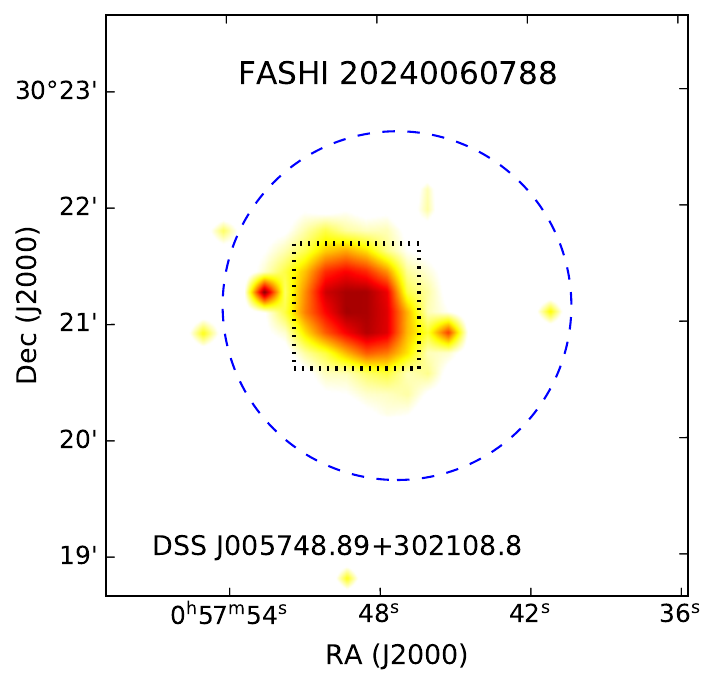}
 \caption{21\,cm H\,{\scriptsize{I}} absorption galaxy FASHI\,005747.27+302108.9 or ID\,20240060788.}
 \end{figure*} 

 \begin{figure*}[htp]
 \centering
 \renewcommand{\thefigure}{\arabic{figure} (Continued)}
 \addtocounter{figure}{-1}
 \includegraphics[height=0.22\textwidth, angle=0]{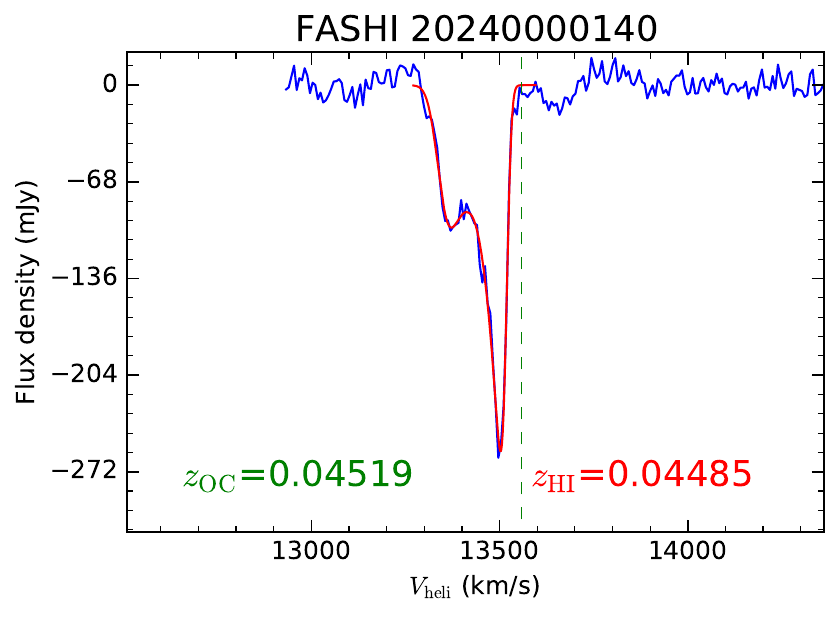}
 \includegraphics[height=0.27\textwidth, angle=0]{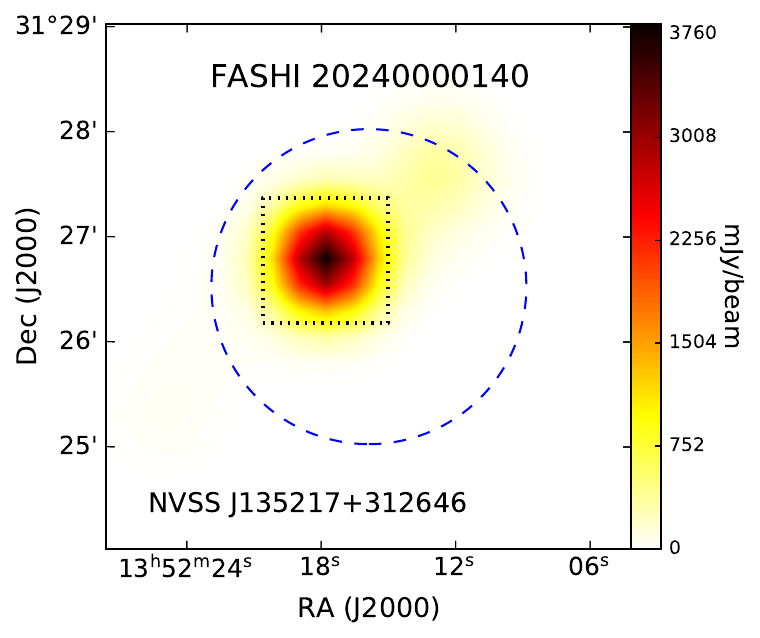}
 \includegraphics[height=0.27\textwidth, angle=0]{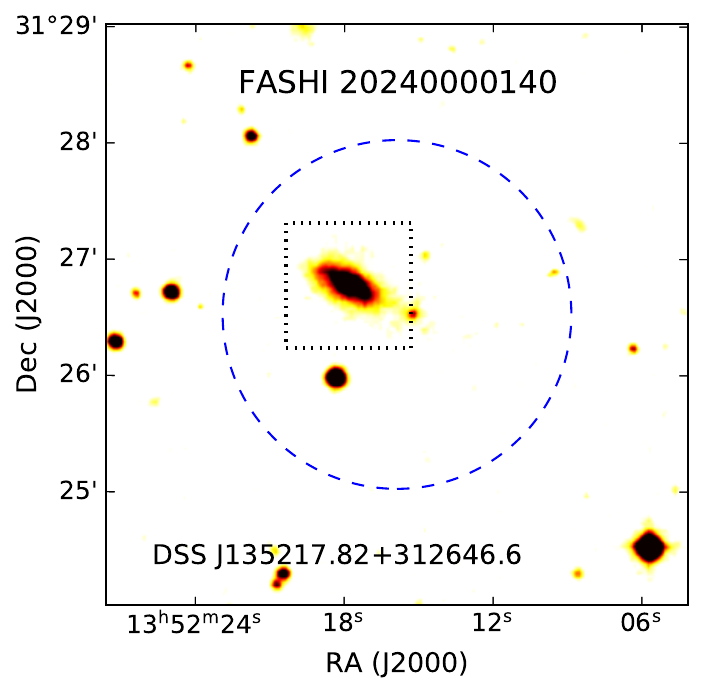}
 \caption{21\,cm H\,{\scriptsize{I}} absorption galaxy FASHI\,135215.88+312631.6 or ID\,20240000140.}
 \end{figure*} 

 \begin{figure*}[htp]
 \centering
 \renewcommand{\thefigure}{\arabic{figure} (Continued)}
 \addtocounter{figure}{-1}
 \includegraphics[height=0.22\textwidth, angle=0]{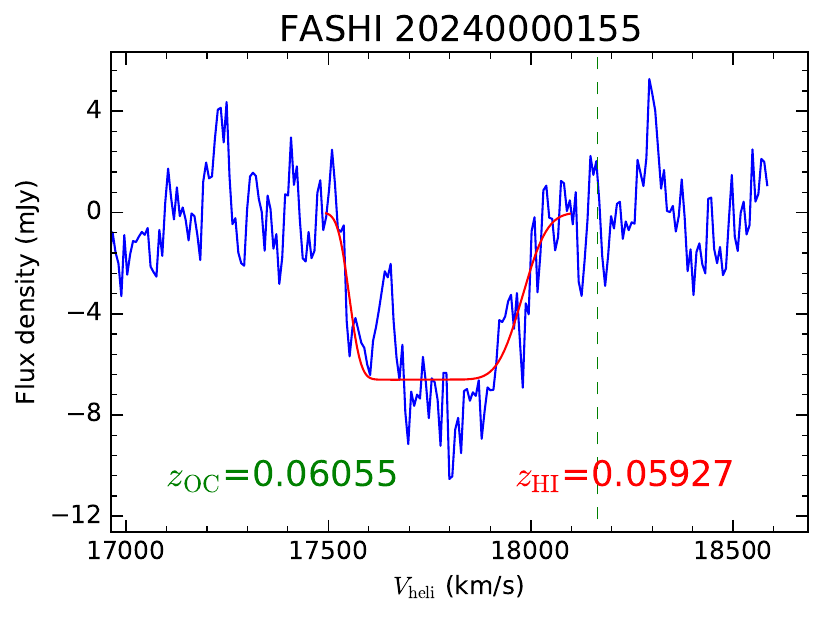}
 \includegraphics[height=0.27\textwidth, angle=0]{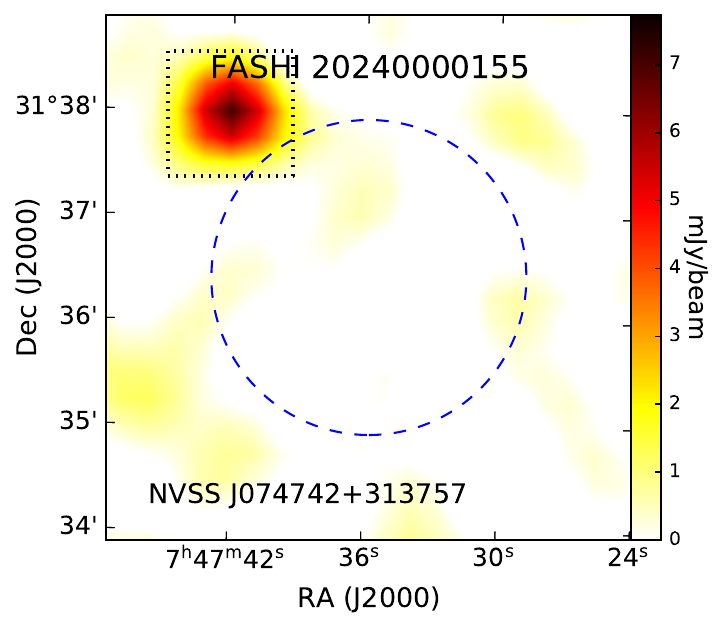}
 \includegraphics[height=0.27\textwidth, angle=0]{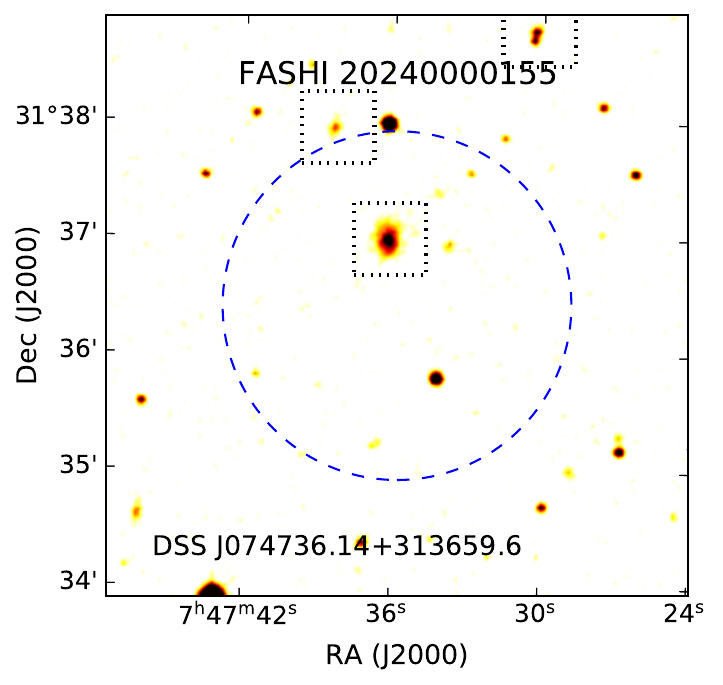}
 \caption{21\,cm H\,{\scriptsize{I}} absorption galaxy FASHI\,074735.82+313625.2 or ID\,20240000155.}
 \end{figure*} 

 \begin{figure*}[htp]
 \centering
 \renewcommand{\thefigure}{\arabic{figure} (Continued)}
 \addtocounter{figure}{-1}
 \includegraphics[height=0.22\textwidth, angle=0]{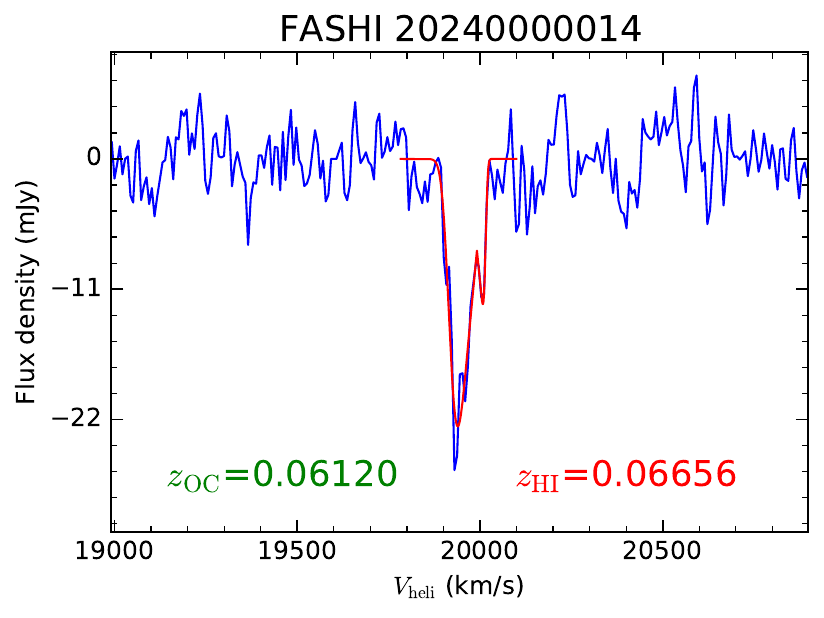}
 \includegraphics[height=0.27\textwidth, angle=0]{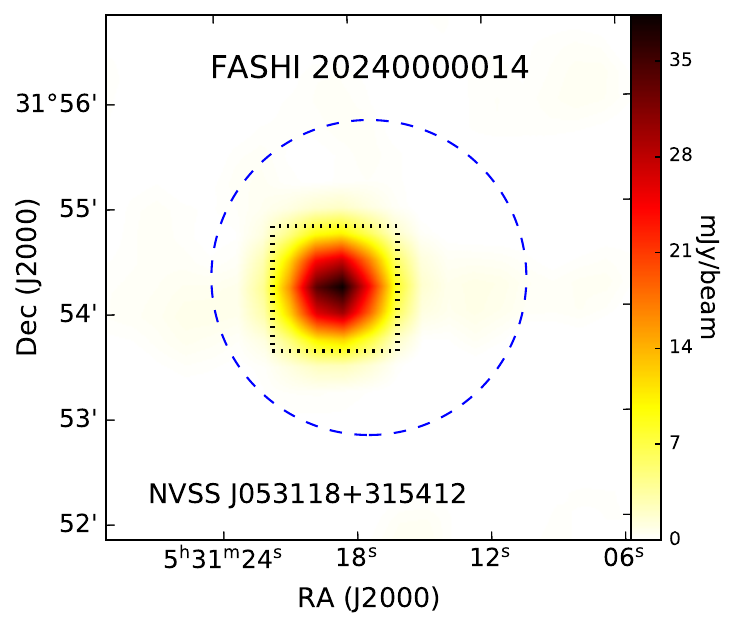}
 \includegraphics[height=0.27\textwidth, angle=0]{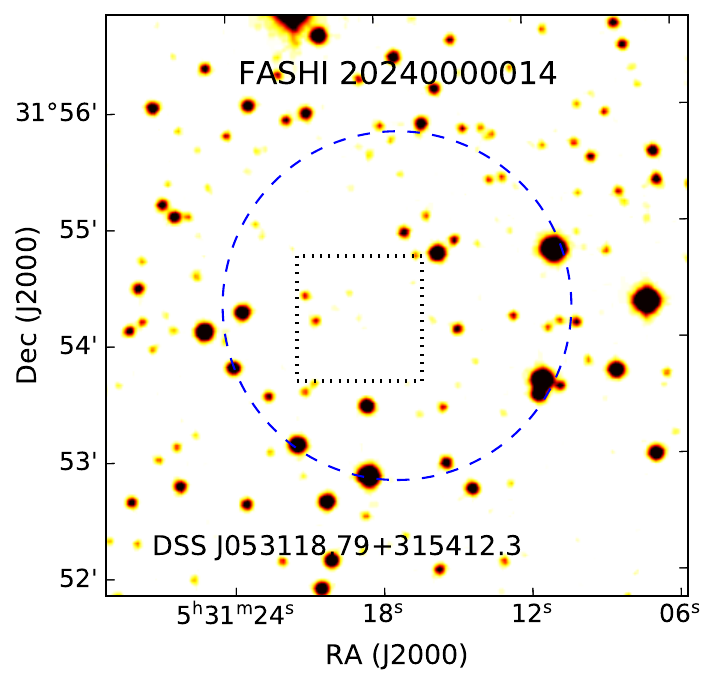}
 \caption{21\,cm H\,{\scriptsize{I}} absorption galaxy FASHI\,053117.26+315418.4 or ID\,20240000014.}
 \end{figure*} 

 \begin{figure*}[htp]
 \centering
 \renewcommand{\thefigure}{\arabic{figure} (Continued)}
 \addtocounter{figure}{-1}
 \includegraphics[height=0.22\textwidth, angle=0]{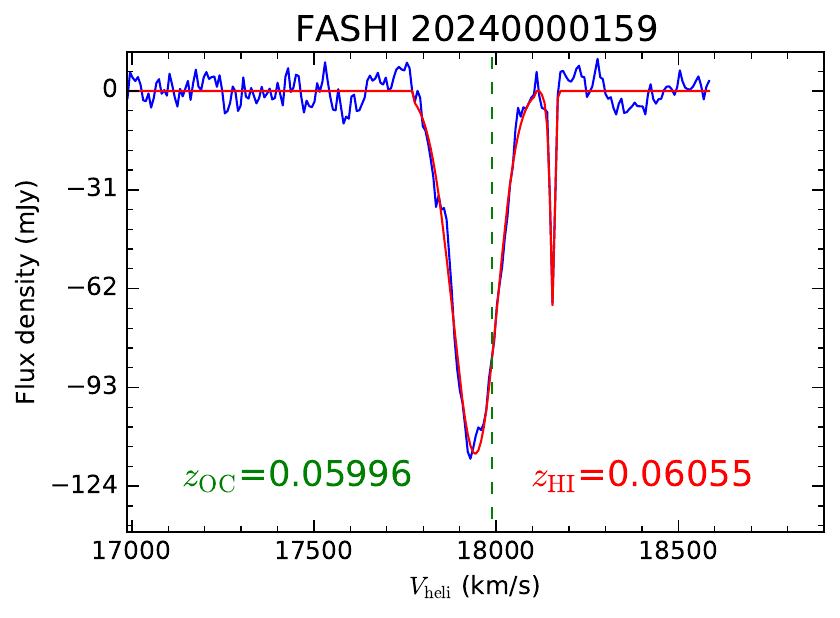}
 \includegraphics[height=0.27\textwidth, angle=0]{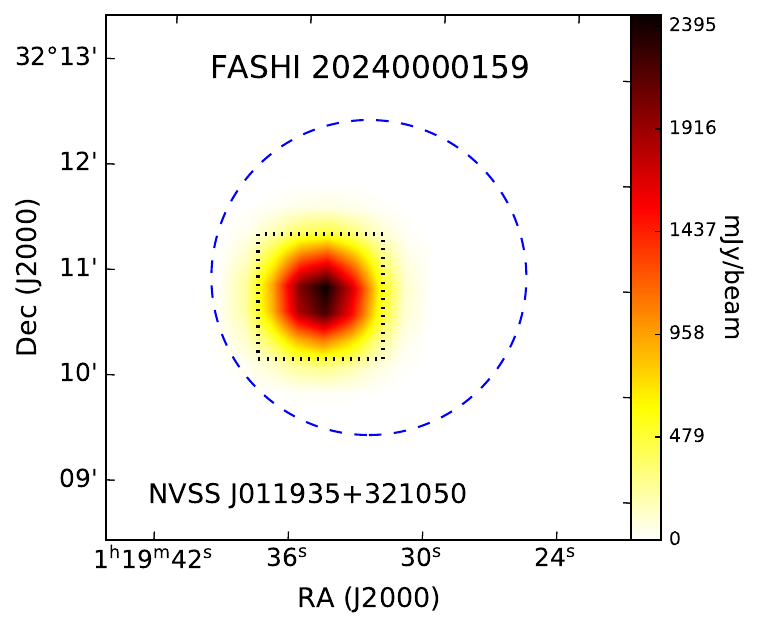}
 \includegraphics[height=0.27\textwidth, angle=0]{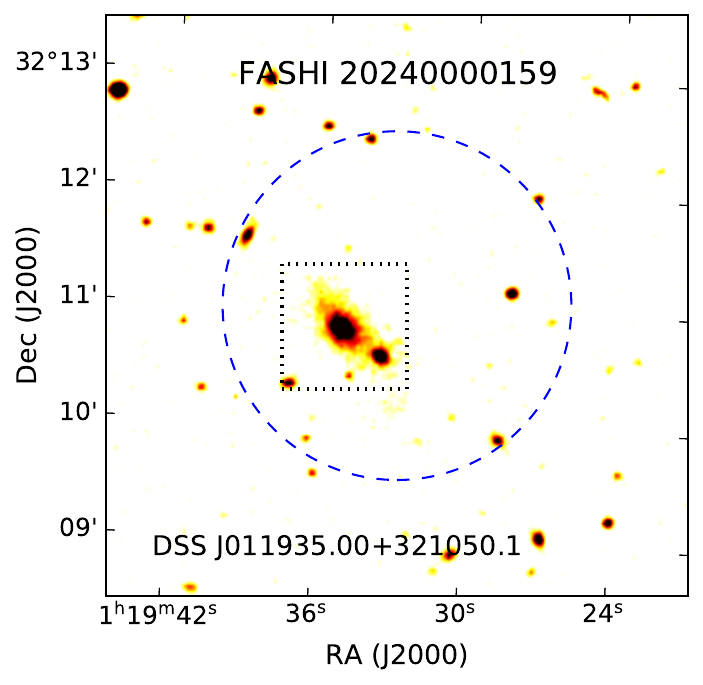}
 \caption{21\,cm H\,{\scriptsize{I}} absorption galaxy FASHI\,011932.90+321101.9 or ID\,20240000159.}
 \end{figure*} 

 \begin{figure*}[htp]
 \centering
 \renewcommand{\thefigure}{\arabic{figure} (Continued)}
 \addtocounter{figure}{-1}
 \includegraphics[height=0.22\textwidth, angle=0]{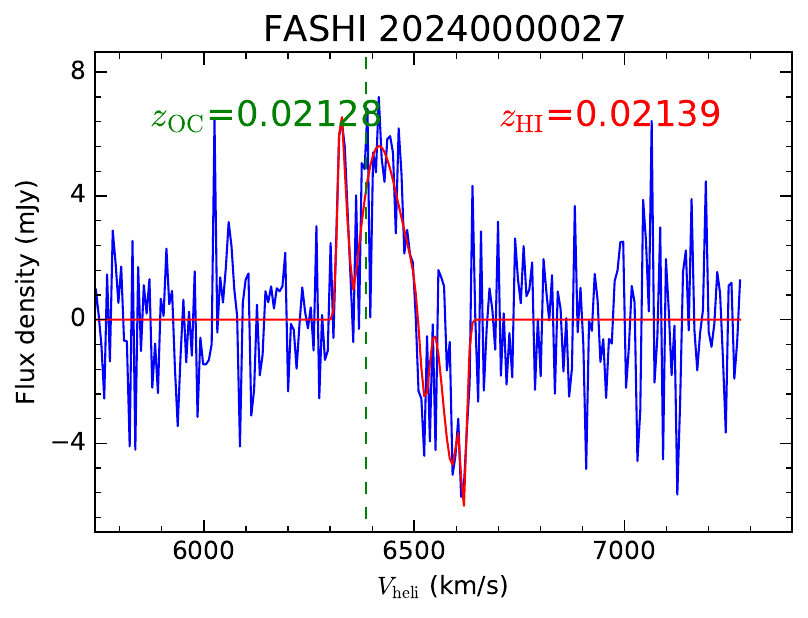}
 \includegraphics[height=0.27\textwidth, angle=0]{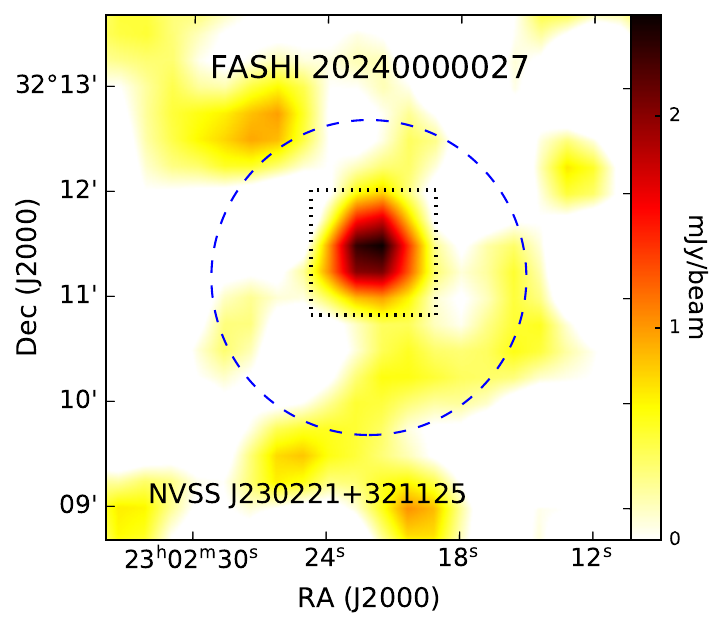}
 \includegraphics[height=0.27\textwidth, angle=0]{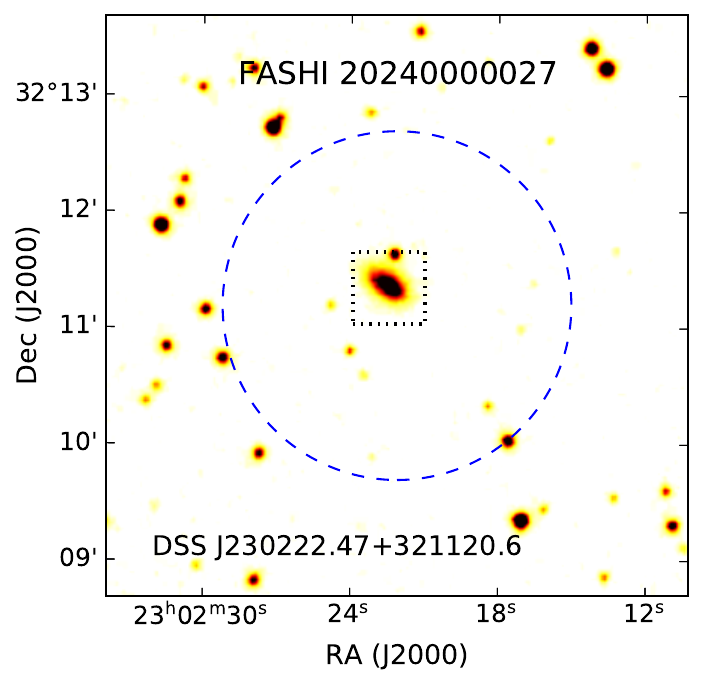}
 \caption{21\,cm H\,{\scriptsize{I}} absorption galaxy FASHI\,230222.13+321111.5 or ID\,20240000027.}
 \end{figure*} 

 \begin{figure*}[htp]
 \centering
 \renewcommand{\thefigure}{\arabic{figure} (Continued)}
 \addtocounter{figure}{-1}
 \includegraphics[height=0.22\textwidth, angle=0]{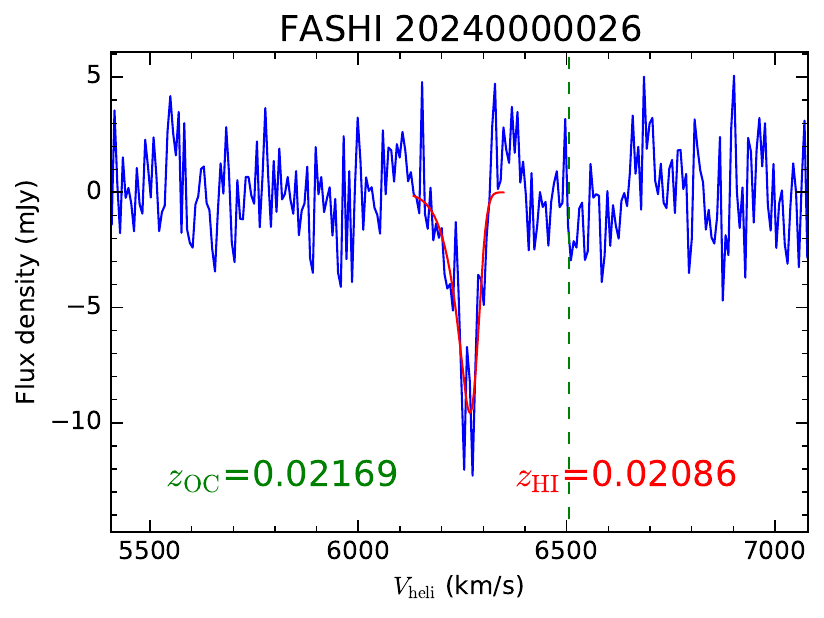}
 \includegraphics[height=0.27\textwidth, angle=0]{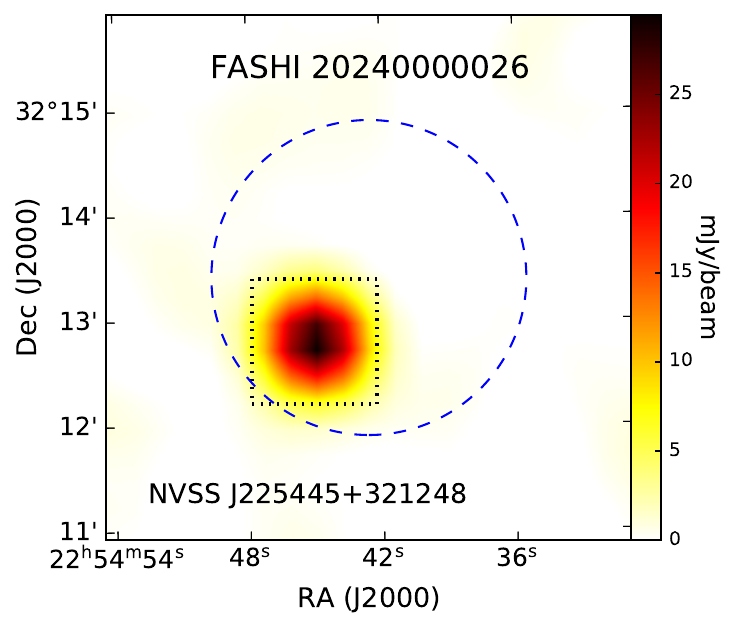}
 \includegraphics[height=0.27\textwidth, angle=0]{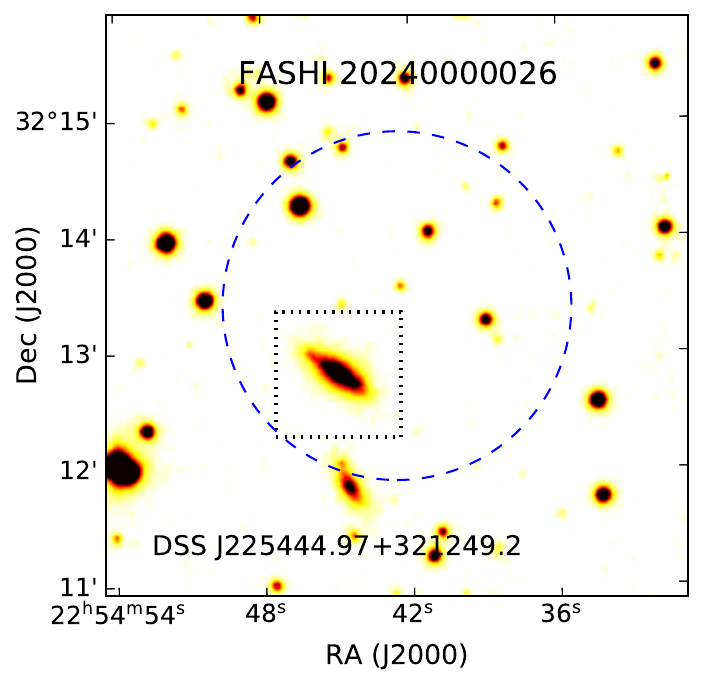}
 \caption{21\,cm H\,{\scriptsize{I}} absorption galaxy FASHI\,225442.56+321324.2 or ID\,20240000026.}
 \end{figure*} 

 \begin{figure*}[htp]
 \centering
 \renewcommand{\thefigure}{\arabic{figure} (Continued)}
 \addtocounter{figure}{-1}
 \includegraphics[height=0.22\textwidth, angle=0]{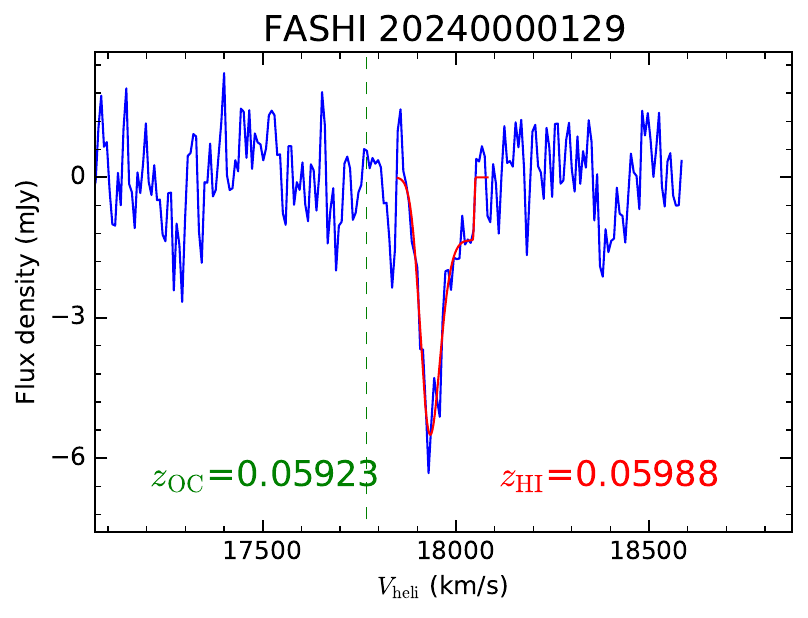}
 \includegraphics[height=0.27\textwidth, angle=0]{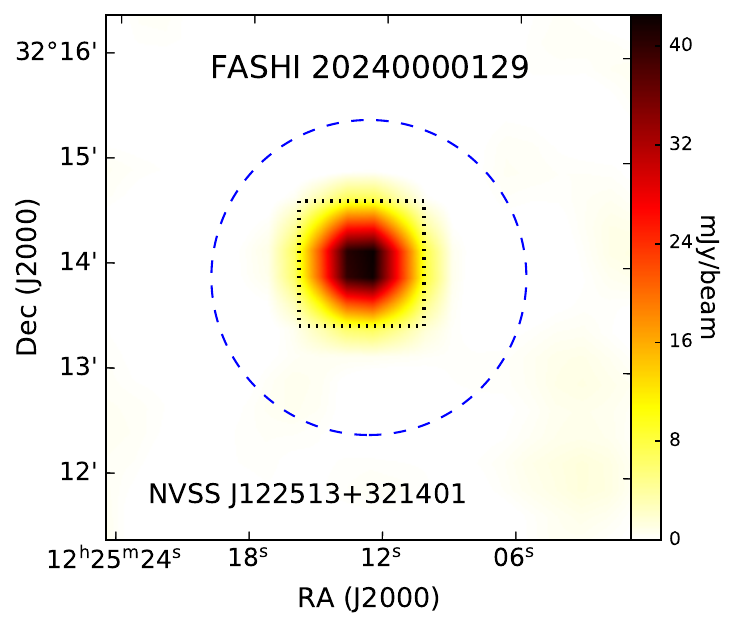}
 \includegraphics[height=0.27\textwidth, angle=0]{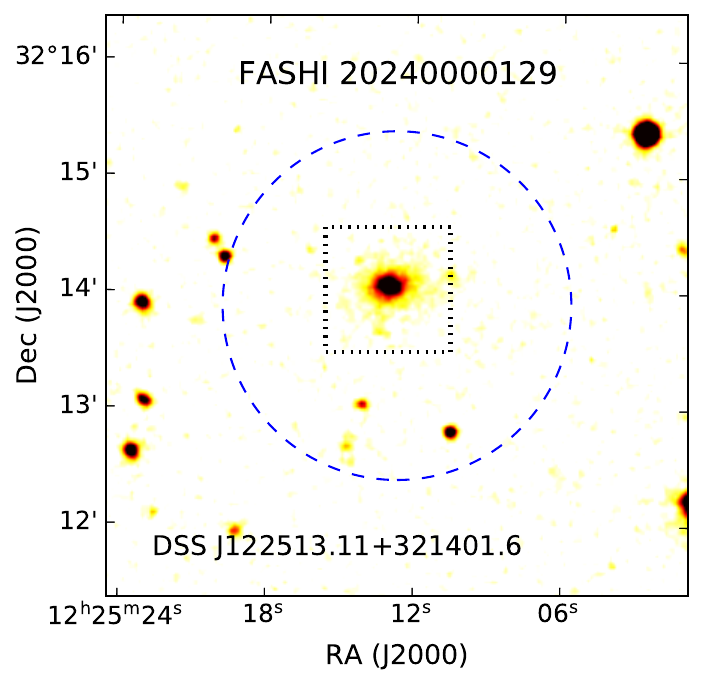}
 \caption{21\,cm H\,{\scriptsize{I}} absorption galaxy FASHI\,122512.74+321353.4 or ID\,20240000129.}
 \end{figure*} 

 \begin{figure*}[htp]
 \centering
 \renewcommand{\thefigure}{\arabic{figure} (Continued)}
 \addtocounter{figure}{-1}
 \includegraphics[height=0.22\textwidth, angle=0]{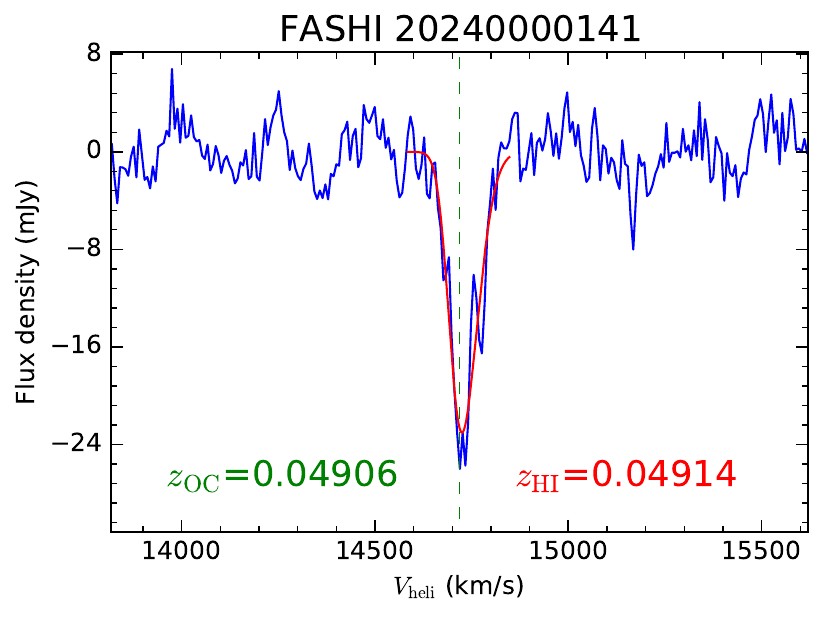}
 \includegraphics[height=0.27\textwidth, angle=0]{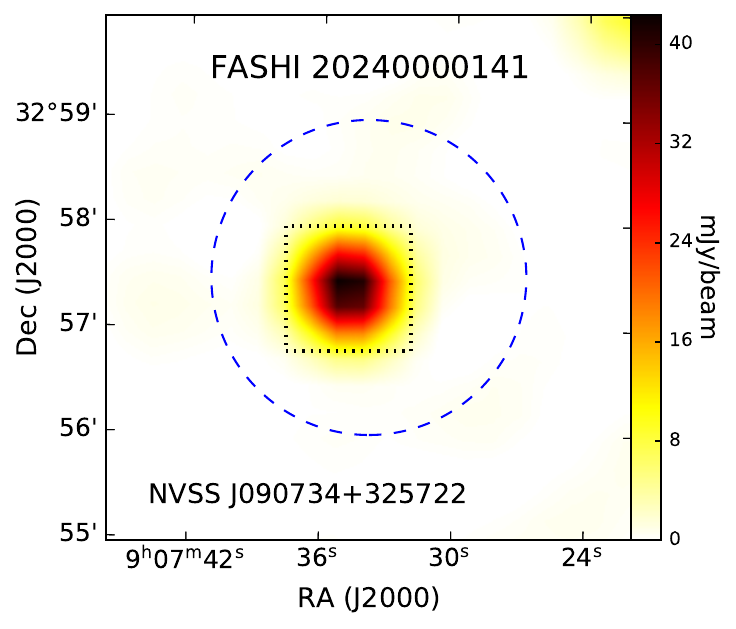}
 \includegraphics[height=0.27\textwidth, angle=0]{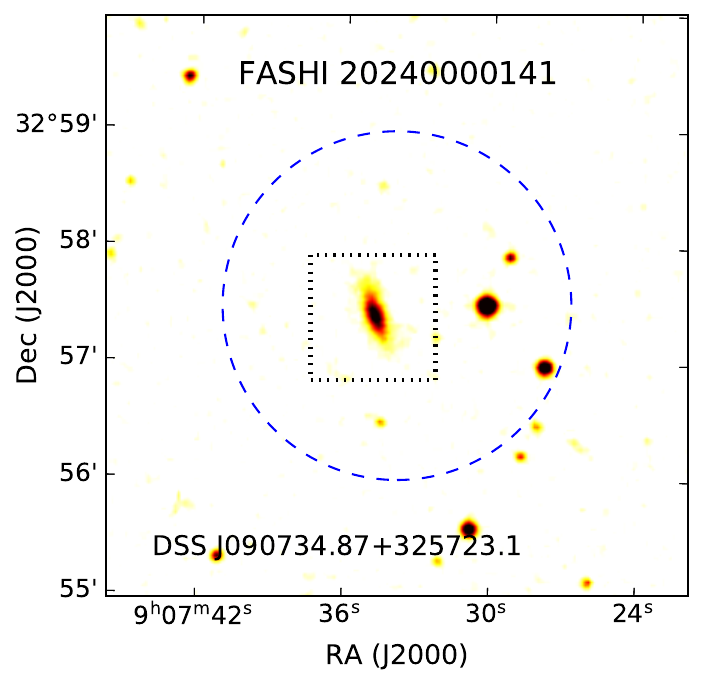}
 \caption{21\,cm H\,{\scriptsize{I}} absorption galaxy FASHI\,090733.90+325729.4 or ID\,20240000141.}
 \end{figure*} 

 \begin{figure*}[htp]
 \centering
 \renewcommand{\thefigure}{\arabic{figure} (Continued)}
 \addtocounter{figure}{-1}
 \includegraphics[height=0.22\textwidth, angle=0]{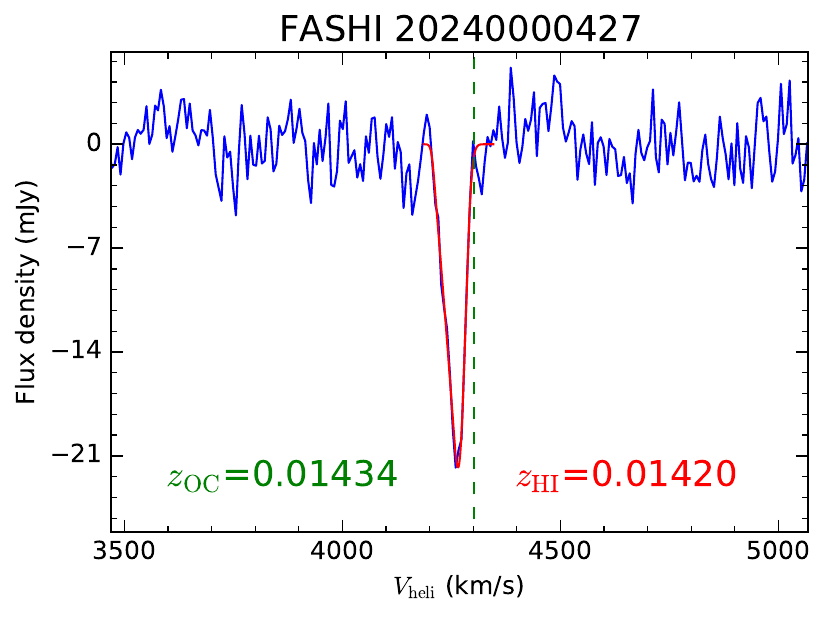}
 \includegraphics[height=0.27\textwidth, angle=0]{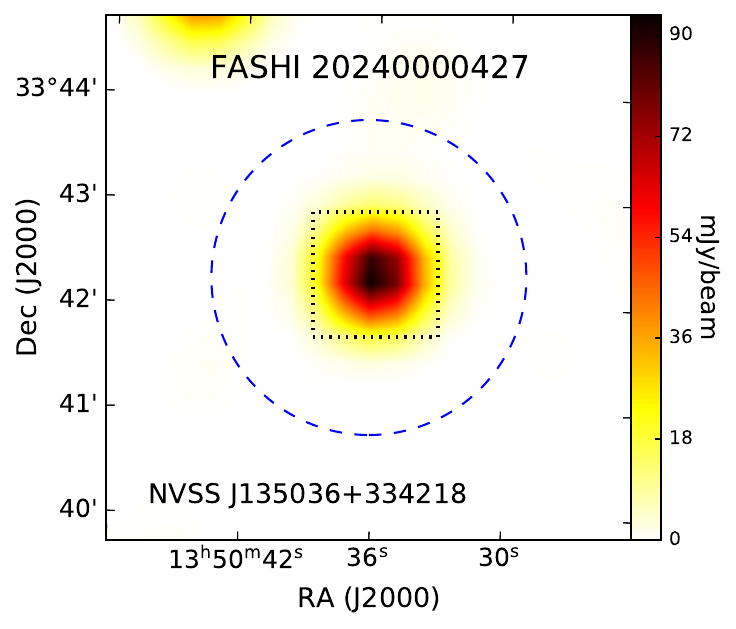}
 \includegraphics[height=0.27\textwidth, angle=0]{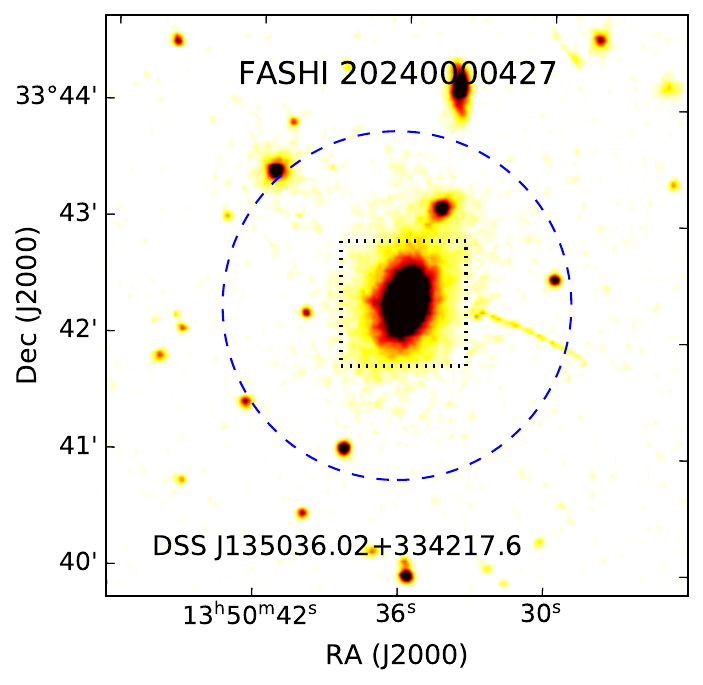}
 \caption{21\,cm H\,{\scriptsize{I}} absorption galaxy FASHI\,135036.30+334216.5 or ID\,20240000427.}
 \end{figure*} 

 \begin{figure*}[htp]
 \centering
 \renewcommand{\thefigure}{\arabic{figure} (Continued)}
 \addtocounter{figure}{-1}
 \includegraphics[height=0.22\textwidth, angle=0]{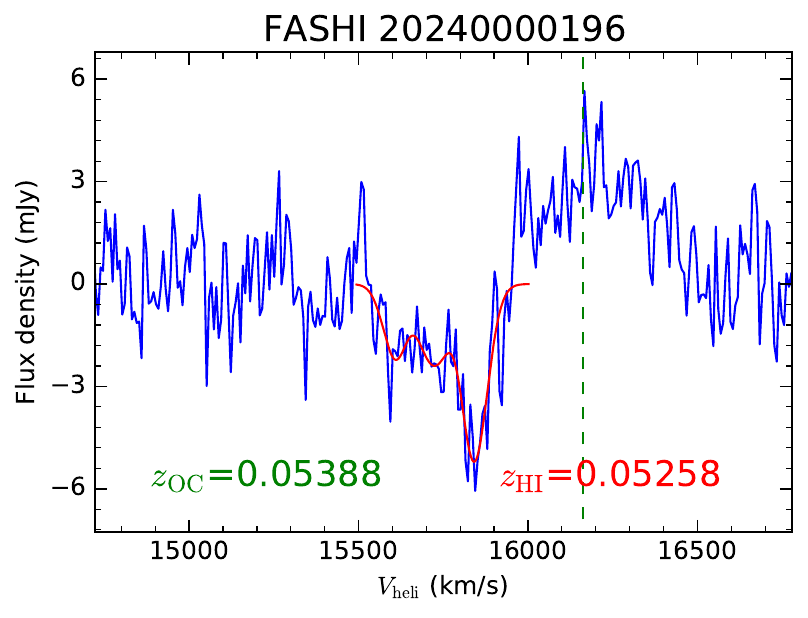}
 \includegraphics[height=0.27\textwidth, angle=0]{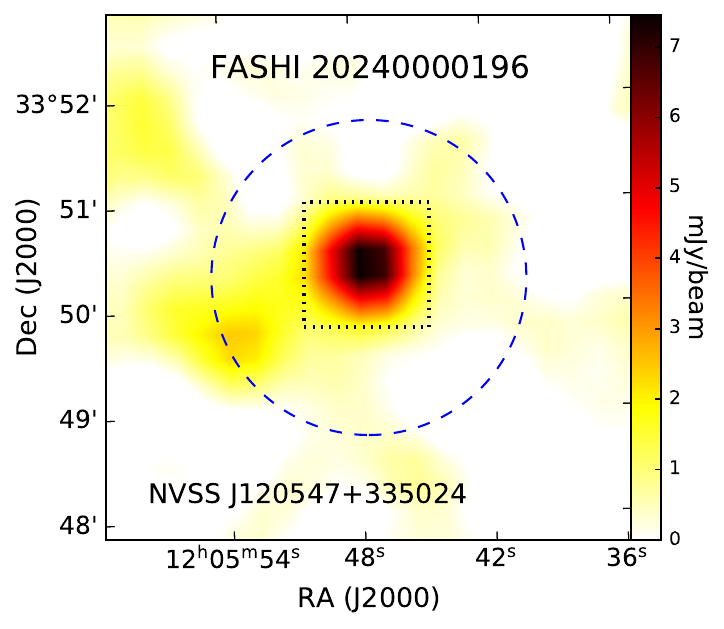}
 \includegraphics[height=0.27\textwidth, angle=0]{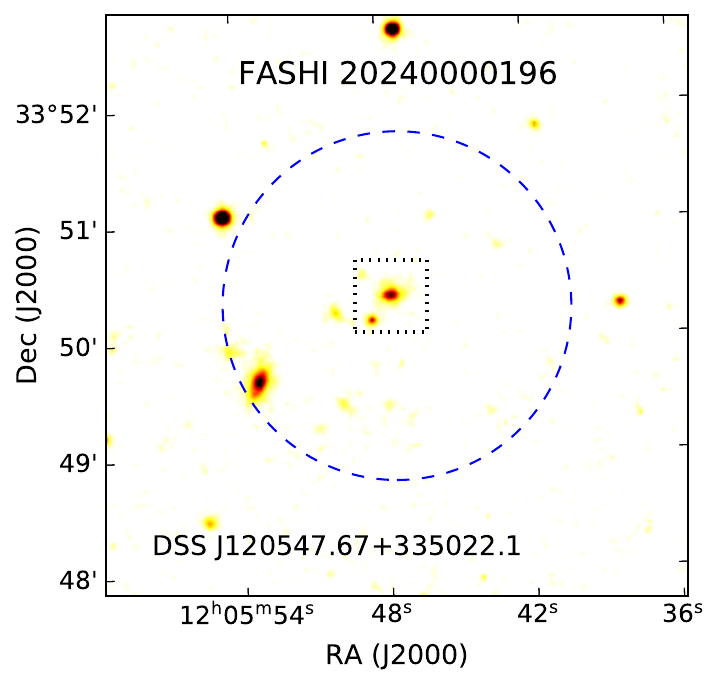}
 \caption{21\,cm H\,{\scriptsize{I}} absorption galaxy FASHI\,120547.43+335016.9 or ID\,20240000196.}
 \end{figure*} 

 \begin{figure*}[htp]
 \centering
 \renewcommand{\thefigure}{\arabic{figure} (Continued)}
 \addtocounter{figure}{-1}
 \includegraphics[height=0.22\textwidth, angle=0]{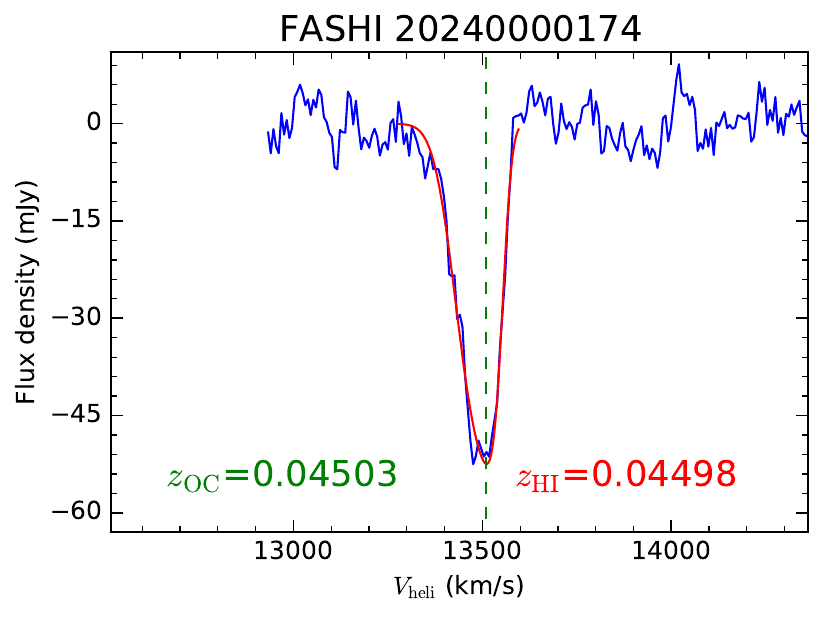}
 \includegraphics[height=0.27\textwidth, angle=0]{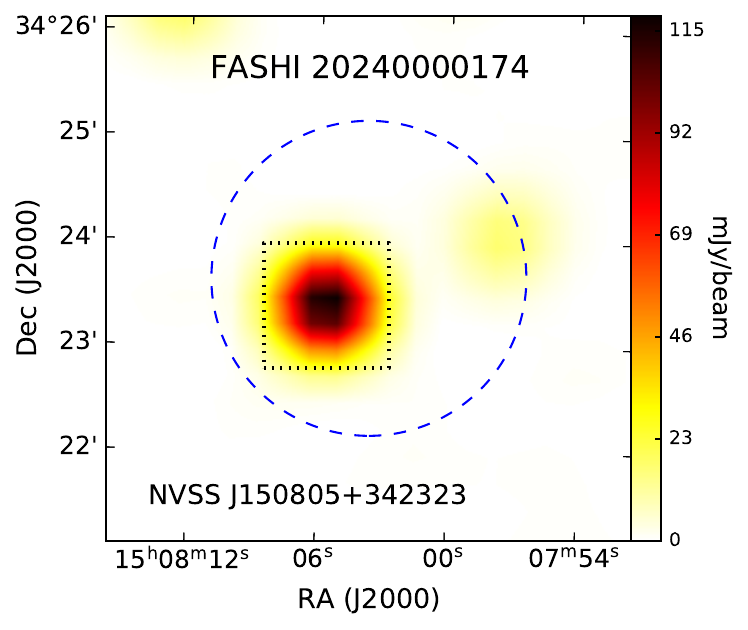}
 \includegraphics[height=0.27\textwidth, angle=0]{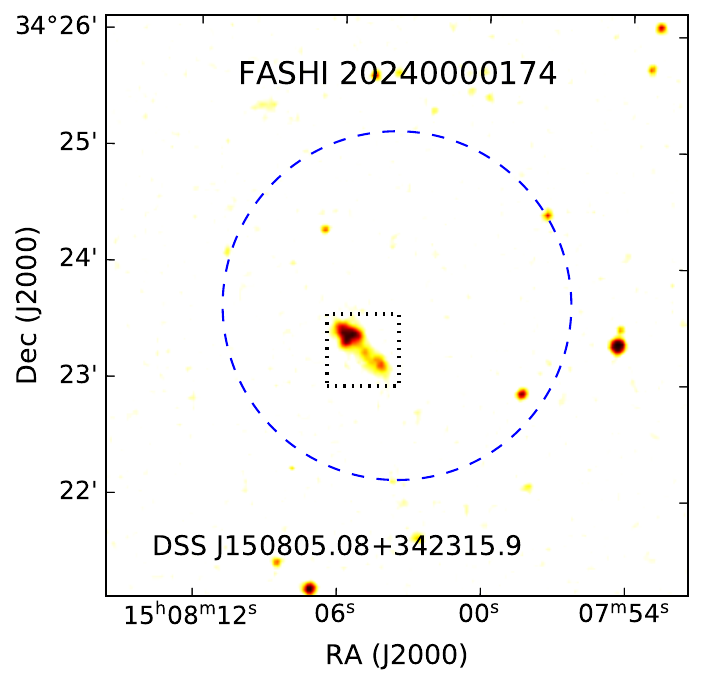}
 \caption{21\,cm H\,{\scriptsize{I}} absorption galaxy FASHI\,150803.69+342339.2 or ID\,20240000174.}
 \end{figure*} 

 \begin{figure*}[htp]
 \centering
 \renewcommand{\thefigure}{\arabic{figure} (Continued)}
 \addtocounter{figure}{-1}
 \includegraphics[height=0.22\textwidth, angle=0]{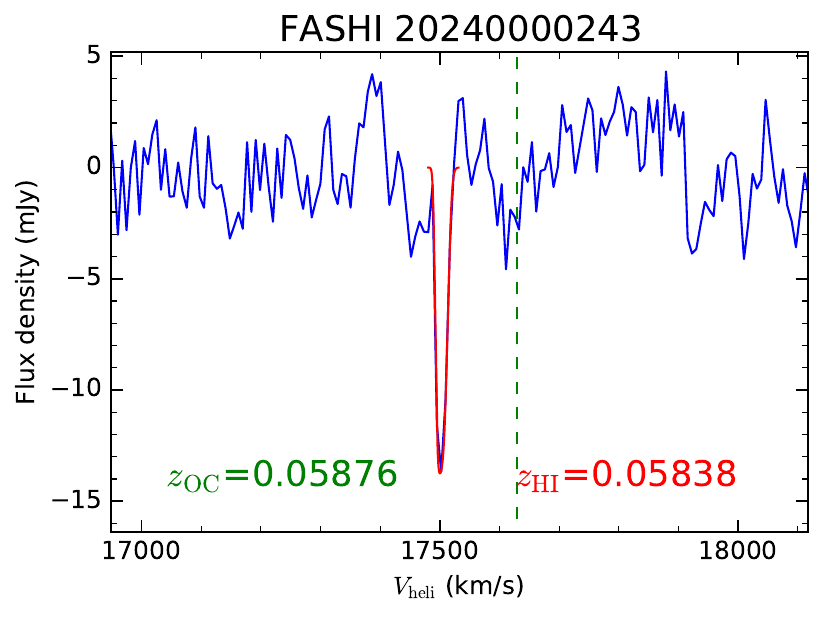}
 \includegraphics[height=0.27\textwidth, angle=0]{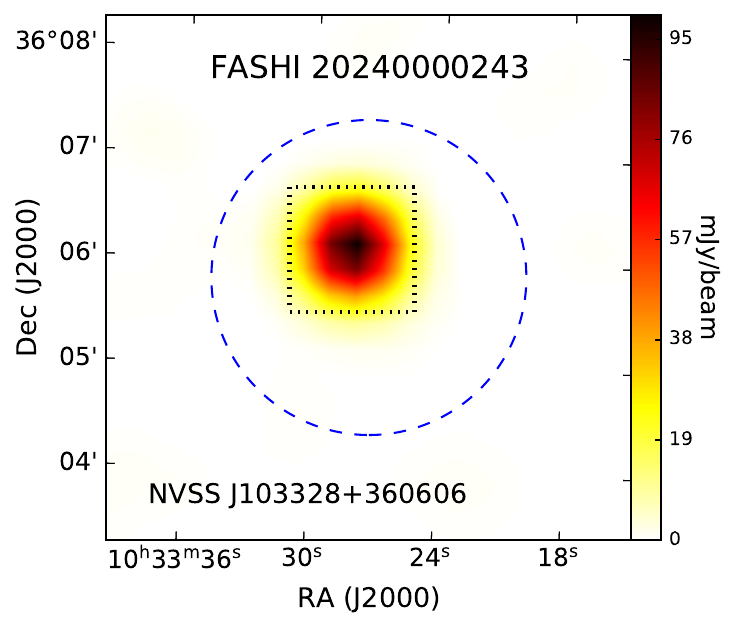}
 \includegraphics[height=0.27\textwidth, angle=0]{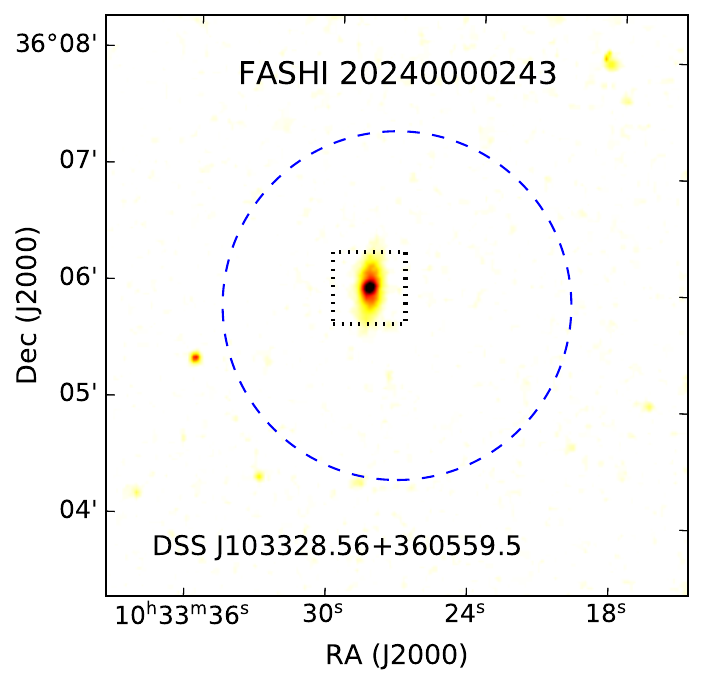}
 \caption{21\,cm H\,{\scriptsize{I}} absorption galaxy FASHI\,103327.36+360550.9 or ID\,20240000243.}
 \end{figure*} 

 \begin{figure*}[htp]
 \centering
 \renewcommand{\thefigure}{\arabic{figure} (Continued)}
 \addtocounter{figure}{-1}
 \includegraphics[height=0.22\textwidth, angle=0]{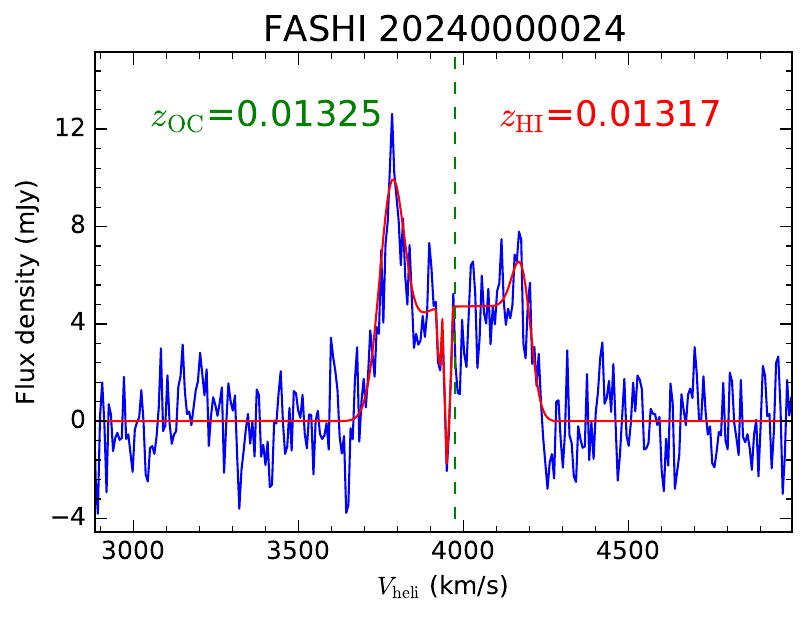}
 \includegraphics[height=0.27\textwidth, angle=0]{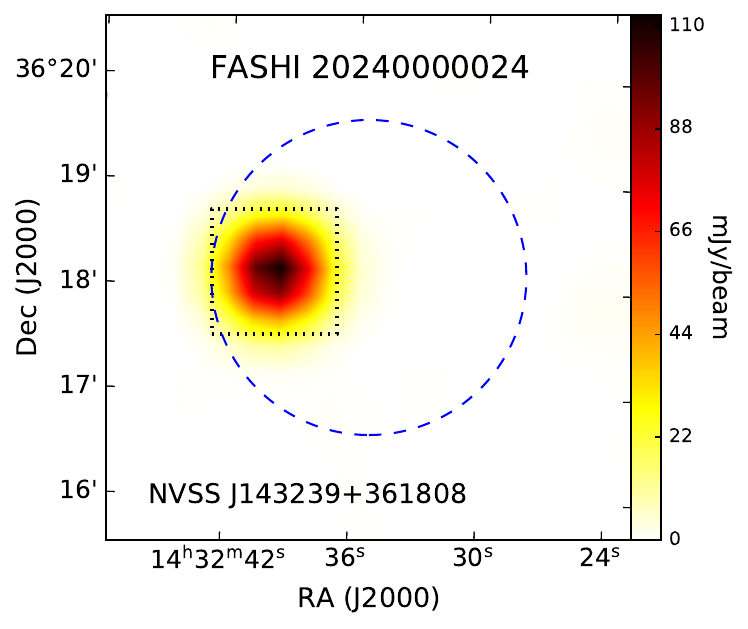}
 \includegraphics[height=0.27\textwidth, angle=0]{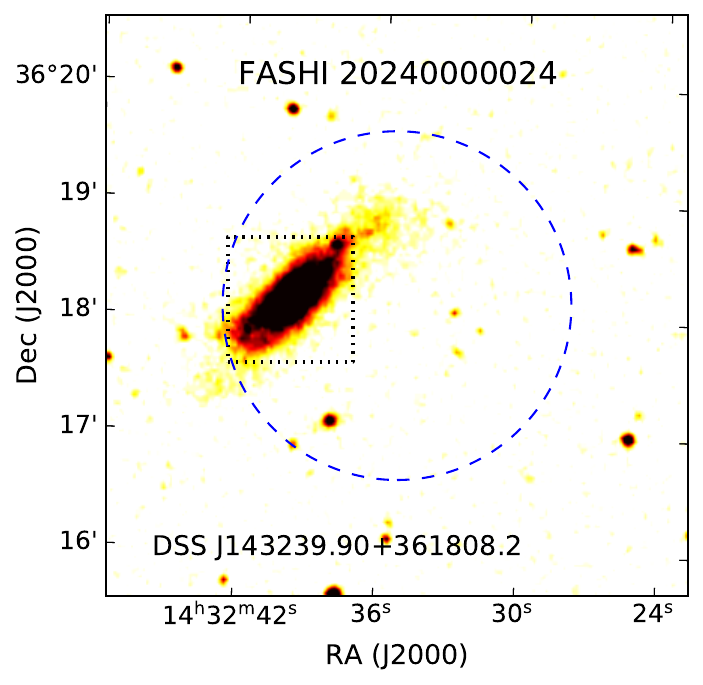}
 \caption{21\,cm H\,{\scriptsize{I}} absorption galaxy FASHI\,143235.35+361806.7 or ID\,20240000024.}
 \end{figure*} 

 \begin{figure*}[htp]
 \centering
 \renewcommand{\thefigure}{\arabic{figure} (Continued)}
 \addtocounter{figure}{-1}
 \includegraphics[height=0.22\textwidth, angle=0]{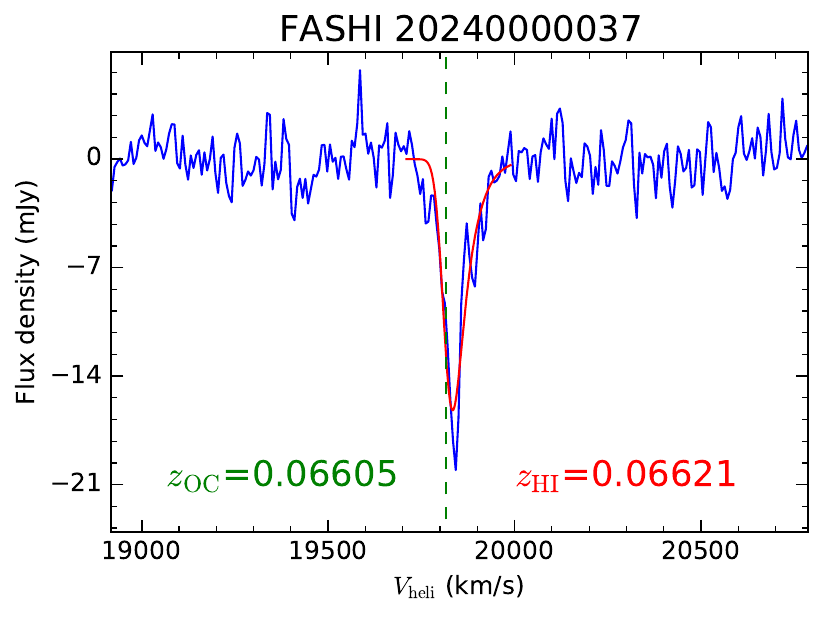}
 \includegraphics[height=0.27\textwidth, angle=0]{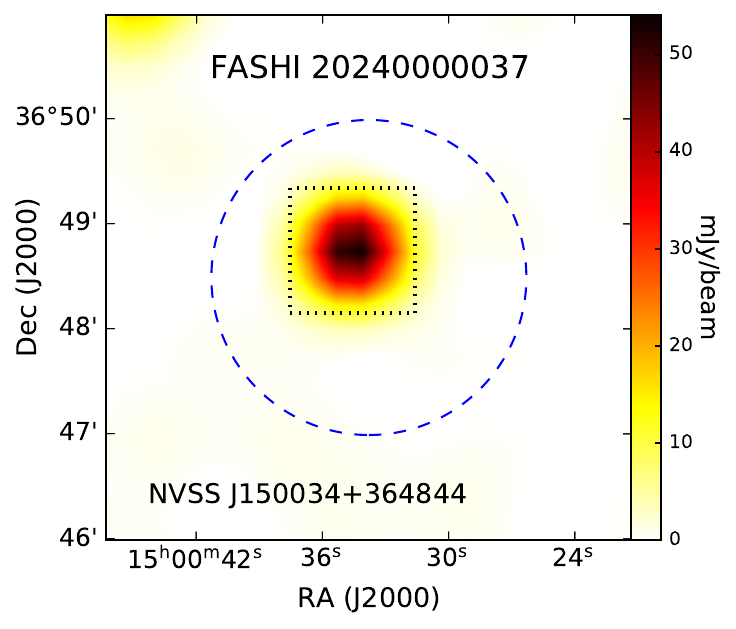}
 \includegraphics[height=0.27\textwidth, angle=0]{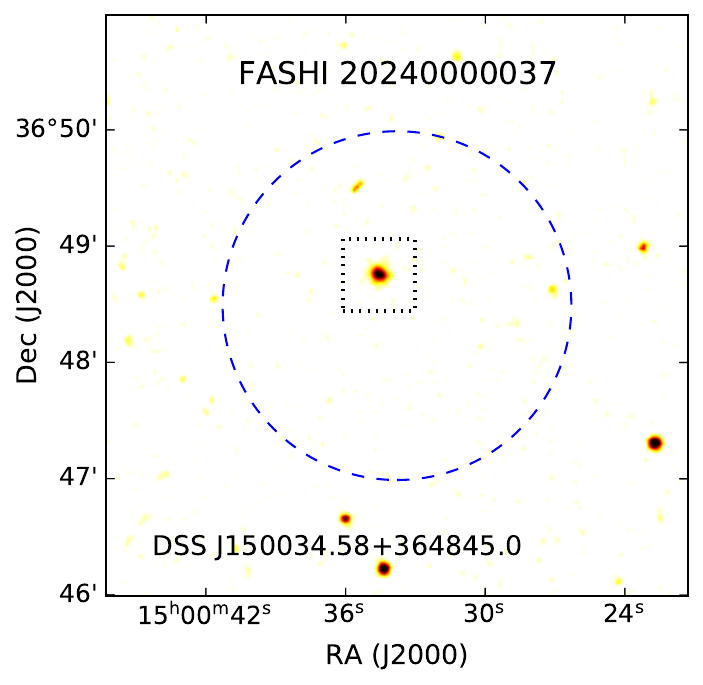}
 \caption{21\,cm H\,{\scriptsize{I}} absorption galaxy FASHI\,150033.79+364829.3 or ID\,20240000037.}
 \end{figure*} 

 \begin{figure*}[htp]
 \centering
 \renewcommand{\thefigure}{\arabic{figure} (Continued)}
 \addtocounter{figure}{-1}
 \includegraphics[height=0.22\textwidth, angle=0]{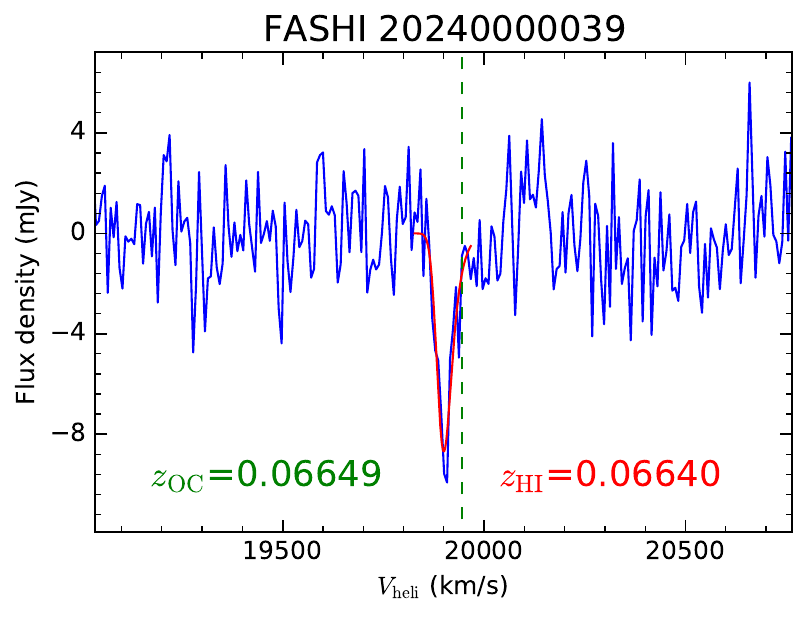}
 \includegraphics[height=0.27\textwidth, angle=0]{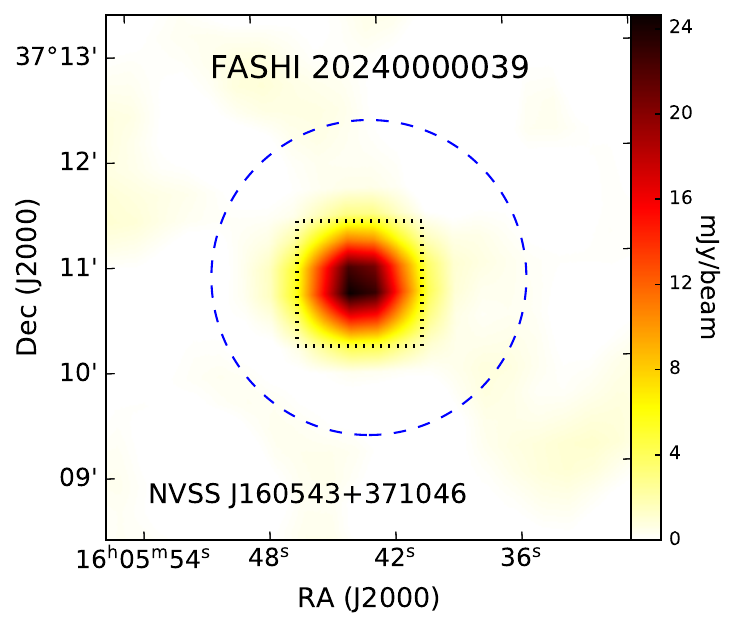}
 \includegraphics[height=0.27\textwidth, angle=0]{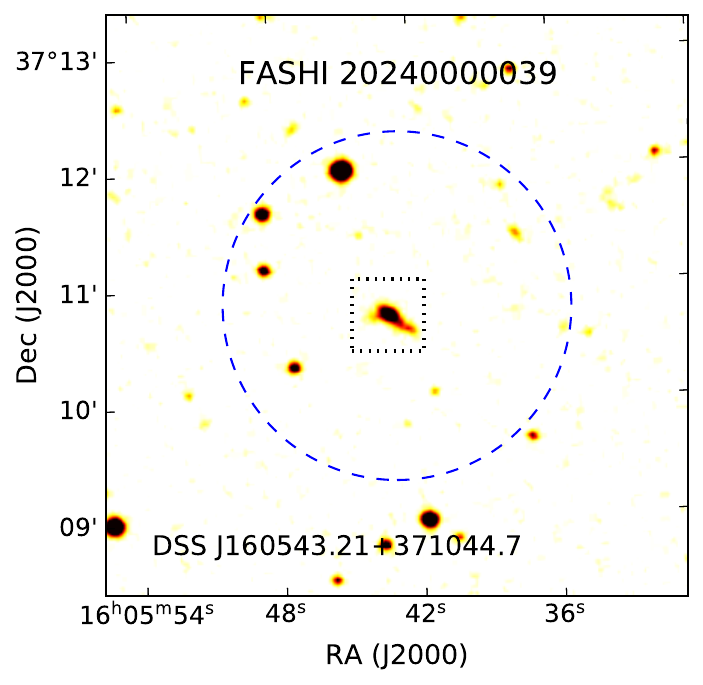}
 \caption{21\,cm H\,{\scriptsize{I}} absorption galaxy FASHI\,160542.81+371049.2 or ID\,20240000039.}
 \end{figure*} 

 \begin{figure*}[htp]
 \centering
 \renewcommand{\thefigure}{\arabic{figure} (Continued)}
 \addtocounter{figure}{-1}
 \includegraphics[height=0.22\textwidth, angle=0]{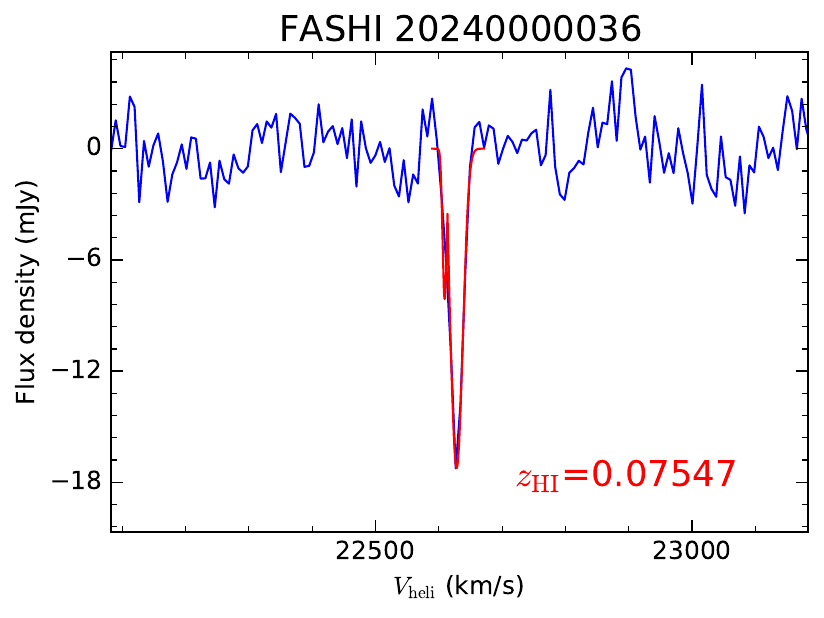}
 \includegraphics[height=0.27\textwidth, angle=0]{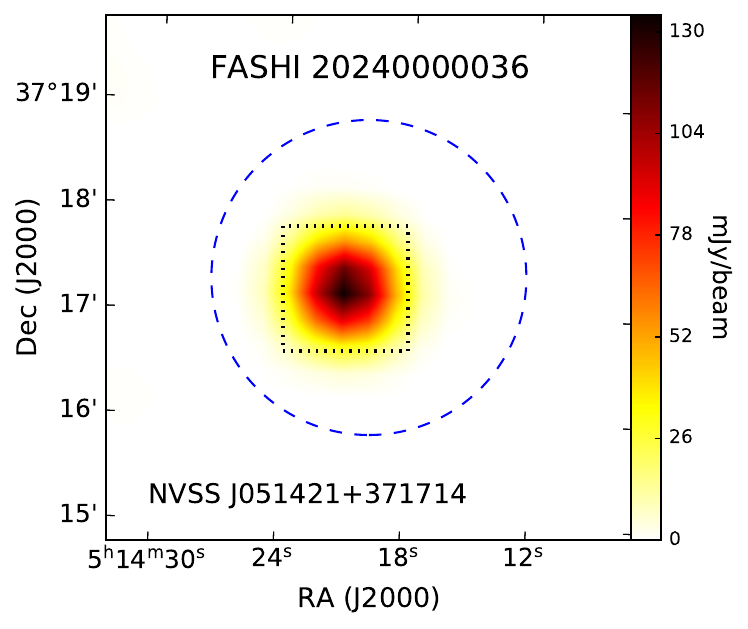}
 \includegraphics[height=0.27\textwidth, angle=0]{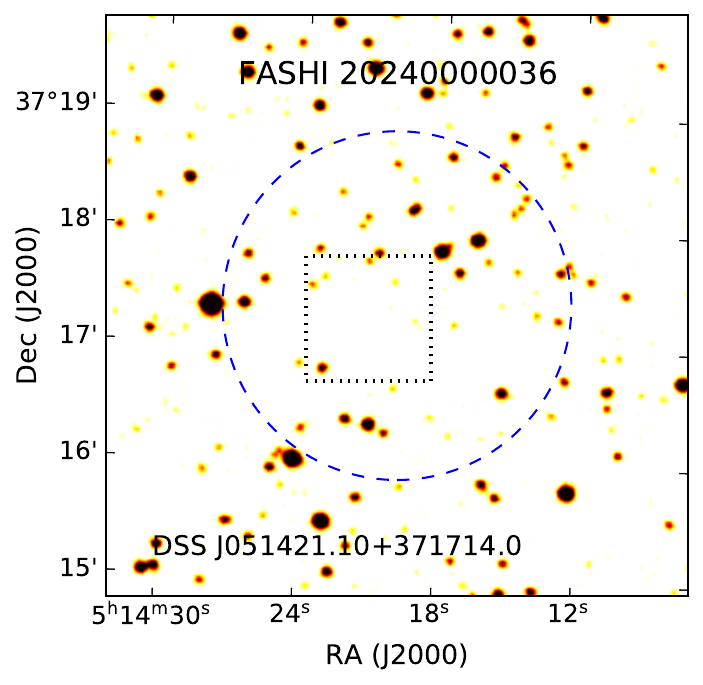}
 \caption{21\,cm H\,{\scriptsize{I}} absorption galaxy FASHI\,051419.90+371721.2 or ID\,20240000036.}
 \end{figure*}

 \begin{figure*}[htp]
 \centering
 \renewcommand{\thefigure}{\arabic{figure} (Continued)}
 \addtocounter{figure}{-1}
 \includegraphics[height=0.22\textwidth, angle=0]{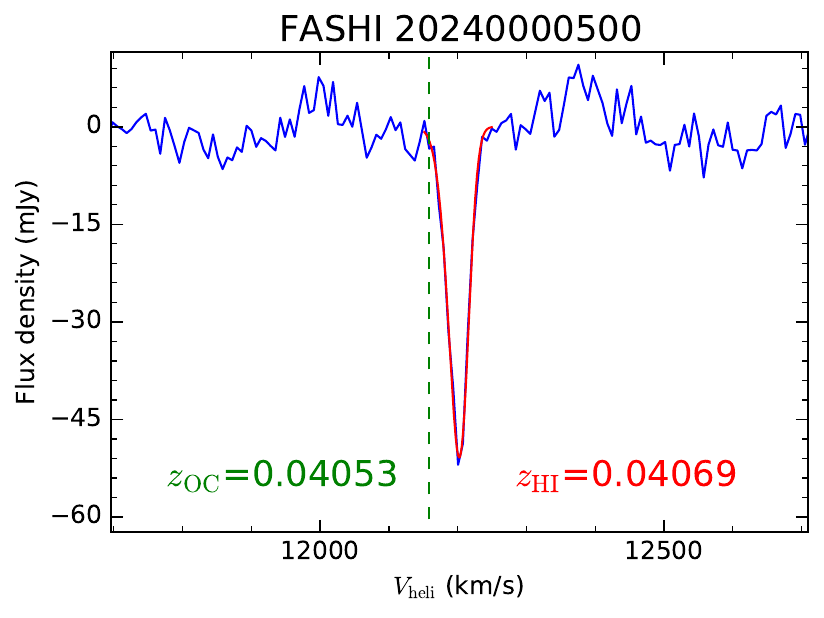}
 \includegraphics[height=0.27\textwidth, angle=0]{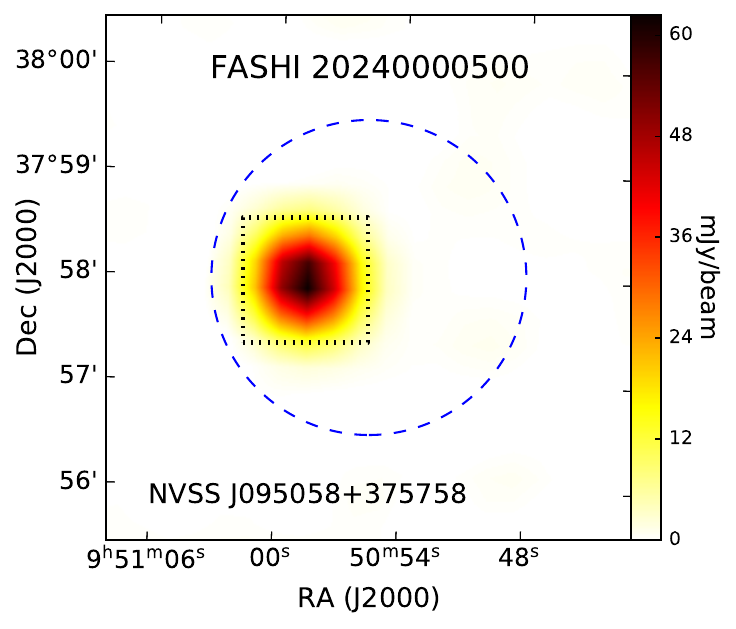}
 \includegraphics[height=0.27\textwidth, angle=0]{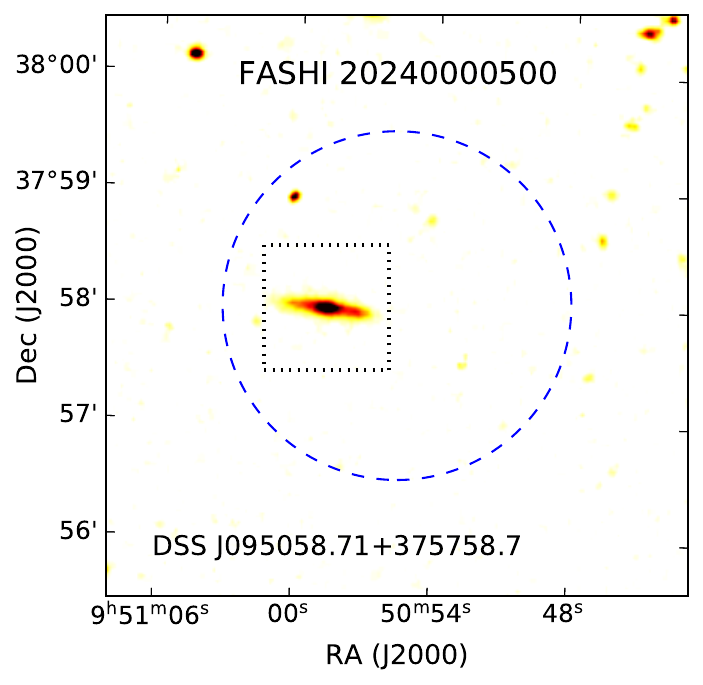}
 \caption{21\,cm H\,{\scriptsize{I}} absorption galaxy FASHI\,095055.65+375800.8 or ID\,20240000500.}
 \end{figure*} 
  
 \clearpage
 
 \begin{figure*}[htp]
 \centering
 \renewcommand{\thefigure}{\arabic{figure} (Continued)}
 \addtocounter{figure}{-1}
 \includegraphics[height=0.22\textwidth, angle=0]{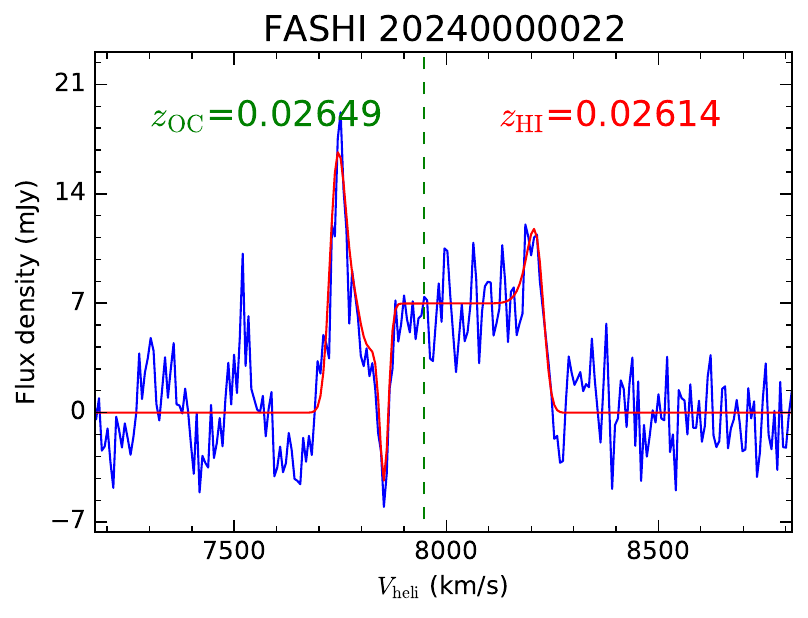}
 \includegraphics[height=0.27\textwidth, angle=0]{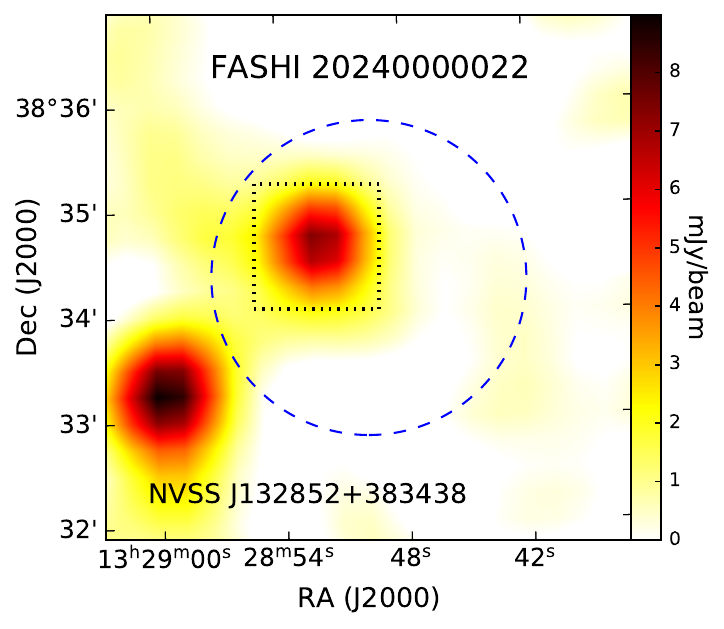}
 \includegraphics[height=0.27\textwidth, angle=0]{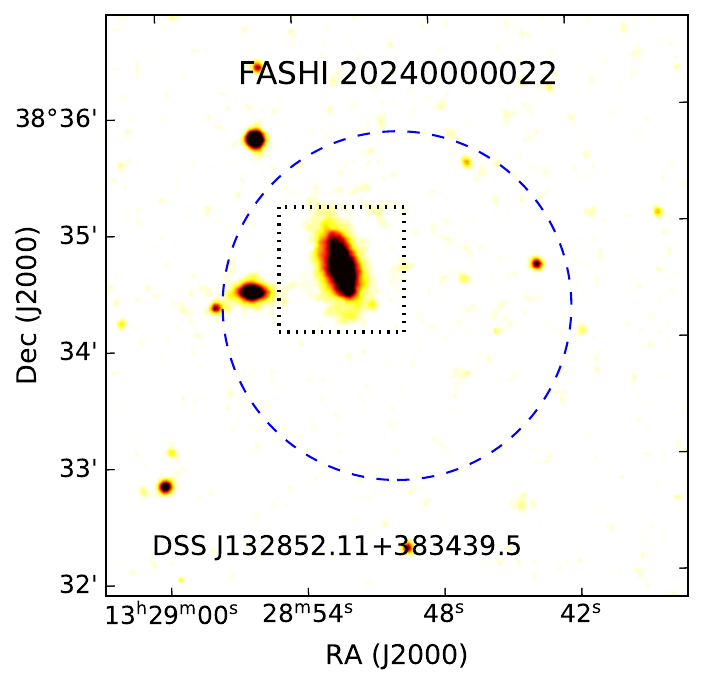}
 \caption{21\,cm H\,{\scriptsize{I}} absorption galaxy FASHI\,132849.73+383419.9 or ID\,20240000022.}
 \end{figure*} 

 \begin{figure*}[htp]
 \centering
 \renewcommand{\thefigure}{\arabic{figure} (Continued)}
 \addtocounter{figure}{-1}
 \includegraphics[height=0.22\textwidth, angle=0]{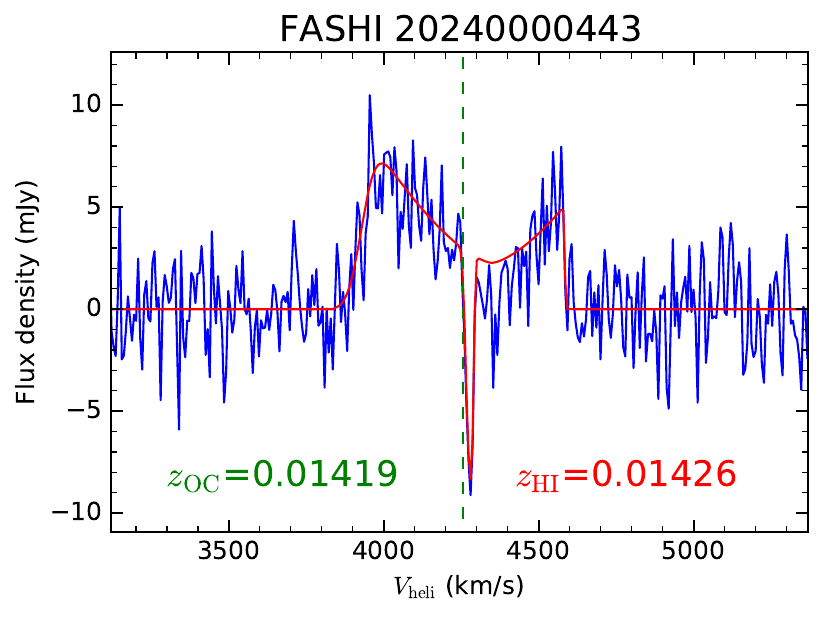}
 \includegraphics[height=0.27\textwidth, angle=0]{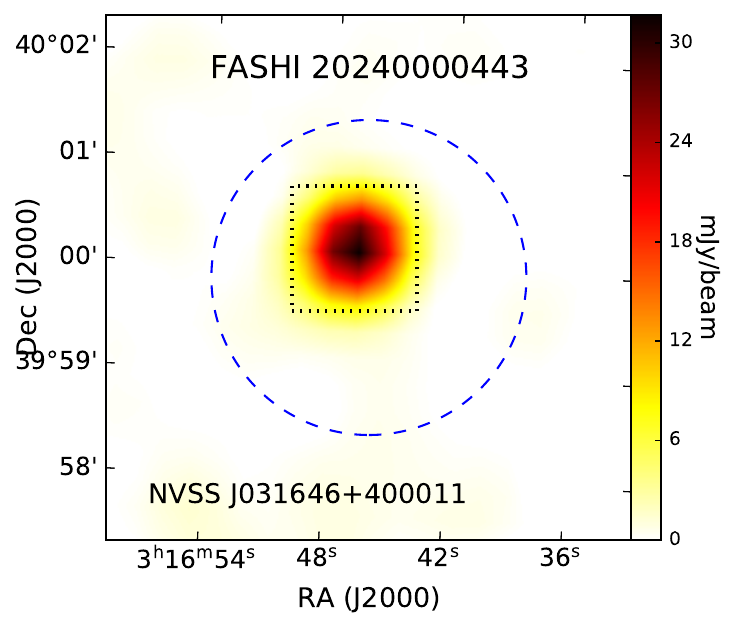}
 \includegraphics[height=0.27\textwidth, angle=0]{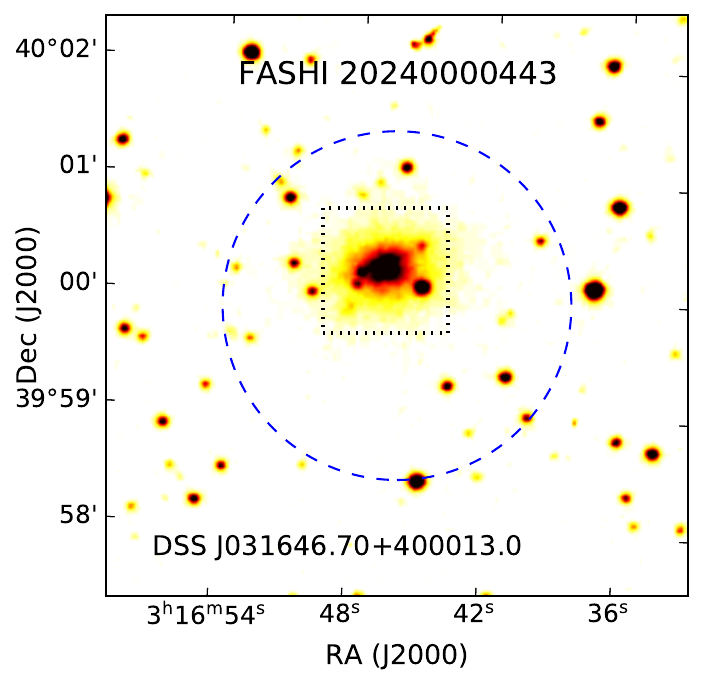}
 \caption{21\,cm H\,{\scriptsize{I}} absorption galaxy FASHI\,031646.12+395955.4 or ID\,20240000443.}
 \end{figure*} 

 \begin{figure*}[htp]
 \centering
 \renewcommand{\thefigure}{\arabic{figure} (Continued)}
 \addtocounter{figure}{-1}
 \includegraphics[height=0.22\textwidth, angle=0]{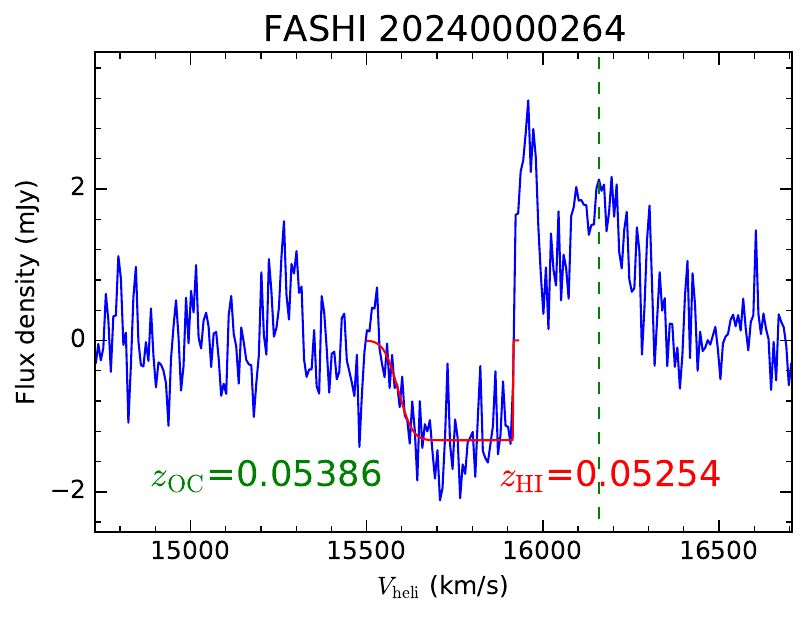}
 \includegraphics[height=0.27\textwidth, angle=0]{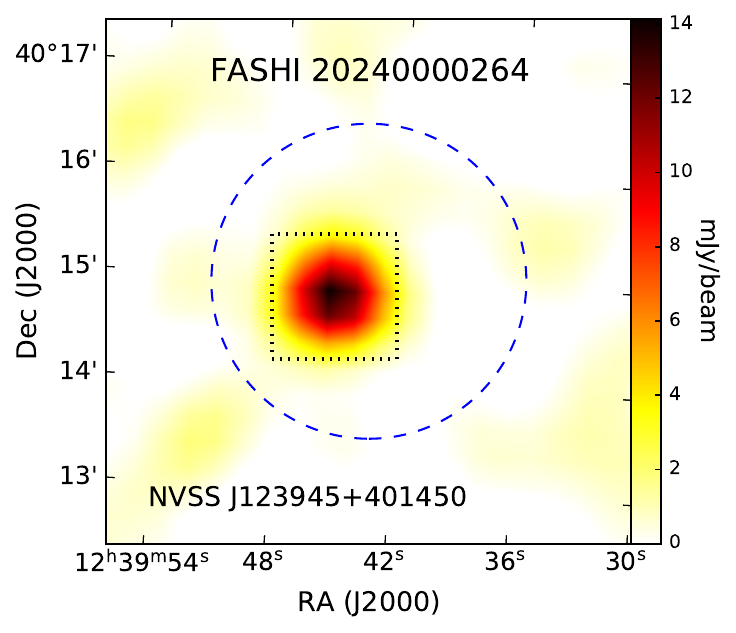}
 \includegraphics[height=0.27\textwidth, angle=0]{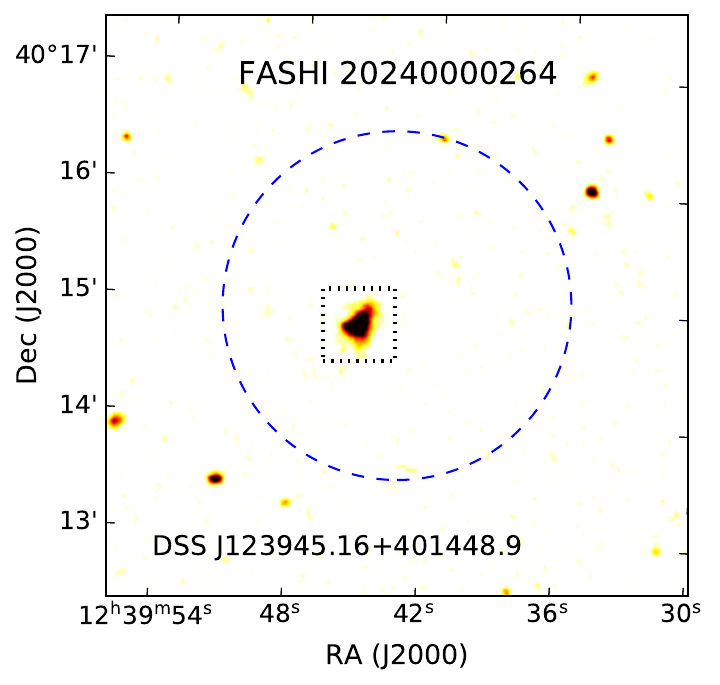}
 \caption{21\,cm H\,{\scriptsize{I}} absorption galaxy FASHI\,123943.52+401459.7 or ID\,20240000264.}
 \end{figure*} 

 \begin{figure*}[htp]
 \centering
 \renewcommand{\thefigure}{\arabic{figure} (Continued)}
 \addtocounter{figure}{-1}
 \includegraphics[height=0.22\textwidth, angle=0]{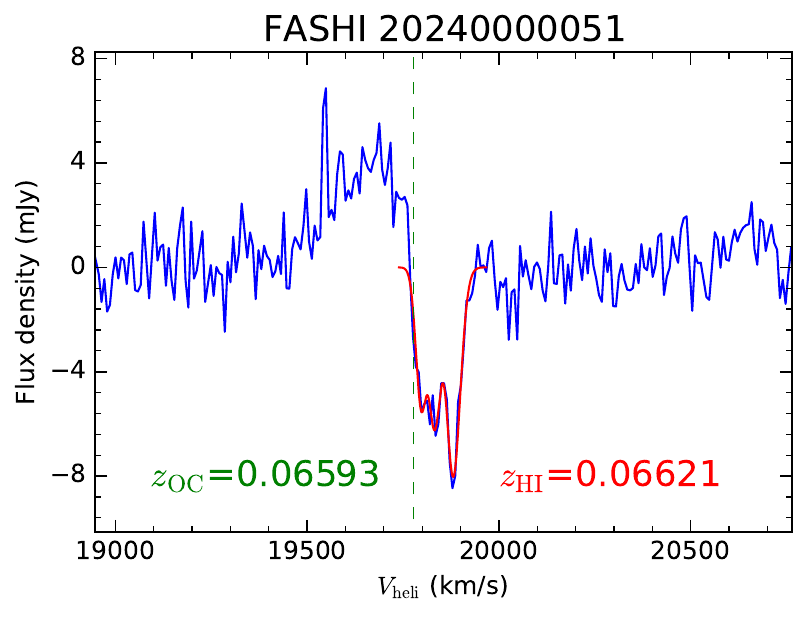}
 \includegraphics[height=0.27\textwidth, angle=0]{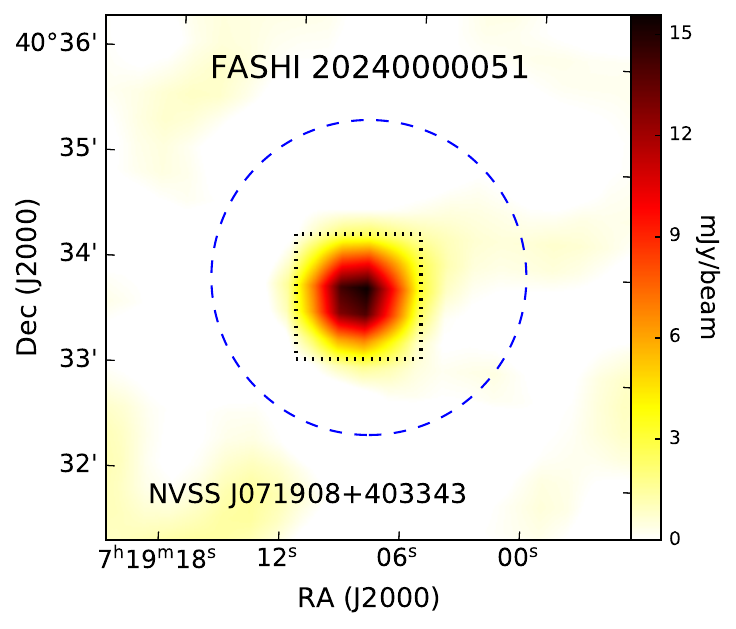}
 \includegraphics[height=0.27\textwidth, angle=0]{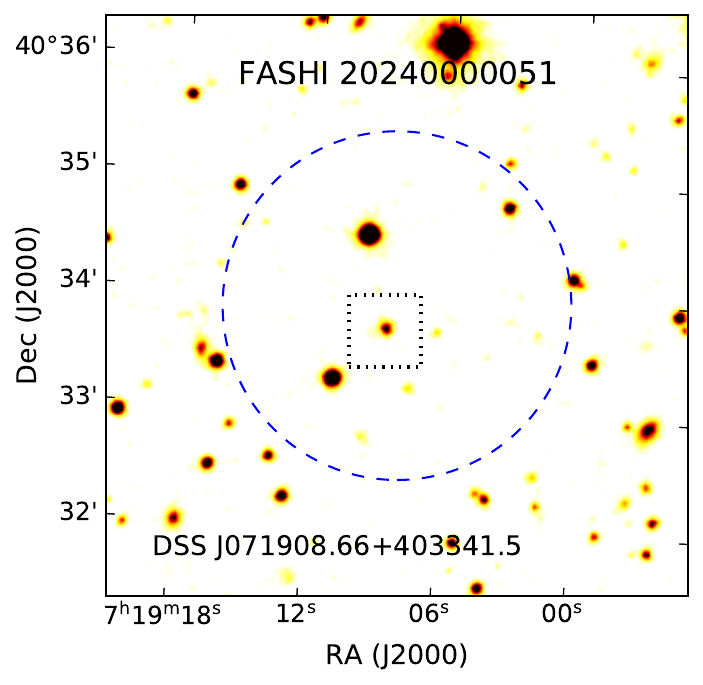}
 \caption{21\,cm H\,{\scriptsize{I}} absorption galaxy FASHI\,071908.19+403355.0 or ID\,20240000051.}
 \end{figure*} 

 \begin{figure*}[htp]
 \centering
 \renewcommand{\thefigure}{\arabic{figure} (Continued)}
 \addtocounter{figure}{-1}
 \includegraphics[height=0.22\textwidth, angle=0]{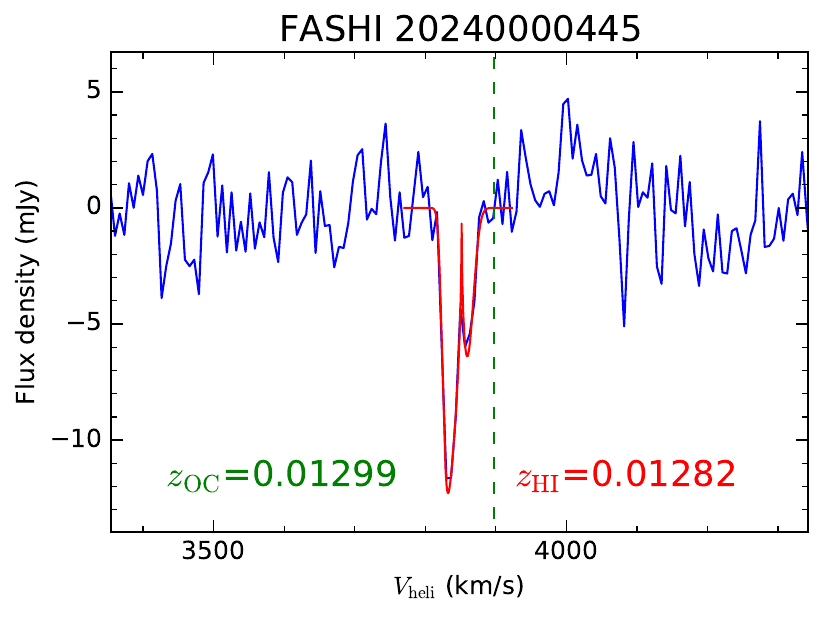}
 \includegraphics[height=0.27\textwidth, angle=0]{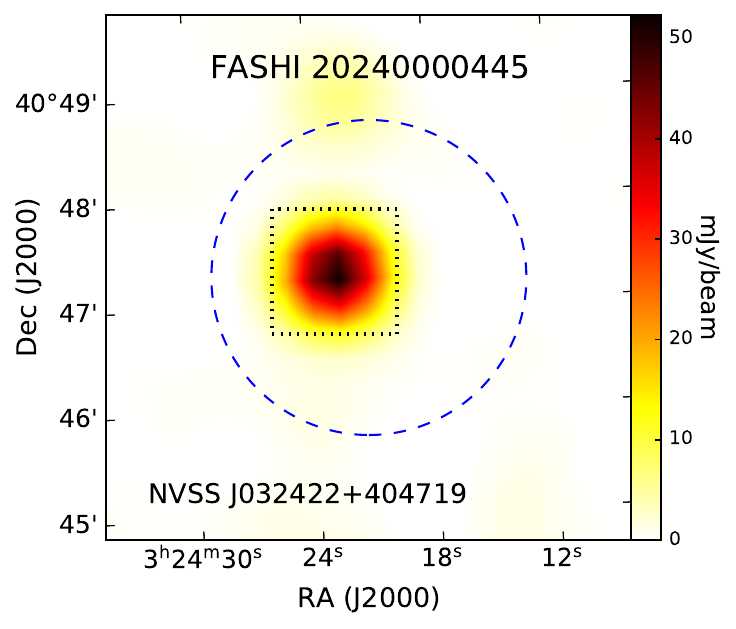}
 \includegraphics[height=0.27\textwidth, angle=0]{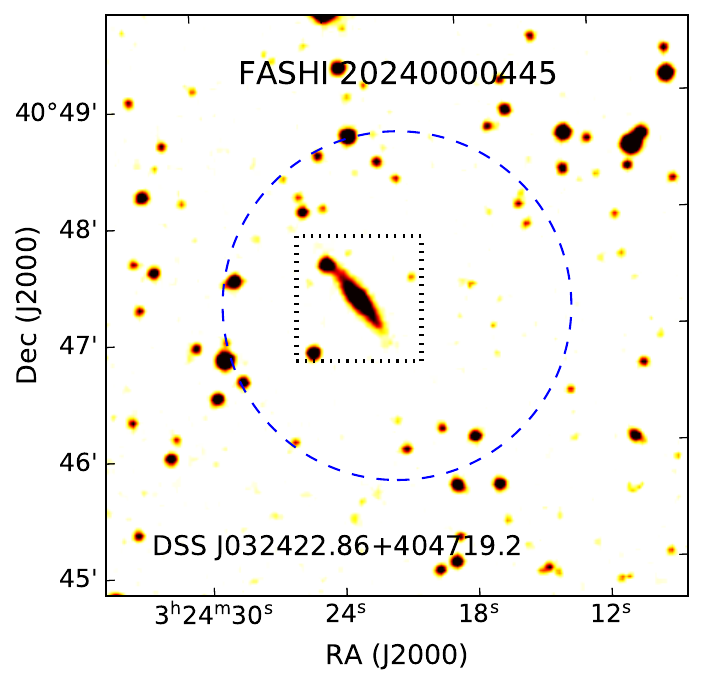}
 \caption{21\,cm H\,{\scriptsize{I}} absorption galaxy FASHI\,032421.15+404714.8 or ID\,20240000445.}
 \end{figure*} 

 \begin{figure*}[htp]
 \centering
 \renewcommand{\thefigure}{\arabic{figure} (Continued)}
 \addtocounter{figure}{-1}
 \includegraphics[height=0.22\textwidth, angle=0]{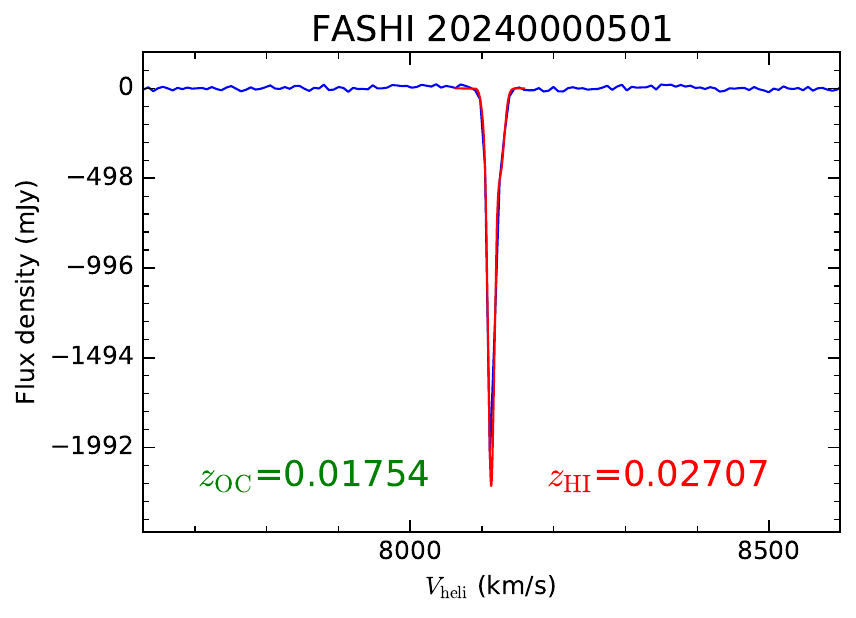}
 \includegraphics[height=0.27\textwidth, angle=0]{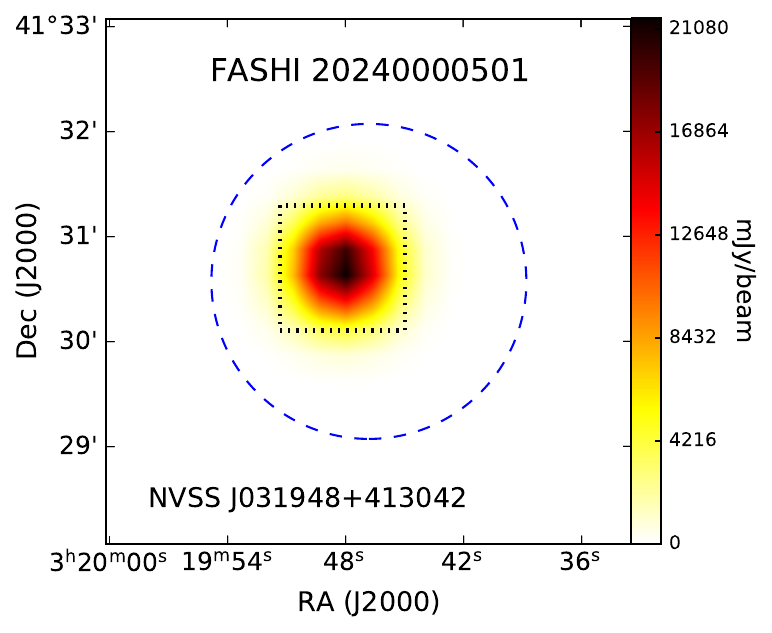}
 \includegraphics[height=0.27\textwidth, angle=0]{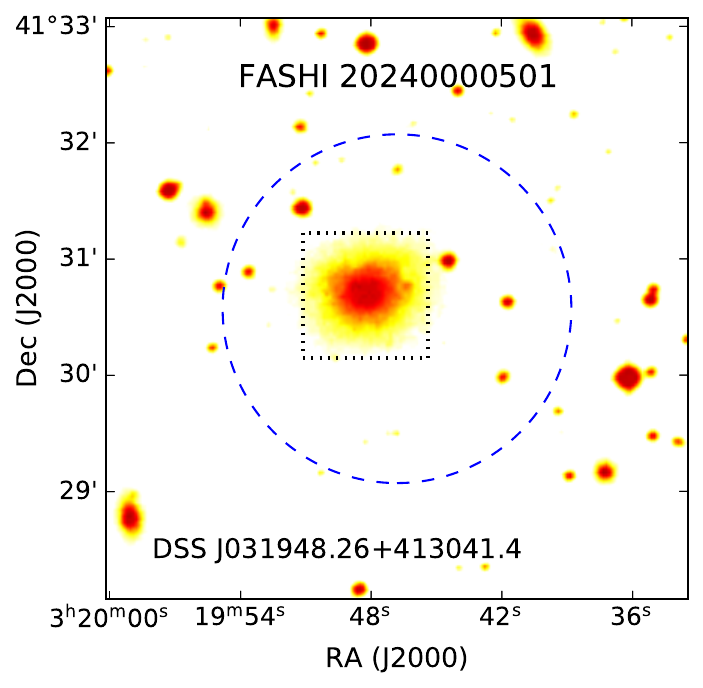}
 \caption{21\,cm H\,{\scriptsize{I}} absorption galaxy FASHI\,031946.81+413034.4 or ID\,20240000501.}
 \end{figure*} 

 \begin{figure*}[htp]
 \centering
 \renewcommand{\thefigure}{\arabic{figure} (Continued)}
 \addtocounter{figure}{-1}
 \includegraphics[height=0.22\textwidth, angle=0]{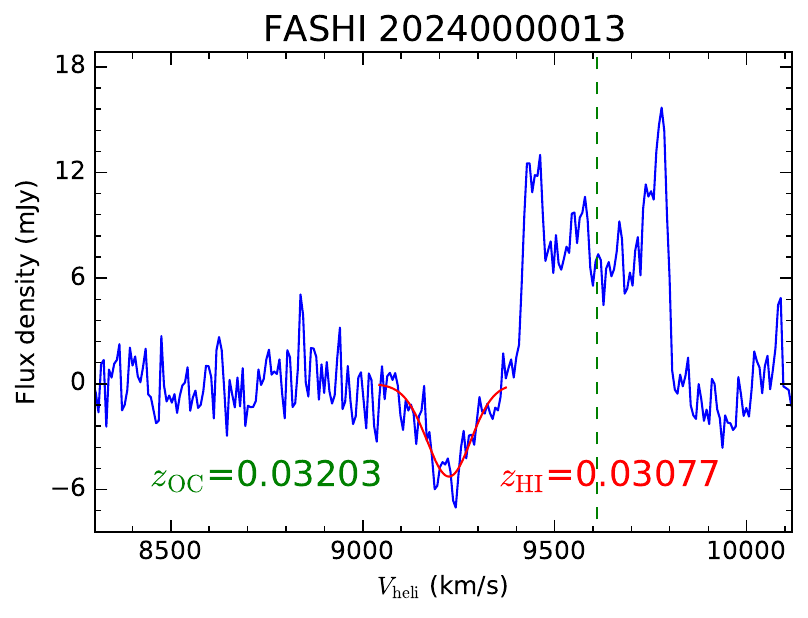}
 \includegraphics[height=0.27\textwidth, angle=0]{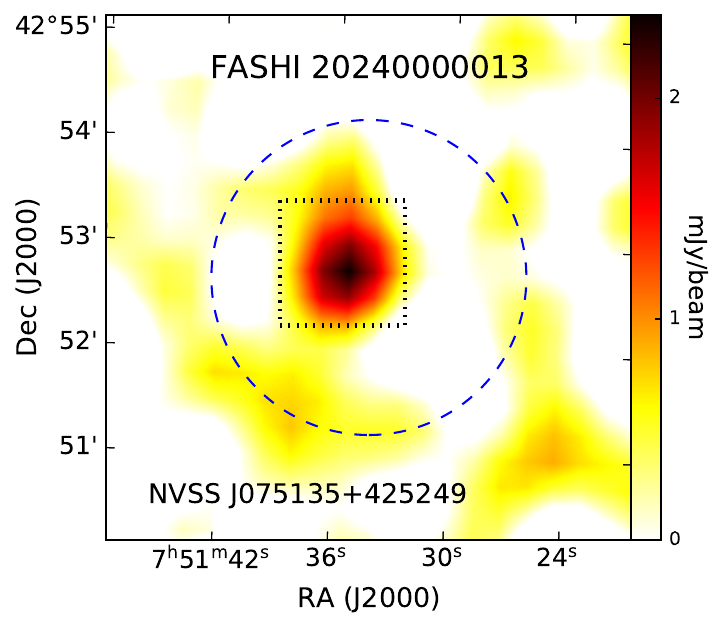}
 \includegraphics[height=0.27\textwidth, angle=0]{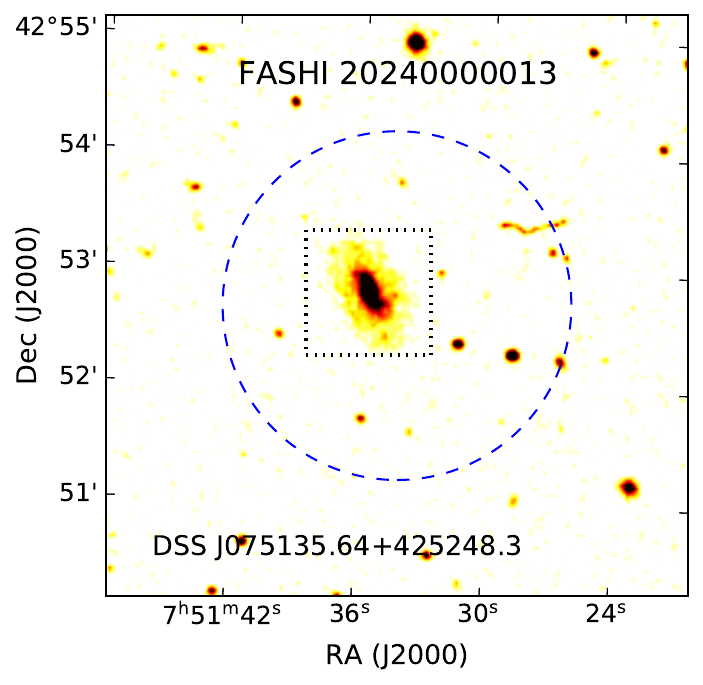}
 \caption{21\,cm H\,{\scriptsize{I}} absorption galaxy FASHI\,075134.30+425242.1 or ID\,20240000013.}
 \end{figure*} 

 \begin{figure*}[htp]
 \centering
 \renewcommand{\thefigure}{\arabic{figure} (Continued)}
 \addtocounter{figure}{-1}
 \includegraphics[height=0.22\textwidth, angle=0]{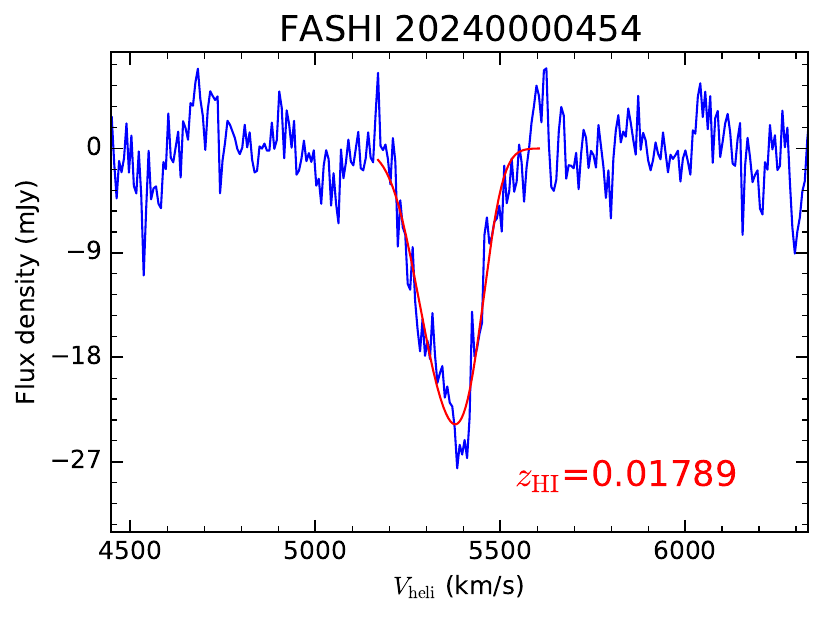}
 \includegraphics[height=0.27\textwidth, angle=0]{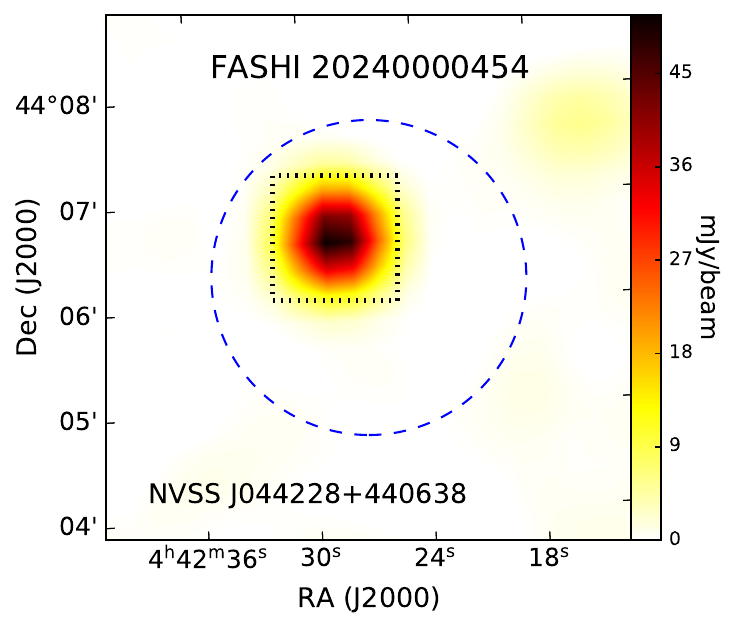}
 \includegraphics[height=0.27\textwidth, angle=0]{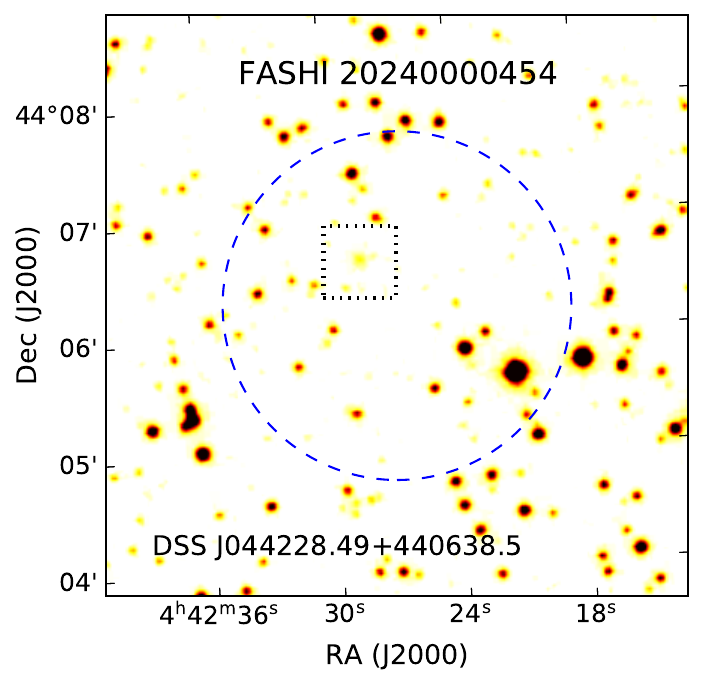}
 \caption{21\,cm H\,{\scriptsize{I}} absorption galaxy FASHI\,044226.82+440615.0 or ID\,20240000454.}
 \end{figure*} 

 \begin{figure*}[htp]
 \centering
 \renewcommand{\thefigure}{\arabic{figure} (Continued)}
 \addtocounter{figure}{-1}
 \includegraphics[height=0.22\textwidth, angle=0]{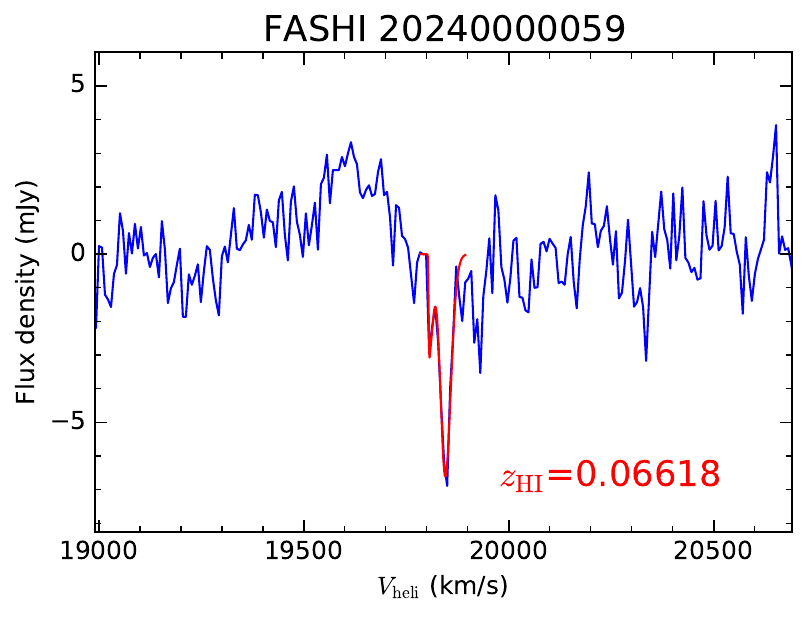}
 \includegraphics[height=0.27\textwidth, angle=0]{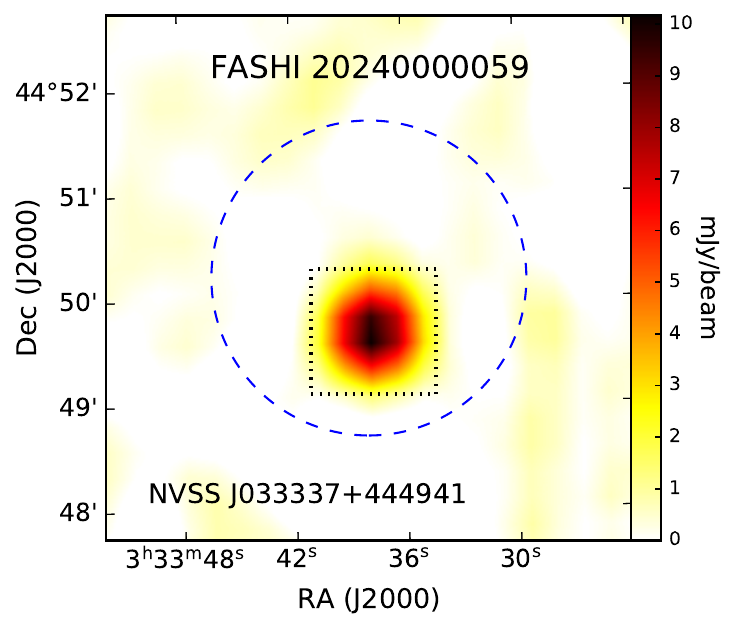}
 \includegraphics[height=0.27\textwidth, angle=0]{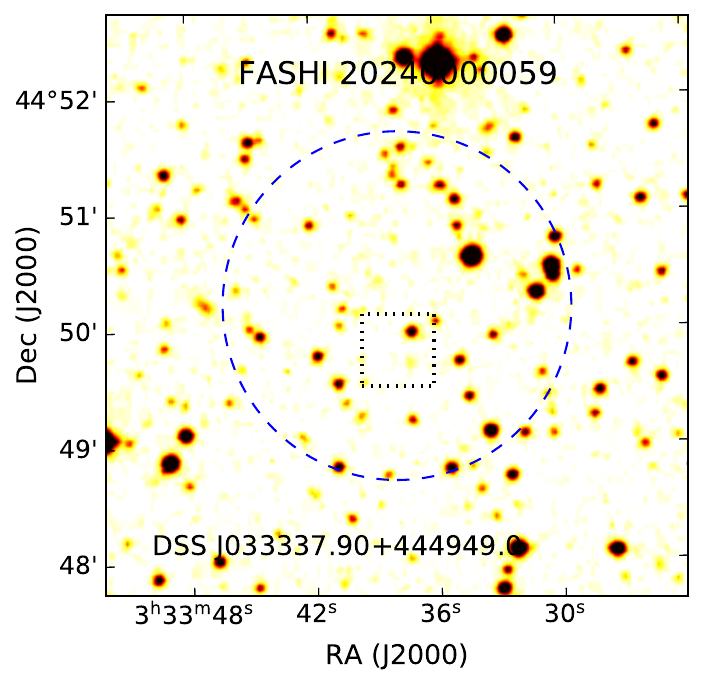}
 \caption{21\,cm H\,{\scriptsize{I}} absorption galaxy FASHI\,033337.92+445011.9 or ID\,20240000059.}
 \end{figure*} 

 \begin{figure*}[htp]
 \centering
 \renewcommand{\thefigure}{\arabic{figure} (Continued)}
 \addtocounter{figure}{-1}
 \includegraphics[height=0.22\textwidth, angle=0]{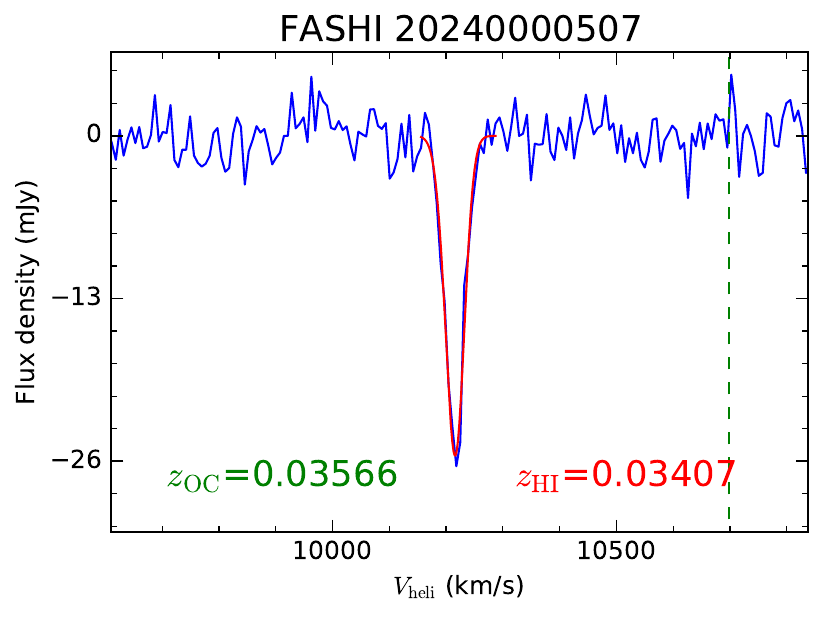}
 \includegraphics[height=0.27\textwidth, angle=0]{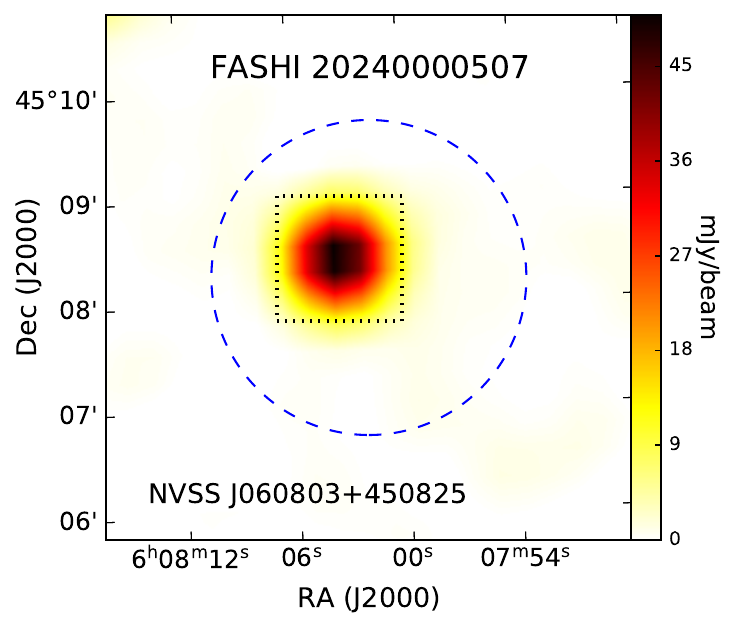}
 \includegraphics[height=0.27\textwidth, angle=0]{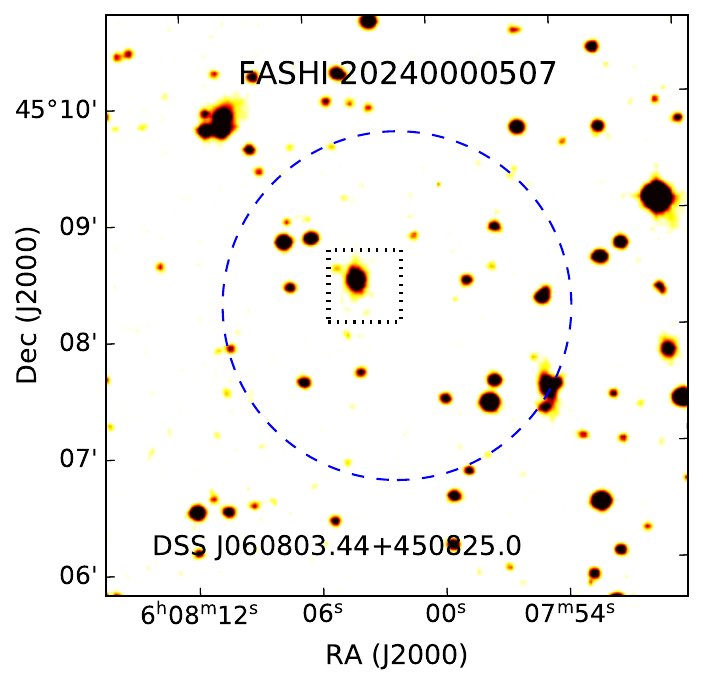}
 \caption{21\,cm H\,{\scriptsize{I}} absorption galaxy FASHI\,060801.90+450814.2 or ID\,20240000507.}
 \end{figure*}

 \begin{figure*}[htp]
 \centering
 \renewcommand{\thefigure}{\arabic{figure} (Continued)}
 \addtocounter{figure}{-1}
 \includegraphics[height=0.22\textwidth, angle=0]{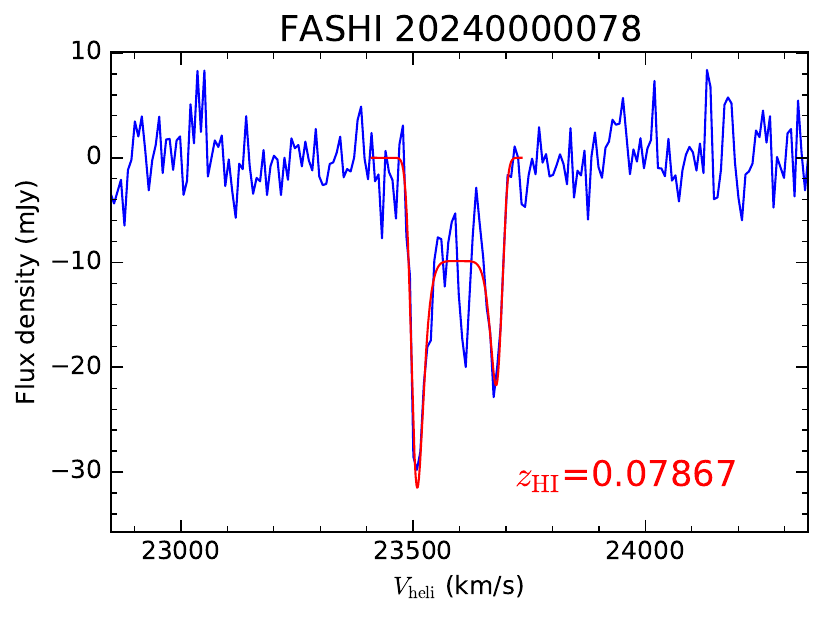}
 \includegraphics[height=0.27\textwidth, angle=0]{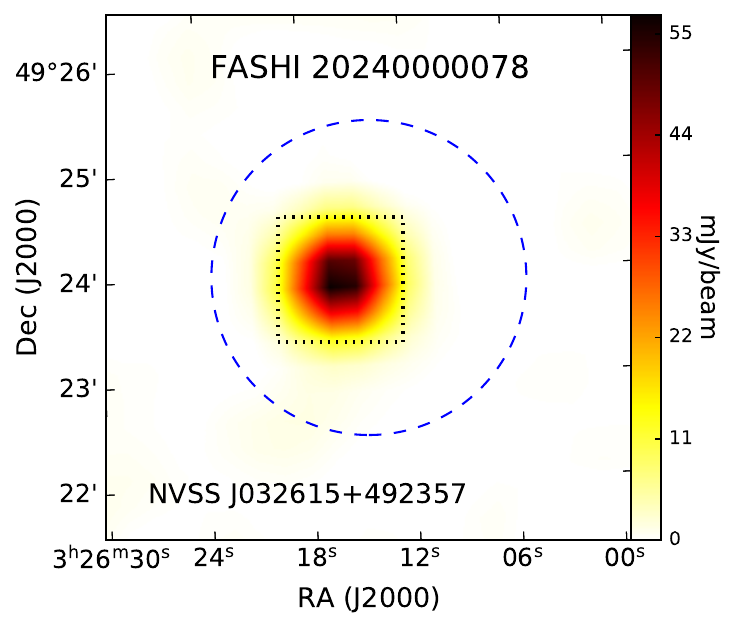}
 \includegraphics[height=0.27\textwidth, angle=0]{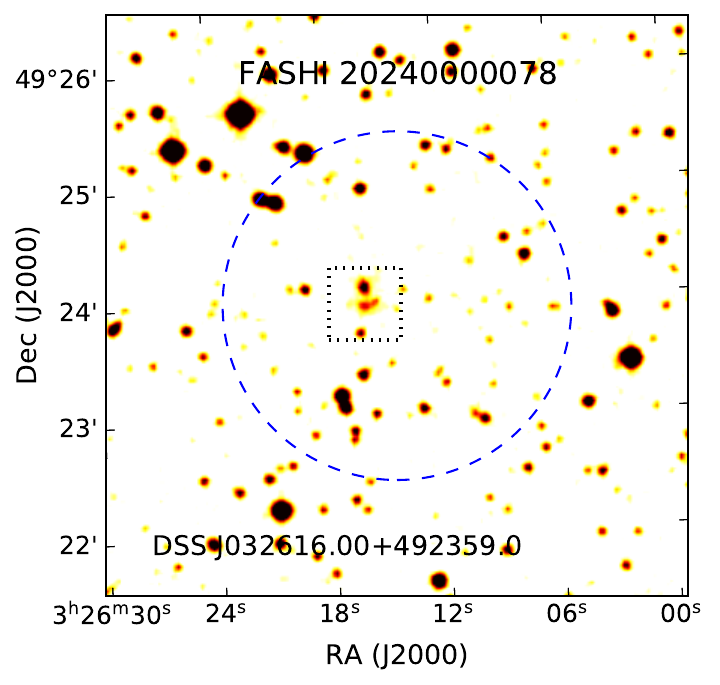}
 \caption{21\,cm H\,{\scriptsize{I}} absorption galaxy FASHI\,032614.32+492357.3 or ID\,20240000078.}
 \end{figure*} 

 \begin{figure*}[htp]
 \centering
 \renewcommand{\thefigure}{\arabic{figure} (Continued)}
 \addtocounter{figure}{-1}
 \includegraphics[height=0.22\textwidth, angle=0]{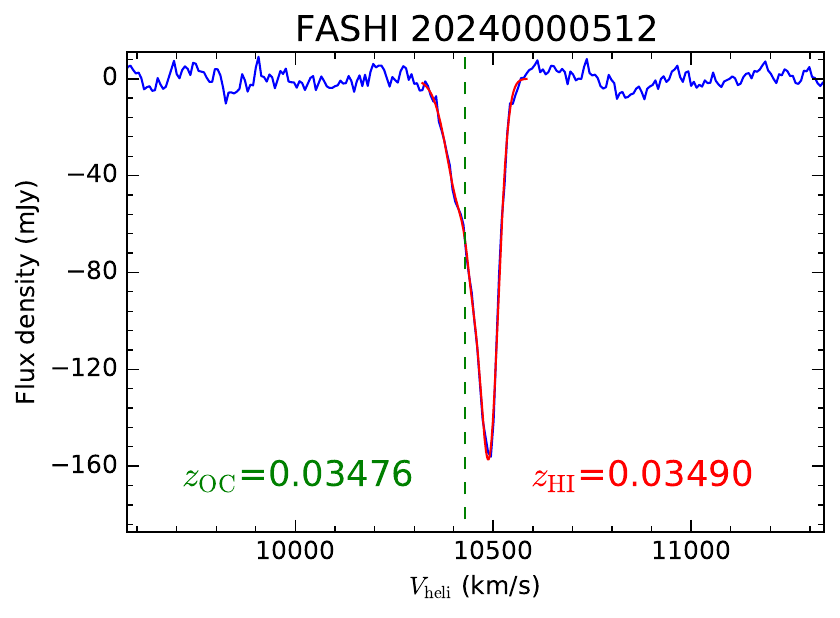}
 \includegraphics[height=0.27\textwidth, angle=0]{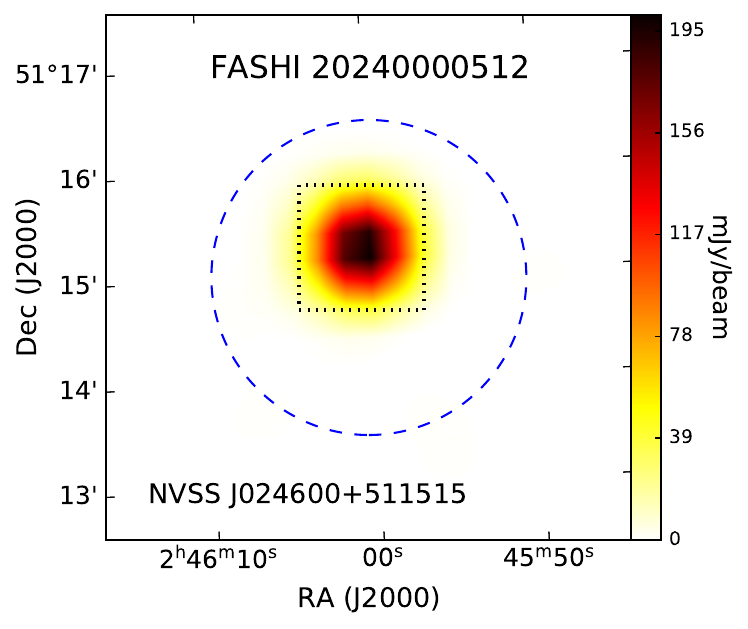}
 \includegraphics[height=0.27\textwidth, angle=0]{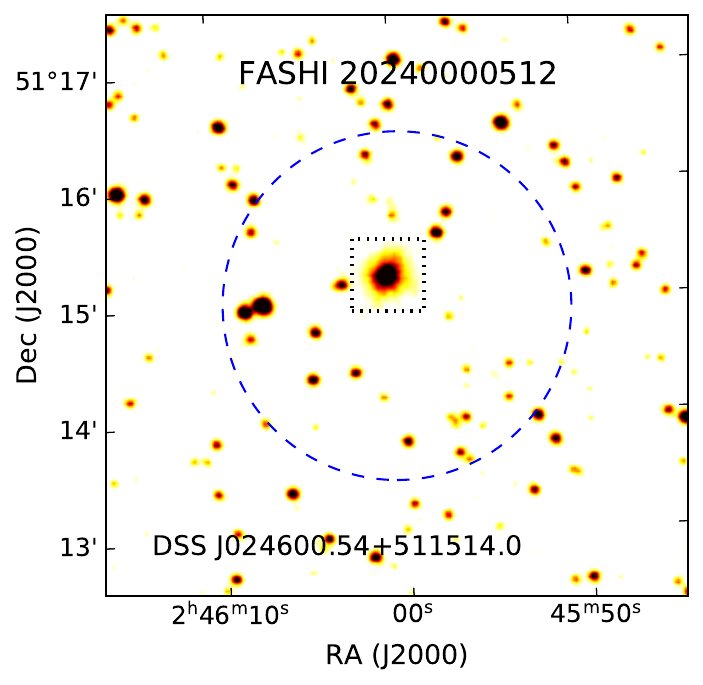}
 \caption{21\,cm H\,{\scriptsize{I}} absorption galaxy FASHI\,024600.14+511458.1 or ID\,20240000512.}
 \end{figure*} 

 \begin{figure*}[htp]
 \centering
 \renewcommand{\thefigure}{\arabic{figure} (Continued)}
 \addtocounter{figure}{-1}
 \includegraphics[height=0.22\textwidth, angle=0]{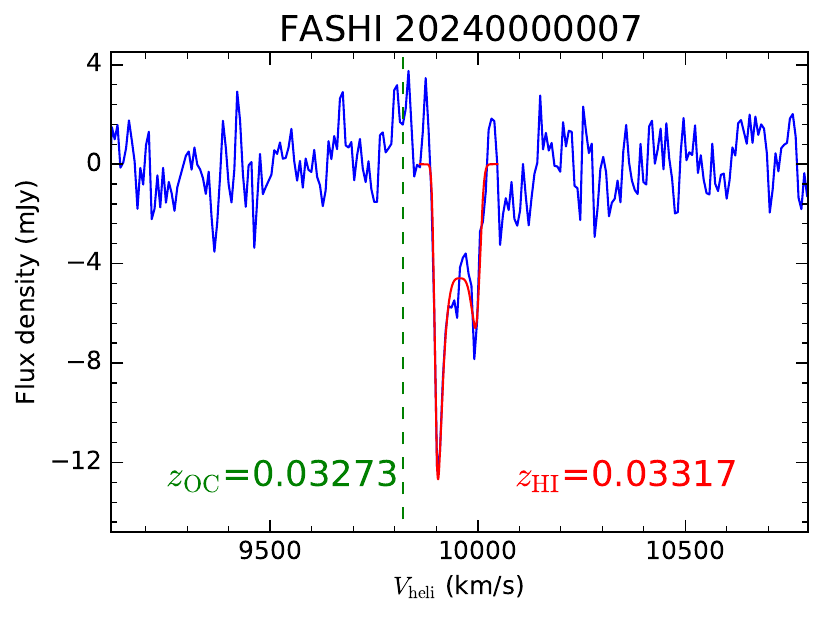}
 \includegraphics[height=0.27\textwidth, angle=0]{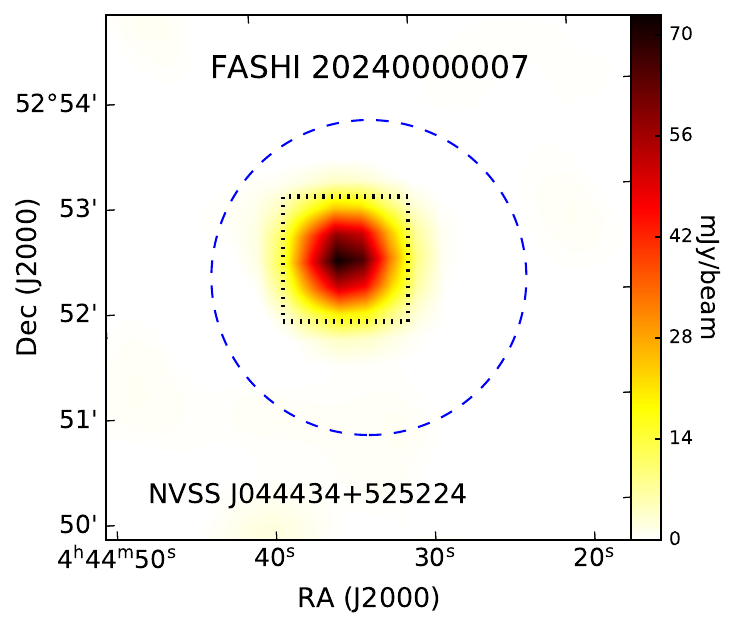}
 \includegraphics[height=0.27\textwidth, angle=0]{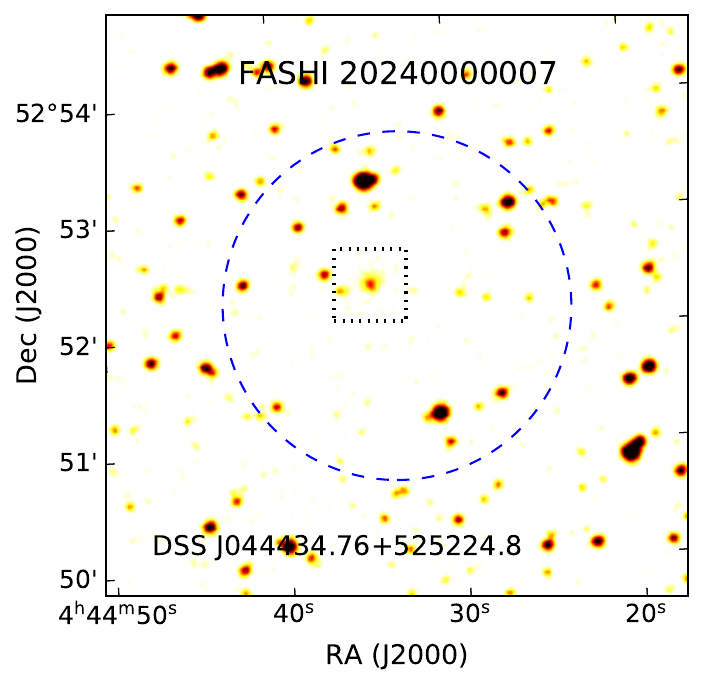}
 \caption{21\,cm H\,{\scriptsize{I}} absorption galaxy FASHI\,044433.30+525213.5 or ID\,20240000007.}
 \end{figure*} 

 \begin{figure*}[htp]
 \centering
 \renewcommand{\thefigure}{\arabic{figure} (Continued)}
 \addtocounter{figure}{-1}
 \includegraphics[height=0.22\textwidth, angle=0]{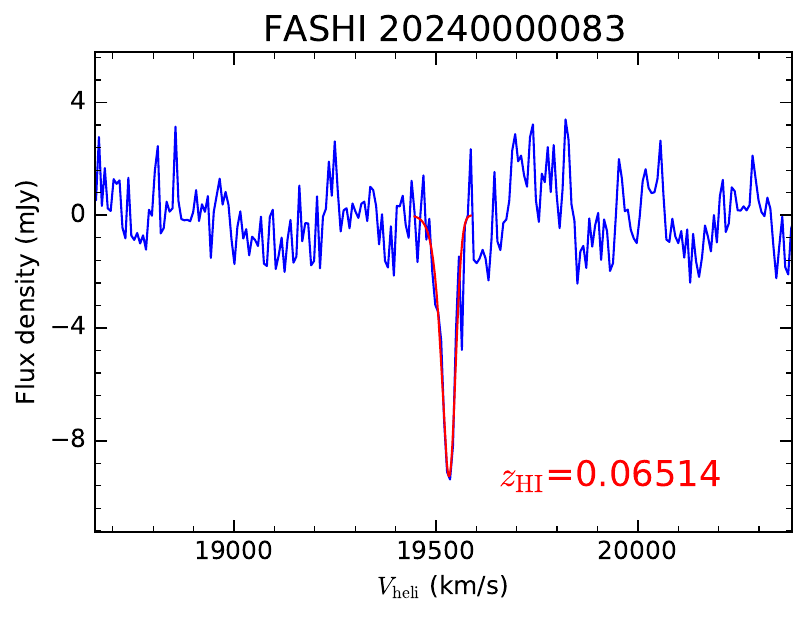}
 \includegraphics[height=0.27\textwidth, angle=0]{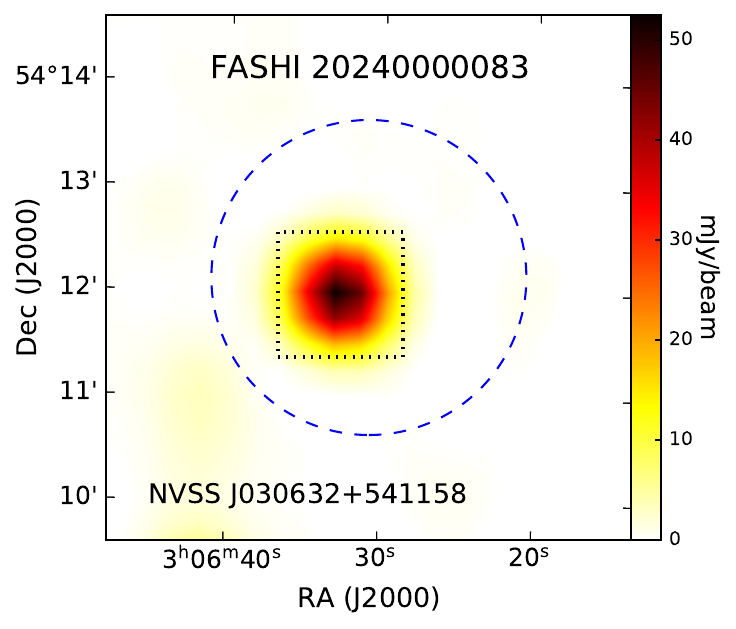}
 \includegraphics[height=0.27\textwidth, angle=0]{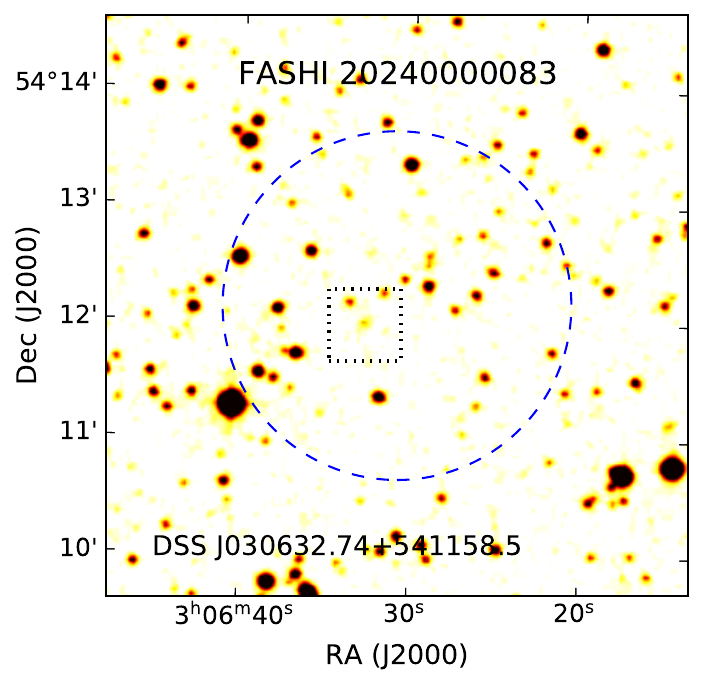}
 \caption{21\,cm H\,{\scriptsize{I}} absorption galaxy FASHI\,030630.87+541208.7 or ID\,20240000083.}
 \end{figure*} 

 \begin{figure*}[htp]
 \centering
 \renewcommand{\thefigure}{\arabic{figure} (Continued)}
 \addtocounter{figure}{-1}
 \includegraphics[height=0.22\textwidth, angle=0]{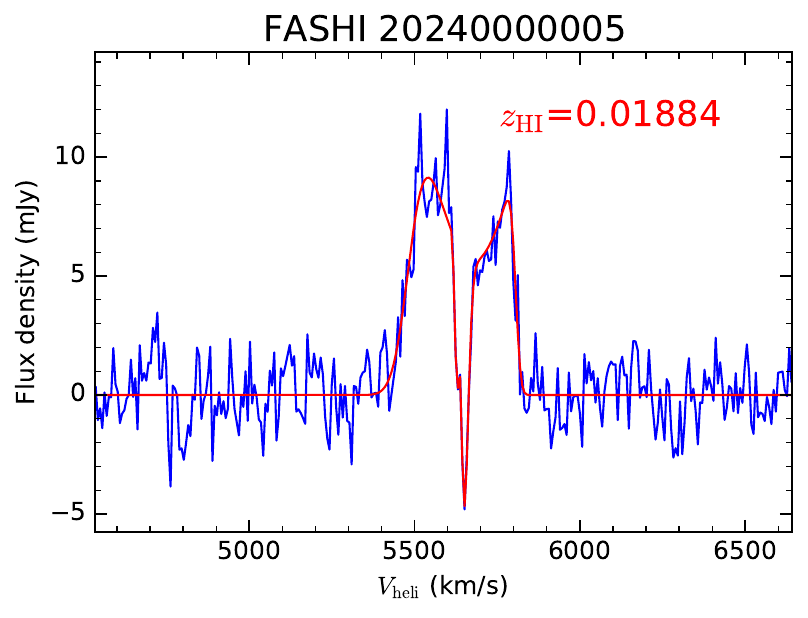}
 \includegraphics[height=0.27\textwidth, angle=0]{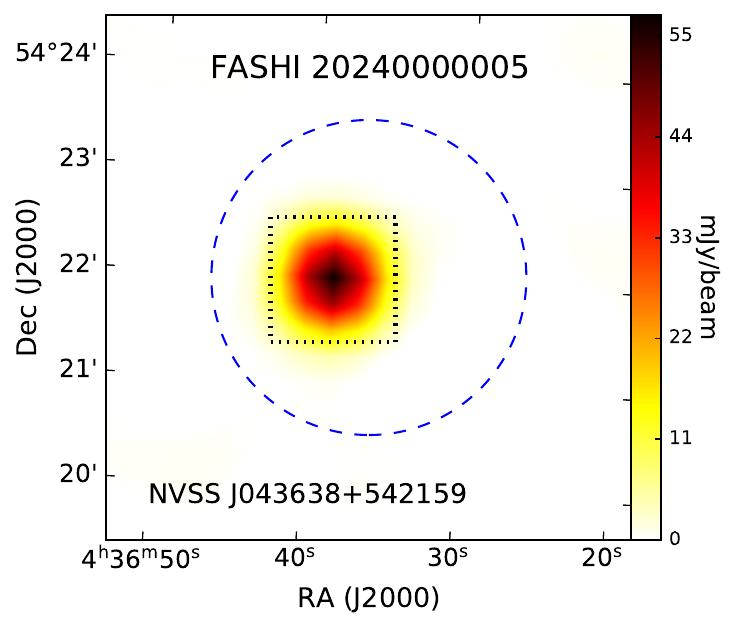}
 \includegraphics[height=0.27\textwidth, angle=0]{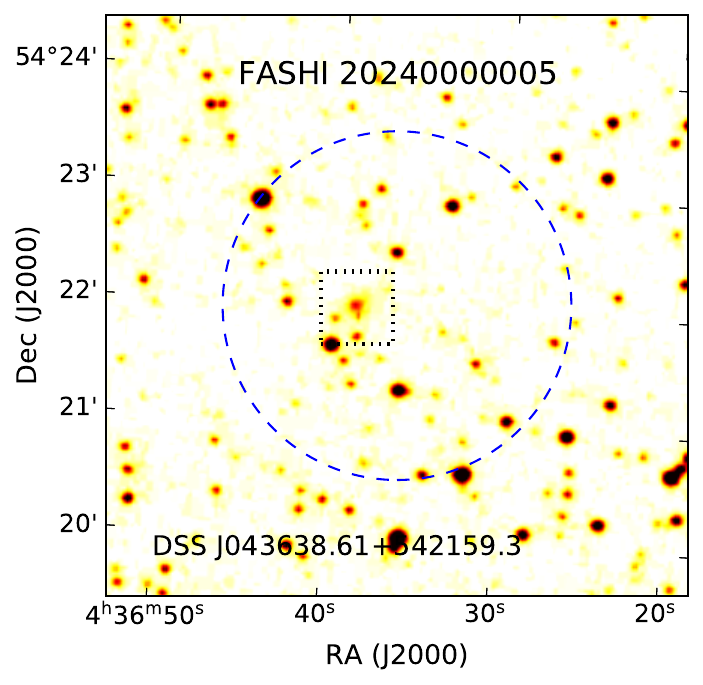}
 \caption{21\,cm H\,{\scriptsize{I}} absorption galaxy FASHI\,043636.25+542201.5 or ID\,20240000005.}
 \end{figure*} 

 \begin{figure*}[htp]
 \centering
 \renewcommand{\thefigure}{\arabic{figure} (Continued)}
 \addtocounter{figure}{-1}
 \includegraphics[height=0.22\textwidth, angle=0]{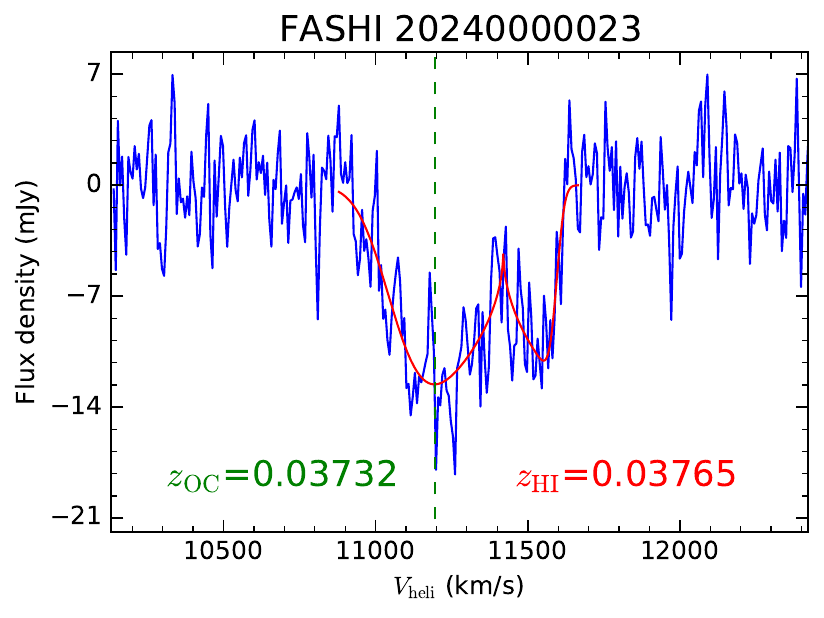}
 \includegraphics[height=0.27\textwidth, angle=0]{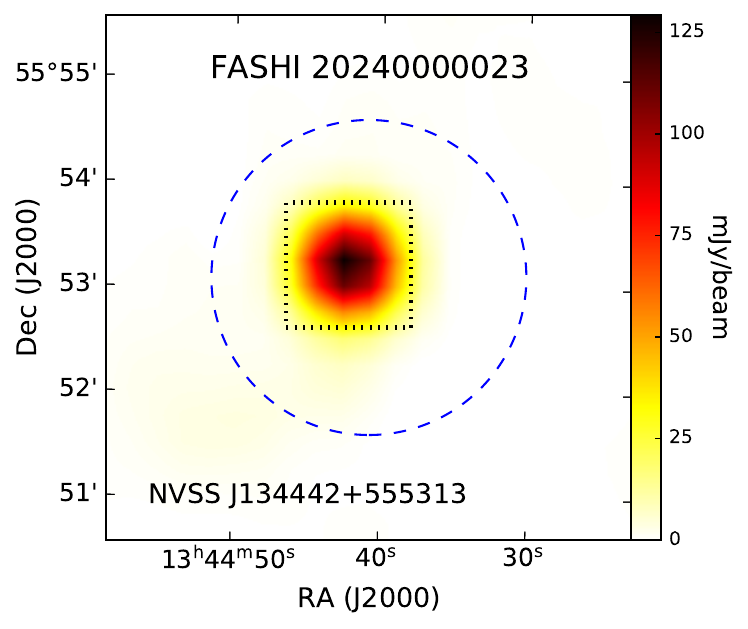}
 \includegraphics[height=0.27\textwidth, angle=0]{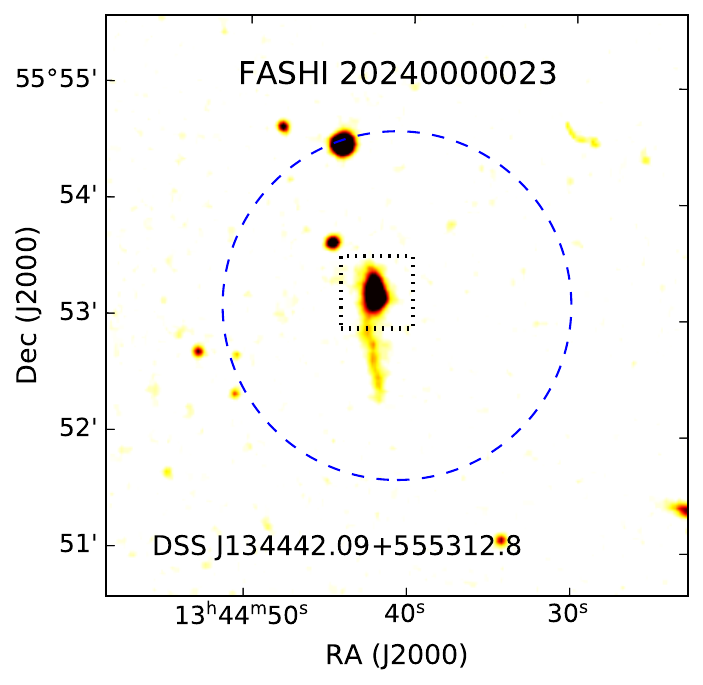}
 \caption{21\,cm H\,{\scriptsize{I}} absorption galaxy FASHI\,134440.84+555306.2 or ID\,20240000023.}
 \end{figure*} 

 \begin{figure*}[htp]
 \centering
 \renewcommand{\thefigure}{\arabic{figure} (Continued)}
 \addtocounter{figure}{-1}
 \includegraphics[height=0.22\textwidth, angle=0]{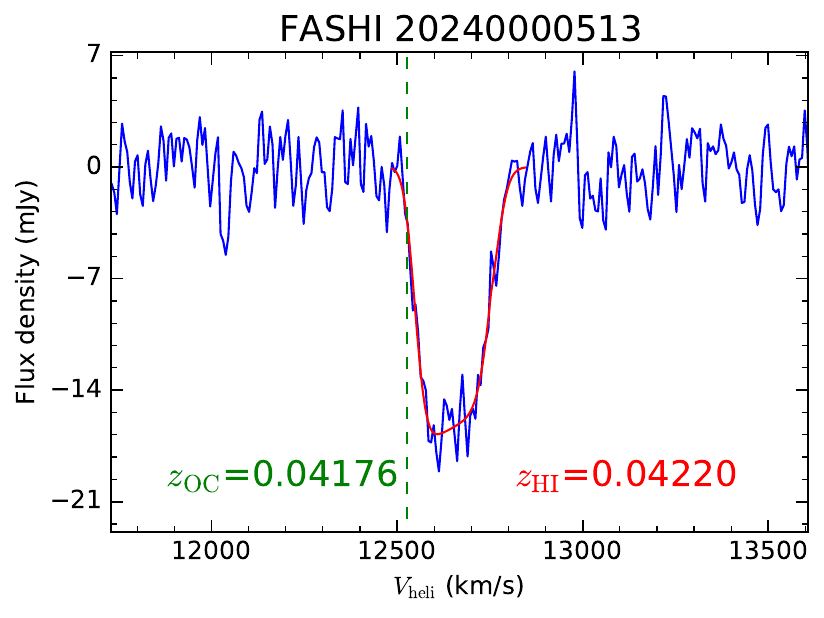}
 \includegraphics[height=0.27\textwidth, angle=0]{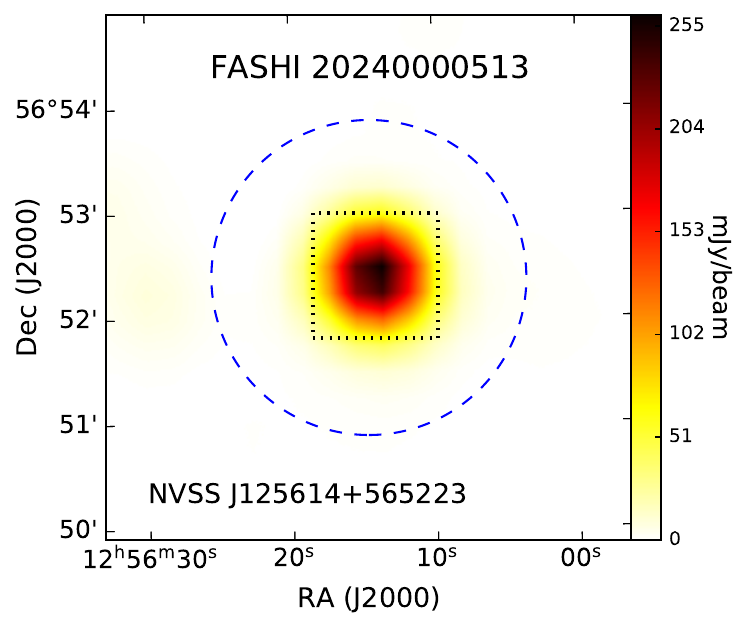}
 \includegraphics[height=0.27\textwidth, angle=0]{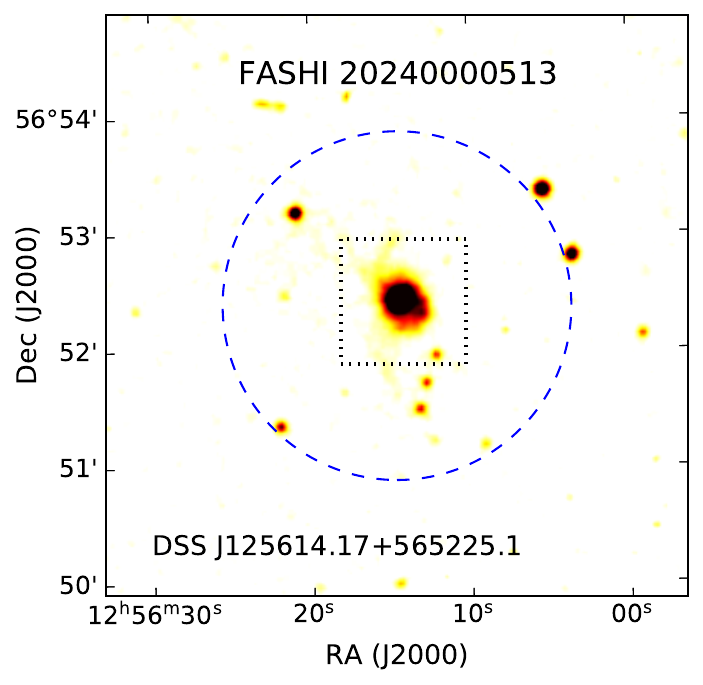}
 \caption{21\,cm H\,{\scriptsize{I}} absorption galaxy FASHI\,125614.58+565222.9 or ID\,20240000513.}
 \end{figure*} 

 \begin{figure*}[htp]
 \centering
 \renewcommand{\thefigure}{\arabic{figure} (Continued)}
 \addtocounter{figure}{-1}
 \includegraphics[height=0.22\textwidth, angle=0]{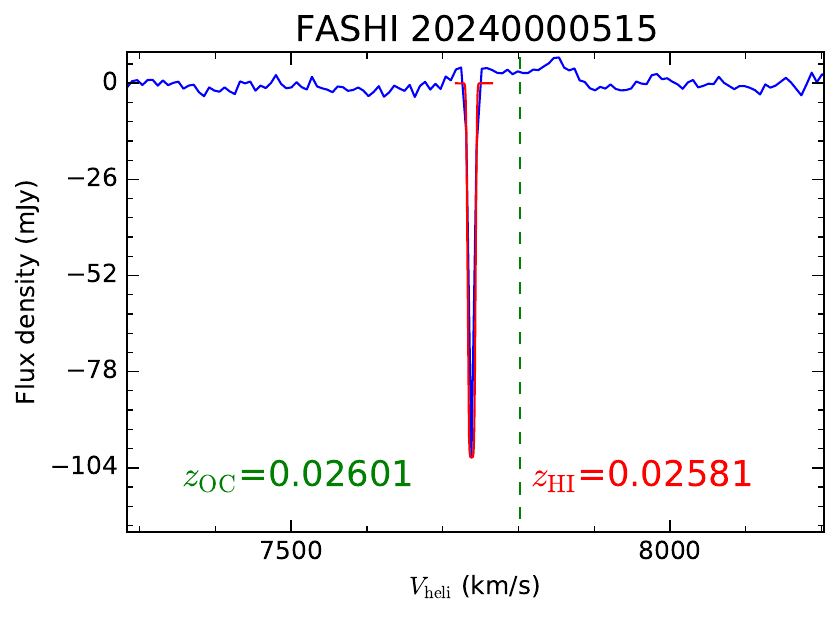}
 \includegraphics[height=0.27\textwidth, angle=0]{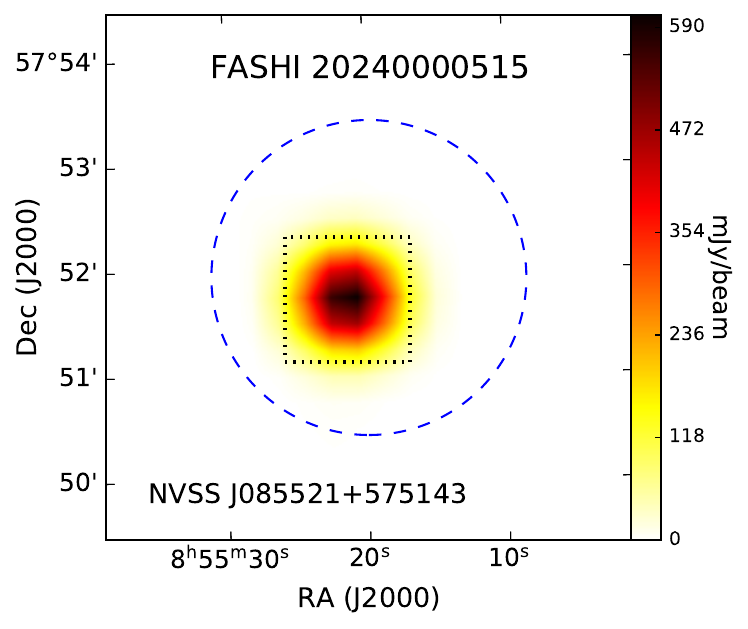}
 \includegraphics[height=0.27\textwidth, angle=0]{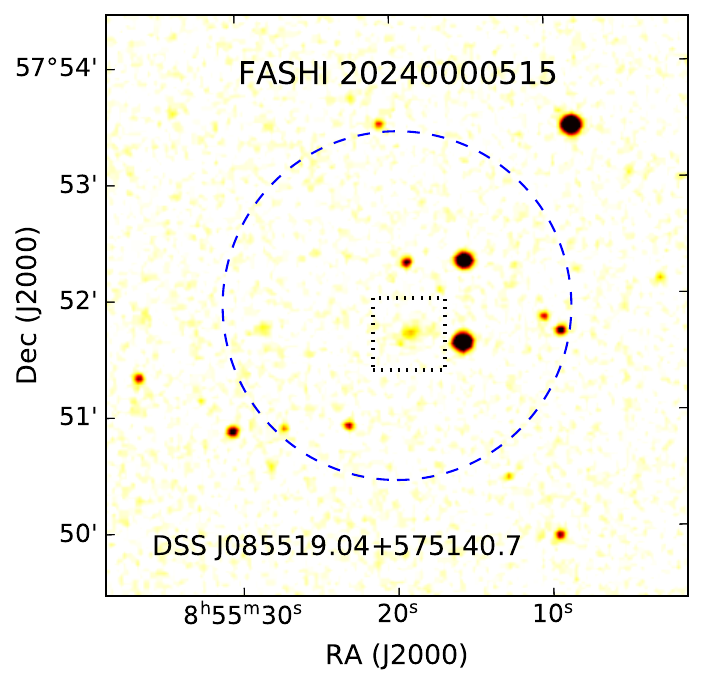}
 \caption{21\,cm H\,{\scriptsize{I}} absorption galaxy FASHI\,085519.79+575155.5 or ID\,20240000515.}
 \end{figure*} 

 \begin{figure*}[htp]
 \centering
 \renewcommand{\thefigure}{\arabic{figure} (Continued)}
 \addtocounter{figure}{-1}
 \includegraphics[height=0.22\textwidth, angle=0]{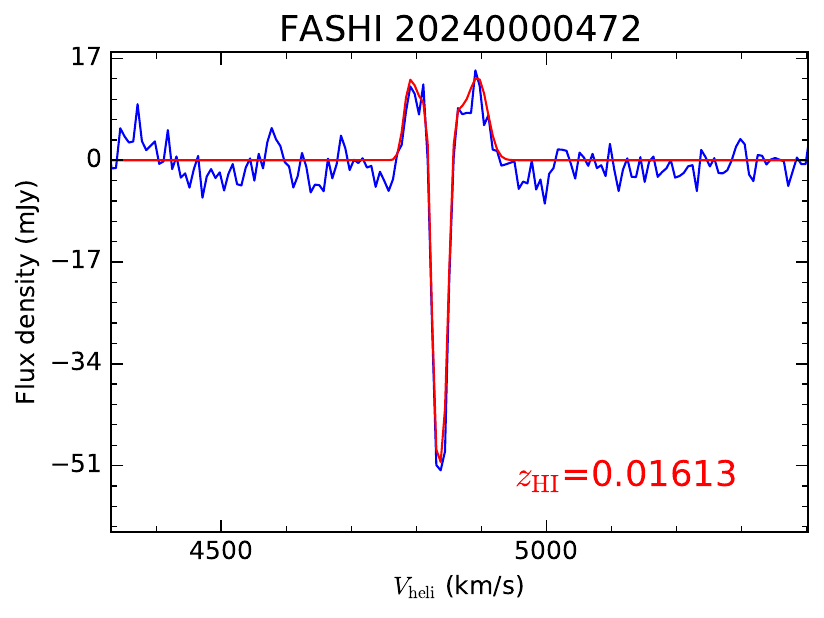}
 \includegraphics[height=0.27\textwidth, angle=0]{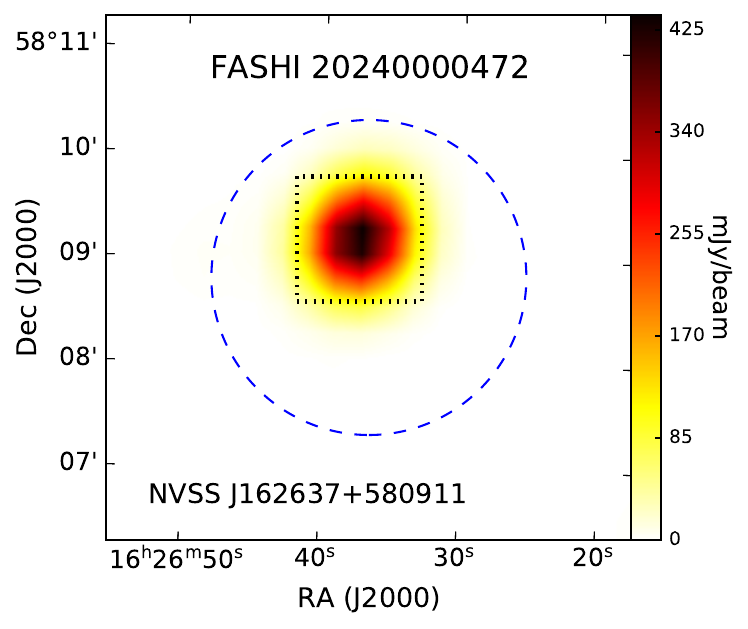}
 \includegraphics[height=0.27\textwidth, angle=0]{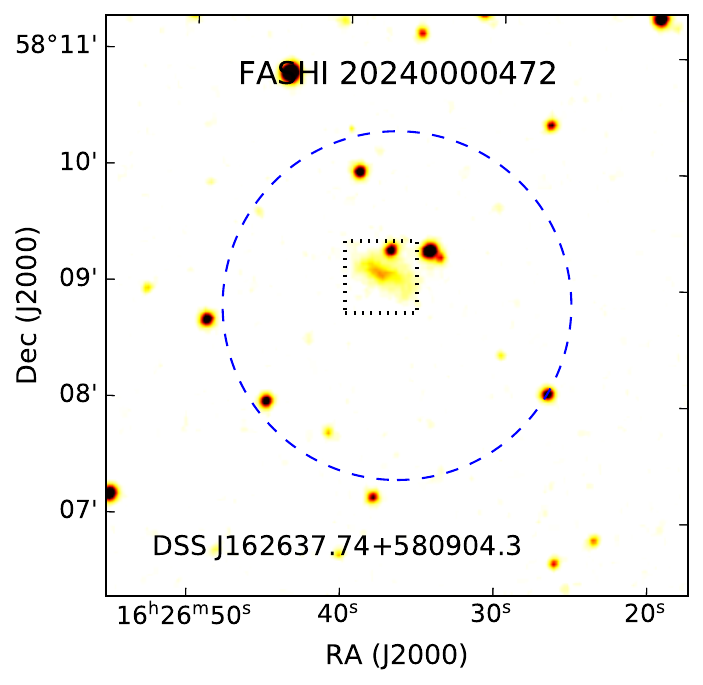}
 \caption{21\,cm H\,{\scriptsize{I}} absorption galaxy FASHI\,162636.67+580849.8 or ID\,20240000472.}
 \end{figure*} 

 \begin{figure*}[htp]
 \centering
 \renewcommand{\thefigure}{\arabic{figure} (Continued)}
 \addtocounter{figure}{-1}
 \includegraphics[height=0.22\textwidth, angle=0]{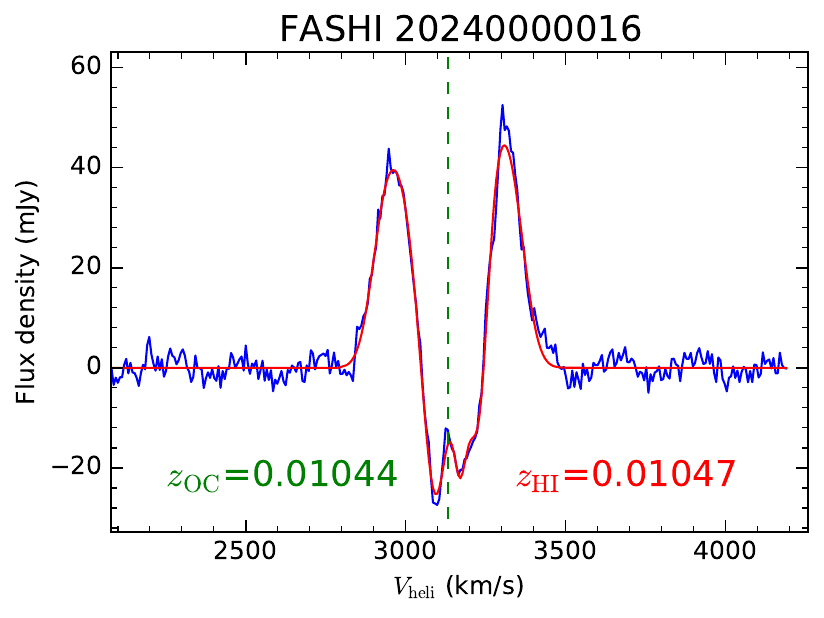}
 \includegraphics[height=0.27\textwidth, angle=0]{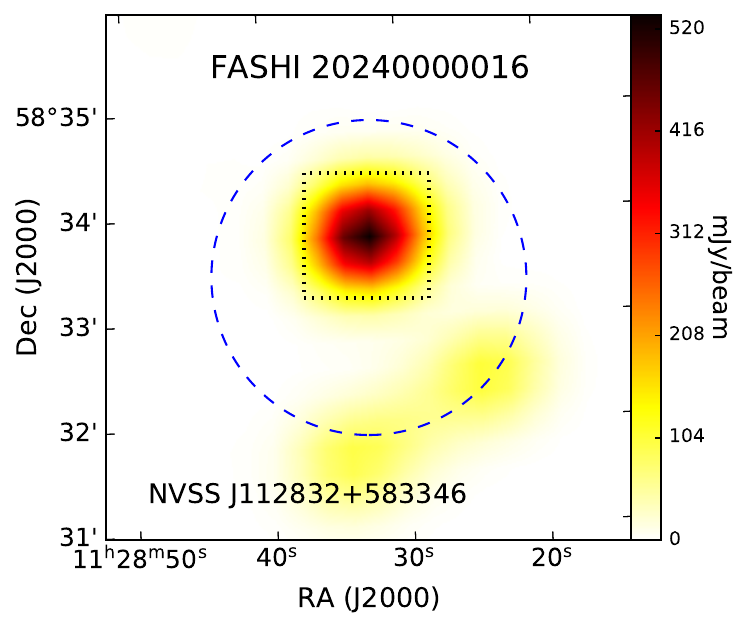}
 \includegraphics[height=0.27\textwidth, angle=0]{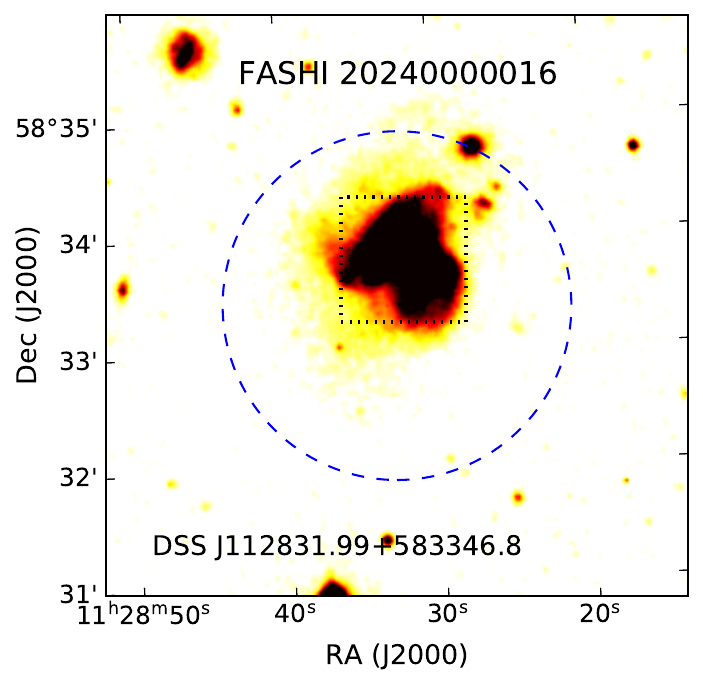}
 \caption{21\,cm H\,{\scriptsize{I}} absorption galaxy FASHI\,112832.57+583323.0 or ID\,20240000016.}
 \end{figure*} 

 \begin{figure*}[htp]
 \centering
 \renewcommand{\thefigure}{\arabic{figure} (Continued)}
 \addtocounter{figure}{-1}
 \includegraphics[height=0.22\textwidth, angle=0]{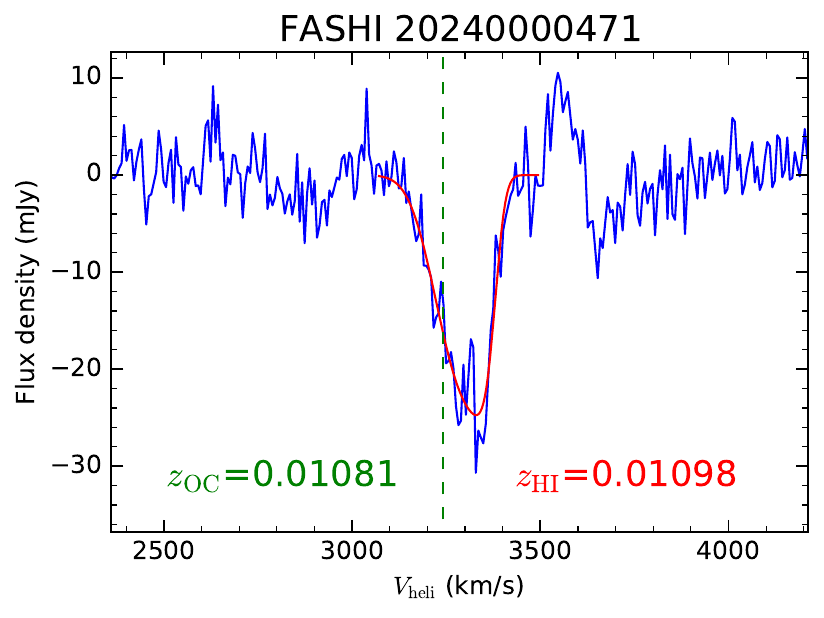}
 \includegraphics[height=0.27\textwidth, angle=0]{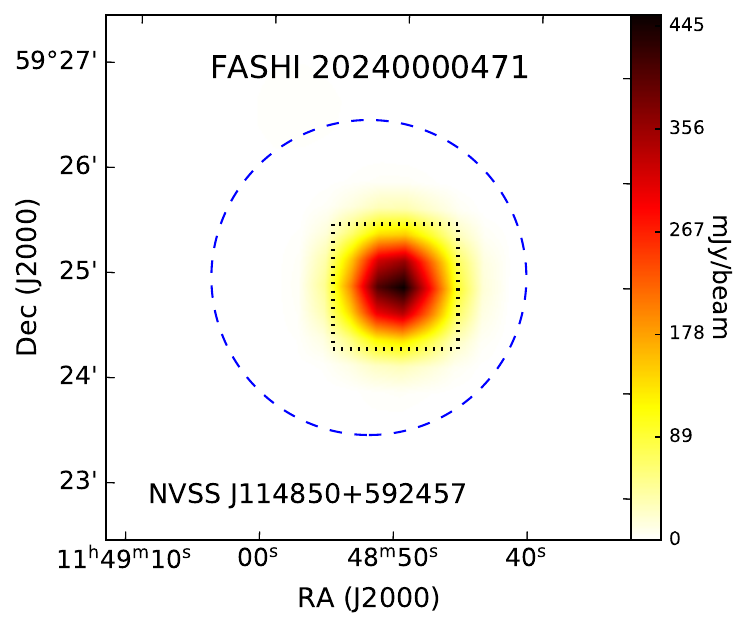}
 \includegraphics[height=0.27\textwidth, angle=0]{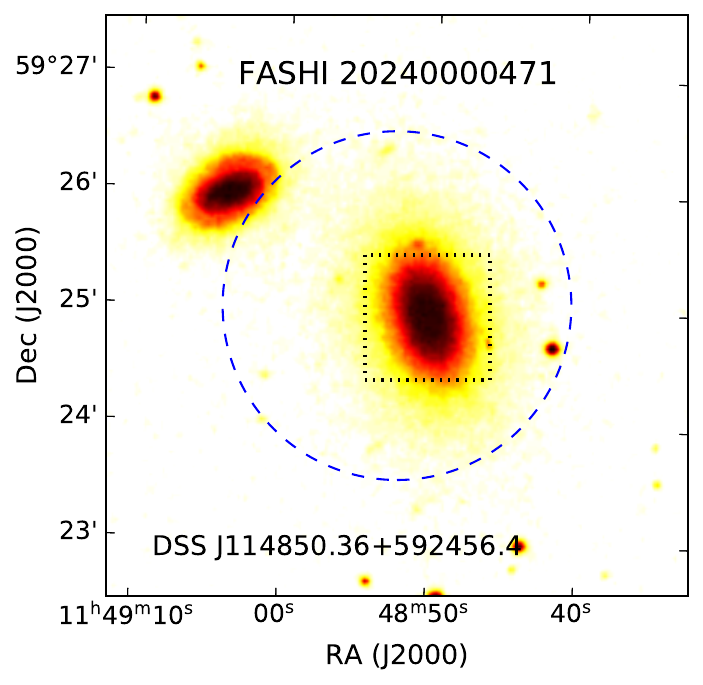}
 \caption{21\,cm H\,{\scriptsize{I}} absorption galaxy FASHI\,114852.43+592501.9 or ID\,20240000471.}
 \end{figure*}

 \clearpage

\end{document}